\documentclass[twocolumn,nofootinbib,aps,prc,amsmath,amssymb,10pt,floatfix,superscriptaddress,floatfix,letterpaper,fleqn]{revtex4-1}

\usepackage{amssymb}
\usepackage{amsfonts}
\usepackage{amsmath}
\usepackage{makeidx}
\usepackage[bookmarksnumbered,colorlinks,bookmarks,breaklinks=true,linkbordercolor={1 1 1},linktocpage,citecolor=blue]{hyperref}
\usepackage{graphicx}
\usepackage{subfigure}
\usepackage{dcolumn}
\usepackage{bm}
\usepackage{fancyheadings}
\usepackage{floatflt}
\usepackage{xcolor}
\usepackage[activate={true,nocompatibility},final,tracking=true,kerning=true,spacing=true,factor=1100,stretch=10,shrink=10]{microtype}
\usepackage{mathrsfs}
\usepackage{supertabular}
\usepackage{array}
\usepackage{multirow}
\usepackage{enumitem}
\usepackage{tabularx}
\usepackage{longtable}
\usepackage{mathrsfs}

\newcommand\T{\rule{0pt}{2.6ex}}       % Top strut
\newcommand\B{\rule[-1.2ex]{0pt}{0pt}} % Bottom strut
\def\etal{\textit{et al.~}}
\def\etals{\textit{et al.}}
\def\ql{``}
\def\qr{''\hspace{0.5mm}}
\def\qrs{''}

\def\a{\color{blue}}
\def\r{\color{red}}
\def\b{\color{black}}
\def\CF{$^{252}$Cf(sf)}
\def\TH{$^{232}$Th}
\def\UT{$^{235}$U}
\def\UF{$^{238}$U}
\def\PU{$^{239}$Pu}
\def\nub{$\overline{\nu}_{\mathrm{tot}}$}
\def\USU{\textit{USU}~}
\def\USUn{\textit{USU}}
\def\GMA{\textit{GMA}}

\newcommand{\eg}{\textit{e.g.}}
\newcommand{\ie}{\textit{i.e.}}

\newcolumntype{P}[1]{>{\centering\arraybackslash}p{#1}}

\renewcommand{\vec}[1]{\boldsymbol{\mathbf{#1}}} % FG 20190711 change of \vec for bold symbol and latin
\newcommand{\eqnarrow}{\setlength{\mathindent}{3mm}}
\newcommand{\eqnormal}{\setlength{\mathindent}{10mm}}

\newcommand{\var}{\text{var}}

\newcommand{\E}{\text{E}}
\newcommand{\der}{\text{d}}
\def\nuc#1#2{\relax\ifmmode{}^{#1}{\protect\text{#2}}\else${}^{#1}$#2\fi} % from elsevier, for example \nuc{233}{U}
\newcommand{\ignore}[1]{}

\begin{document}
\title{Unrecognized Sources of Uncertainties (\USUn) in Experimental Nuclear Data}

\author{R.~Capote}\email[Corresponding author: ]{r.capotenoy@iaea.org}
\affiliation{NAPC--Nuclear Data Section, International Atomic Energy Agency, A-1400 Vienna, Austria}

\author{S.~Badikov}
\affiliation{Atomstandard, Rosatom State Corporation, Moscow, Russia}

\author{A.~Carlson}
\affiliation{National Institute of Standards and Technology, Gaithersburg, MD 20899, USA}

\author{I.~Duran}
\affiliation{Universidade de Santiago de Compostella, 15782 Santiago de Compostela, La Coruna, Spain}

\author{F.~Gunsing}
\affiliation{CEA Irfu, University Paris-Saclay, F-91191 Gif-sur-Yvette, France }

\author{D.~Neudecker}
\affiliation{Los Alamos National Laboratory, Los Alamos, NM 87545, USA}

\author{V.G.~Pronyaev}
\affiliation{Contractor, NAPC--Nuclear Data Section, International Atomic Energy Agency, A-1400 Vienna, Austria}
%\affiliation{Atomstandard, Rosatom State Corporation, Moscow, Russia}

\author{P.~Schillebeeckx}
\affiliation{Joint Research Centre - Geel, 2440 Geel, Belgium}

\author{G.~Schnabel}
\affiliation{Uppsala University, Lagerhyddsvagen, 75120 Uppsala, Sweden}

\author{D.L.~Smith}
\affiliation{Argonne National Laboratory, 1710 Avenida del Mundo, No. 1506, Coronado, CA 92118, USA}

\author{A.~Wallner}
\affiliation{Australian National University, Canberra ACT, Australia}

\received{9 August 2019}

\begin{abstract}
\vspace{2mm}
Evaluated nuclear data uncertainties reported in the literature or archived in data libraries are often perceived as unrealistic, most often because they are thought to be too small. The impact of this issue in applied nuclear science has been discussed widely in recent years. Commonly suggested causes are: poor estimates of specific error components, neglect of uncertainty correlations, and overlooked known error sources. However, instances have been reported where very careful, objective assessments of all known error sources have been made with realistic error magnitudes and correlations provided, yet the resulting evaluated uncertainties still appear to be inconsistent with observed scatter of predicted mean values. % It has been suggested that 
These discrepancies might be attributed to significant unrecognized sources of uncertainty (USU) that limit the accuracy to which these physical quantities can be determined, and %that 
in some way they need to be incorporated into the evaluation procedures. The objective of our work has been %to investigate this topic in considerable detail and 
to develop procedures for revealing and including USU estimates in nuclear data evaluations involving experimental input data. This effort led to our suggesting that the presence of USU indeed may be revealed, and estimates of magnitudes made, through quantitative analyses.  This paper identifies several specific clues that can be explored by evaluators in identifying the existence of USU. It then describes some numerical procedures we have introduced to generate quantitative estimates of USU magnitudes. Key requirements for these procedures to be viable are that sufficient numbers of data points be available, for statistical reasons, and that additional supporting information about the measurements be provided by the experimenters. Several realistic examples are described here to illustrate these procedures and demonstrate their outcomes. Limitations of these procedures are also discussed and illustrated by realistic examples. Our work strongly supports the view that USU is an important issue in nuclear data evaluation, with significant consequences for applications, and that this topic warrants further investigation by the nuclear science community.
\newpage
\end{abstract}

%\topmargin 50mm
%\vspace*{50mm}
\vspace*{40mm}
\maketitle

%\clearpage
%\addtocontents{toc}{\protect\newpage}
\tableofcontents

\vspace*{9mm}
\lhead{Unrecognized sources of uncertainties...}
\chead{NUCLEAR DATA SHEETS} \rhead{R.~Capote \textit{et al. }} \lfoot{} \rfoot{} %

%\newpage
\section{INTRODUCTION}

Gathering and analysis of information with the objectives of reaching conclusions and generating recommendations is a characteristically human endeavor. However, flawed outcomes attributable to inadequate or incomplete information, or to distortions arising from biased treatment of the data, unfortunately are far too common.

The Greek philosopher Aristotle (384{\color{red}{--}}322 B.C.E)~\cite{Aristotle} expressed the view that {{\textit{... a mark of the educated man is to demand accuracy only to the degree that the subject matter permits ...}} . Over 2,300 years later, humans still grapple with finding the best objective, practical ways to achieve this goal in various circumstances.

Although in some dispute, C.F.~Gauss (1777{\color{red}{--}}1855)~\cite{Gauss} supposedly first proposed the concept of un-weighted least-squares estimation as a means to provide the best quantitative value for a single variable from a collection of many representative values.
%This idea came to him when he was 18 years old. He employed the method in celestial calculations that were reported when he was 24 years old.
The evolution of quantitative data evaluation (\ie, uncertainty quantification -- UQ) methodologies from Gauss's time onward has tracked the development of statistical methods to a considerable degree. The literature is rich with expositions of various proposed methods and applications in many fields, \eg, refer to~\cite{Smith:1991}.

This paper provides a brief overview of contemporary evaluation methodology, including complete UQ. The emphasis is on deficiencies related to frequent insufficiency of required input information, with the possible need for considering unrecognized sources of uncertainties in nuclear data experiments. Some methods for compensating for these deficiencies are suggested and examples are provided in this paper.

%\section{UNCERTAINTIES IN METROLOGY AND NUCLEAR DATA EVALUATION}\label{sect:metrology}
\section{UNCERTAINTIES IN DATA EVALUATION}\label{sect:metrology}
%\subsection{\r Estimating the weighted mean from observables\b }

\subsection{Parameter Estimation}\label{ssect:param}
This section offers a brief overview of the fundamental principles of statistical parameter estimation.

A set of measured observables $\vec{y}$ with reported measurement
uncertainties $\vec{u}$ is assumed to be described by a model
function $f(\vec{x}; \vec{\theta})$ of independent variables
$\vec{x}$ and parameters $\vec{\theta}$. The statistical model that applies
for this case can be written as
\begin{equation} \label{eq:model0}
    y_i = f(x_i ; \vec{\theta}) + \varepsilon_i + \eta_i
\end{equation}
where $\varepsilon_i$ and $\eta_i$ are realizations of random variables,
uncorrelated between $i$ for $\varepsilon_i$, and correlated for $\eta_i$.
The sum  $\varepsilon_i + \eta_i$ is again a random variable, often, but
not necessarily, considered to be jointly normally distributed, with
mean zero and a combined covariance matrix $\vec{C}$, and $f(x_i ; \vec{\theta})$
is the calculated equivalent of the experimental value $y_i$.

The likelihood function $\mathcal{L}(\vec{\theta}|\vec{y},\vec{C})$,
of the model parameters $\vec{\theta}$, given the observed data
$\vec{y}$ and the covariance matrix $\vec{C}$, is the joint
probability density function of $\vec{\varepsilon} + \vec{\eta}$.
Multiplied by the prior probability $P(\vec{\theta})$, and after
normalization, it is equivalent to the Bayesian posterior distribution.
The maximum of the likelihood $\mathcal{L}$ corresponds to those parameters
$\vec{\hat{\theta}}$ that make the data $\vec{y}$ most
probable.

A non-informative prior $P(\vec{\theta}) = 1$ is often used in
adjustment procedures where all data are used simultaneously. Inclusion of prior
information is an important aspect of the procedures commonly used in nuclear data evaluation, where
datasets are added to a previous adjustment procedure. In such a case
the posterior of the previous dataset evaluation can be used as a prior.
Then, only the new dataset, in combination with this prior, is employed
in forming the new posterior function that defines the revised evaluation..

The maximum likelihood method can be used to estimate any parameter in a
model describing the data, including parameters in the covariance
matrix. In a frequently used case, we want to estimate only parameters of
a model (or proxy) function. Then in the case of Gaussian distributions
for $\varepsilon_i$ and  $\eta_i$, the likelihood function $\mathcal{L}$ is proportional to
$\exp(-\chi^2/2)$, with
\begin{equation} \label{eq:chi2}
 \chi^2 = \left( \vec{y} - \vec{f(\vec{x} ; \vec{\theta})} \right)^\textnormal{T} \vec{C}^{-1}  \left( \vec{y} - \vec{f(\vec{x} ; \vec{\theta})} \right)\,.
\end{equation}
Bold $\vec{f}$ is a vector generated by model function $f$.
Finding those values $\vec{\hat{\theta}}$ of the parameters
$\vec{\theta}$ which maximize the likelihood $\mathcal{L}$ is the same as
minimizing $\chi^2$. This minimum is found by solving
$\der{\chi^2}/\der{\vec{\theta}} = 0$, and the lower bound of its
variance, if the estimator $\vec{\hat{\theta}}$ is unbiased, is
given by the expectation value % $1/\E\{\der^{2}\chi^2 /\der\vec{\theta}^2 \} $.
$\E\{-\frac{1}{2}(\der^{2}\chi^2 / \der\vec{\theta}^2) ^{-1} \}$.
%If the expectation value is difficult to
%determine, the square root of the variance can also be found by searching
%for the value where the logarithm of the likelihood $\mathcal{L}$ drops to half its
%maximum. This corresponds to the standard deviation in case of a normal
%distribution.

If the likelihood function $\mathcal{L}$ is a multivariate Gaussian, and in addition
the model function $f(\vec{x};\vec{\theta})$ is linear in its parameters
$\vec{\theta}$, the chi-square minimization becomes the method of {\it
linear least squares minimization} \cite{Smith:1991}. Equating the first derivatives to
zero, as required to find the maximum likelihood solution
$\vec{\hat{\theta}}$, results in a solvable set of equations with a
unique solution.

% FG20190711: general case for sect:linear, but too much distraction here
%
%%%   \begin{equation}
%%%      \vec{\hat{\theta}} = \left( \vec{D}^\textnormal{T}\vec{C}^{-1}\vec{D} \right)^{-1}  \vec{D}^\textnormal{T} \vec{C}^{-1} \vec{y}
%%%   \end{equation}
%%%   where the matrix of derivatives or design matrix is defined as
%%%    $\vec{D} = \partial f(\vec{x} ; \vec{\theta})/\partial \vec{\theta}$.
%
Also the second derivatives %, needed to obtain the variances of $\vec{\hat{\theta}}$,
can be obtained in a way that lends itself readily to
straightforward computations, resulting in the covariance matrix $\vec{C_\theta}$ of the fitted
parameters.
%
%%%   \begin{equation}
%%%      \vec{C_\theta} = \left( \vec{D}^\textnormal{T} \vec{C}^{-1} \vec{D} \right)^{-1}
%%%   \end{equation}
%
Further simplifications are possible if $\vec{C}$ is a
diagonal matrix (uncorrelated uncertainties) since derivation of
the inverted covariance matrix $\vec{C}^{-1}$ is trivial.

\subsection{Estimation of the Mean, Outliers and Uncertainties}
In this section, we forgo the more general treatment of Sect.~\ref{ssect:param} in order to illustrate certain important mathematical features that are encountered in parameter estimation.

A case that is frequently encountered in practice is the estimation of a single physical
quantity $\mu$ from a collection of measured values $y_i$ with reported
uncertainties $u_i$. Here the model function does not depend on $\vec{x}$ and
the parameters $\vec{\theta}$ consist of a single parameter $\mu$, \ie, $f(\vec{x};
\vec{\theta}) \equiv \mu$. Assuming no correlations in the measurement
uncertainties, \ie, $\eta_i = 0$, the statistical model describing the
relation between the observations $y_i$ and the real true value $\mu$ then assumes the following
simple form:
\begin{equation} \label{eq:model1}
    y_i = \mu + \varepsilon_i \,.
\end{equation}
If normal distributions $\mathcal{N}(0,u^2_i)$ are assumed for the random
variables $\varepsilon_i$, we can immediately obtain  an estimate
$\hat{\mu}$ for $\mu$ by equating the derivative of
Eq.~\eqref{eq:chi2} to zero. This estimate is the well-known
weighted mean $\hat{\mu} = \frac{\sum y_i/u_i^2}{\sum 1/u_i^2}$. For the
variance we obtain $\var{(\hat{\mu})} = \frac{1}{\sum 1/u_i^2}$. And if all
$n$ values of $u_i$ are equal to the same single value $u$, these
expressions become $\hat{\mu} = \sum y_i/n$, and  $\var{(\hat{\mu})} =
u^2/n$. Since this estimator is biased, usually the unbiased estimator
$\var{(\hat{\mu})} = u^2/(n-1)$ is used.

In both cases the variance of the mean tends to zero if $n$ tends to infinity. For a non-diagonal
matrix $\vec{C}$, analytical expressions can still be found for
$\hat{\mu}$ and $\var{(\hat{\mu})}$, but solving
Eq.~\eqref{eq:chi2} by numerical means is often less cumbersome.

When the degree of correlation exceeds a critical value, depending on
the data, it may cause the well-known effect that the resulting mean
lies outside the data interval and its uncertainty decreases, see for
example \cite{cox:2006}. In the field of nuclear data this phenomenon is also known
as Peelle's Pertinent Puzzle (PPP) \cite{Peelle:1988}.  This effect is
mainly related to an incorrectly constructed covariance matrix from
measurement observables, which may suggest that the used statistical
model is not adequate to describe the data
\cite{INDC192,D_Agostini:1994,Froehner:2003,Neudecker:2012}. As a practical matter, in nuclear data evaluation physically unreasonable evaluated results may be generated by the least-squares method in certain extreme cases (strong correlations and discrepant input data) if no compensation for PPP is applied. % Note that the PPP is discussed in more detail in Sect.~\ref{sect:defic}.

Several comments are worthwhile mentioning: The expectation value
for the $\chi^2$ quantity of Eq.~\eqref{eq:chi2} equals the number of degrees of freedom $n-1$.
Therefore, a strong deviation of $\chi^2$ from $n-1$, especially for larger
values of $n$, is a strong indication that either the uncertainties
$u_i$ are over- ($\chi^2/(n-1) \ll 1$) or under-estimated ($\chi^2/(n-1) \gg 1$). The
expectation value of $\chi^2$ is also the basis of a method suggested in this work to
account for unrecognized uncertainties, either by
scaling the measurement uncertainties $u_i$ by a common factor \cite{Tanabashi:2018}, or by
adding in quadrature a common uncertainty, in such a way that $\chi^2/(n-1)$
is artificially forced to become 1. A more frequently used approach is to split the distributions $\varepsilon$ into groups that can be parameterized by a normal distribution directly from experimental uncertainties $u_i$.
As an example we cite Ref.~\cite{Badikov:2012}, where such a procedure is used to describe one-parameter discrepant
data.

A second point to consider is the fact that the variances of the
distributions of $\varepsilon^2_i$ and $\eta^2_i$, or more generally, the
elements of the covariance matrix $\vec{C}$, are not known. Commonly
the measurement uncertainties $u^2_i$, which are usually the only
available estimates, correspond only to the diagonal elements, but these
estimates are biased, as pointed out in, for example Ref.~\cite{Zhang:2006}.
The unbiased estimate $u^2_i (n-1)/(n-3)$ can be used
instead, so the difference is sizeable only for small $n$. Nevertheless,
it is important to be aware that in practice the covariance matrix
$\vec{C}$ can serve only to approximate the statistical information of the underlying joint
probability distribution since the higher moments of the probability density function are neglected.

But the most important issue to stress is that experimental datasets
considered for nuclear data evaluation are often not statistically
ideal. The process of UQ is usually not straightforward. Also, not all uncertainties (known or unknown)
are reported. In addition $u$ may vary over the years,
between laboratories, or with respect to the used experimental methodologies. However, it has been demonstrated that in specially
designed interlaboratory comparisons, often for single physical
quantities, such conditions can be optimized and mastered
\cite{Kossert:2004,Rukhin:2009,Zimmerman:2012,Ratel:2015,Lepy:2015}.

Before undertaking a nuclear data evaluation, it is of utmost importance
that values, uncertainties and correlations are adjusted and corrected
where possible and warranted. The identification of outliers is an essential part of
such a procedure. For a single quantity, outliers are usually defined by considering
a standard score which is a relative difference between $y_i$ and $\hat{\mu}$. The
ratio $ r_i = (y_i - \hat{\mu}) / \sqrt(u^2_i + \var{(\hat{\mu})}) $,
see for example Ref.~\cite{James:1992} and similar definitions
\cite{Willink:2002,Steele:2005,Ellison:2018}, is easy to use. A score
beyond this definition for an outlier must be documented if
values are discarded. Only when data are still inconsistent after rejecting such outlier data points does it make sense to search for hidden or unrecognized uncertainties.

\section{DEFICIENCIES IN CONTEMPORARY DATA EVALUATION}\label{sect:defic}

While the mathematical methods (algorithms and formulas) used in conventional metrology for establishing best values and uncertainties are similar to those used for nuclear data evaluation work, \eg, as discussed in Sect.~\ref{sect:metrology}, and they are both based on discerning properties of underlying probability distributions, the conceptual underpinnings of the work is quite different in these two fields. Conventional metrology, for the most part, assumes that the statistical properties of data sets are to be based entirely on objective observations, in particular on sufficiently many observations, under tightly controlled conditions, so that the populations from which the statistical samples are drawn can be assumed to be stable and the usual rules of statistics to be fully applicable. Adherence to the requirement of fulfilling such conditions is an approach often labeled as Frequentist. On the other hand, nuclear data evaluators rarely have the luxury of enjoying such ideal conditions. They must rely on databases that may involve both observational and theoretical input information as well as limited sample sizes, incompleteness, and indirect information. This real world situation in data evaluation is consistent with a conceptual approach to statistics known as Bayesian, \ie, one where the worth and quality of available information is assessed in terms of perceived degrees of confidence.

Evaluation techniques employed by the nuclear data community tended to be relatively unsophisticated until around the mid-1970's when motivation to improve these methods was stimulated by the need for the nuclear energy and radiation dosimetry data user communities to satisfy emerging stricter quantitative control of cost, reliability, and safety factors. There are two fundamental aspects of contemporary nuclear data evaluation that must be examined in this context: evaluation methods and features of the included input data to be evaluated. Consequences of deficiencies in both areas must be considered. These issues are discussed briefly in this paper.

\subsection{Method Deficiencies}

In addition to the discussion in Sect.~\ref{sect:metrology}, expositions on nuclear data evaluation methodology during preceding decades can be found in a monograph by Smith~\cite{Smith:1991}, in review papers by Capote \etal\cite{Capote:2010} and by Smith and Otuka~\cite{Smith-Otuka:2012}, and elsewhere, \eg, as mentioned in a comprehensive report of the NEA WPEC SG-24~\cite{WPEC-SG24}.

It was not unusual for evaluated data sets in the early years of nuclear data evaluation (prior to the 1970's) to consist of numbers read visually from smooth eye guides drawn through experimental data points that were hand plotted on graph paper, with limited consideration given to their relative uncertainties (if such information was even available). No statistical analyses were incorporated in producing such subjective results.

The most commonly employed nuclear data evaluation approach in use today is the generalized least-square method, \eg, see~\cite{Smith:1993,GLUCS,KALMAN,Rising:2013,GANDR}. One manifestation of the mathematical formulation of this method is summarized briefly in Sect.~\ref{sect:metrology} of this paper, so this will not be repeated in this section. A variant of this approach to fitting smooth curves to data that is based on Pad\'e polynomials~\cite{Pade:1892,Graves-Morris:1973,Baker:1975} has also been employed, especially in Russia, \eg, see~\cite{Vinogradov:1987,Badikov:1992,Hermanne:2018}. This method is described further in Sect.~\ref{sect:pade-res}.

More recently, stochastic evaluation methods that rely on Monte Carlo sampling have been suggested and investigated by several researchers in the nuclear data field, \eg, see UMC-G~\cite{UMCG,UMCG1}, BFMC~\cite{BFMC}, UMC-B~\cite{UMCB}, and BMC~\cite{BMC}. For the reason mentioned above, nowadays, nuclear data evaluation is largely based on Bayesian approaches that enable simultaneous consideration of both model-predicted values (if needed for completeness) and experimental data, as well as correlations among all these data. Many usages of these evaluation techniques have been published, including applications to large amount of data (\eg, see Refs.~\cite{HFB-mass,TENDL-2017}). Practical implementation of these sophisticated quantitative evaluation procedures has been facilitated by the rapid growth of readily-accessible, inexpensive computer power following Moore's Law~\cite{Moore:1965}.

One least-squares approach that appeared in 1981 originated from the work of Poenitz~\cite{Poenitz:1981,PoenitzAumeier:1997}. It was directed primarily towards evaluation of Neutron Cross-section Standards, which is based on a comprehensive experimental database. In addition to introducing a method that relies on weighted least-squares averaging, including correlations, Poenitz developed the original version of computer code \GMA~to fit all types of cross sections (absolute and shape), their ratios, spectrum-averaged cross sections and thermal constants in one full analysis. The \GMA~code and its accompanying database of archived experimental values are applicable to evaluation of the Neutron Cross-section Standards. The original code \GMA~was utilized to generate the ENDF/B-VI Standards~\cite{ENDFB-VI-standards,Carlson:1993}. This computational capability has been updated over the ensuing years and it remains in use today. The \GMA~input database continues to expand steadily as new and updated experimental data become available. A revised version of \GMA~(denoted by \GMA\textit{P}) was developed by V.G. Pronyaev~\cite{Pronyaev:2003} at the IAEA.
%An empirical ``fix'' (algorithm) that compensates for PPP in a practical way was suggested by Chiba and Smith~\cite{Chiba-Smith:1991}. This algorithm has been incorporated in the code \GMA\textit{P}~\cite{Pronyaev:2003}.
It has been employed in performing the two most recent widely adopted evaluations of the neutron cross-section standards library, the IAEA Standards 2006~\cite{Badikov:2007} (adopted as ENDF/B-VII Standards~\cite{Badikov:2007,Standards:2009}) and the IAEA Standards 2017~\cite{Standards:2018}, which were adopted as ENDF/B-VIII Standards~\cite{ENDFB-VIII:2018}. %The IAEA 2017 standards evaluation is included as a component of the U.S. evaluated nuclear data library ENDF/B-VIII.0~\cite{ENDFB-VIII:2018}.

An empirical ``fix'' (algorithm) that compensates for PPP (discussed in Sect.~\ref{sect:metrology}) in a practical way was suggested by Chiba and Smith~\cite{Chiba-Smith:1991}. This algorithm was incorporated in code \GMA\textit{P}. Since this revised version has been used exclusively in the Neutron Standards evaluation work since 2003, henceforth in this paper the code and accompanying database will be referred to by the new name \GMA\textit{P}. However, the reader should be aware that the name \GMA~is still widely used by some individuals in reference to even the revised version.
%Several modifications to the basic least-square method (to address mathematical issues that were found to influence the outcomes of evaluations in a negative way) have been investigated and implemented as the method became widely used, \eg, see~\cite{INDC192}. One such refinement addresses an effect known as Peelle's Pertinent Puzzle (PPP)~\cite{Peelle:1988}. Physically unreasonable evaluated results may be generated by the least-squares method in extreme cases (strong correlations and discrepant input data) if no compensation for PPP is applied. The PPP effect is also discussed in Sect.~\ref{sect:metrology} of this paper.
As is evident from the discussion in Sect.~\ref{sect:metrology}, the least-squares method is an inherently linear formalism that can generate significantly biased results when large input data uncertainties are involved and there exist non-linear relationships between the input data and derived values for observable quantities, \eg, as may happen when ratio data are included~\cite{UMCG}. Suggestions for dealing with nonlinear effects within the framework of Bayesian evaluation theory are being explored, but they have been implemented only in a few cases, \eg, see~\cite{UMCB}.

In most practical situations, most contemporary evaluation methods yield acceptable results if adequate, reliable input data are available, \ie, when modest uncertainties and mostly non-discrepant data are incorporated. So, deficiencies in evaluation methodology are not the most serious concerns facing contemporary nuclear data evaluators.

\subsection{Data Deficiencies}

The most pressing concern faced by contemporary nuclear data evaluators is how to deal with inadequacy of the input data to be evaluated, both with respect to reliability and completeness. This unfortunate situation persists in spite of considerable effort expended by the nuclear science community over many decades to acquire and compile comprehensive databases of experimental values~\cite{EXFOR}. While this concern encompasses both theoretical and experimental information, the focus in this paper is on experimental data.

Experimental input data used in evaluations need to be statistically consistent (\ie, largely free of problematic outliers), and generated from analyses of measured results that incorporate accurate and inclusive modeling of the relationships between the quantities that actually are measured and those that are sought to be obtained indirectly from the raw data (derived results). Estimated uncertainty magnitudes and reported correlations need to be reasonable as well as comprehensive. Data provided by experimenters, and archived in original form in publications or computer-accessible data repositories, \eg, in EXFOR~\cite{EXFOR}, are considered to be raw data. Raw data usually need to be adjusted to be consistent with contemporary standards and fundamental nuclear properties such as decay parameters \cite{ENSDF,Livechart,NUDAT} (half lives and branching ratios, \textit{etc}.). Additionally, a comprehensive UQ exercise should be undertaken. This is a point that will be stressed in the whole paper.
As mentioned in Sect.~\ref{sect:metrology}, rarely can experimental data published by original authors be accepted by evaluators directly as reported, especially if many years have transpired since the measurements were performed. Consequently, it is the responsibility of evaluators to revise the input data used in their evaluations, as needed, and to insure that the information is comprehensive. Otherwise, the results of these evaluations are likely to be flawed and misleading.

While evaluators are limited in what they can do if they consider only reported experimental data uncertainties, which often are incomplete, progress can be made if they are able to supplement this information with estimates of required uncertainties and correlations obtained from templates of suggested values generated through the collective experiences of other experimenters and evaluators in the field, \eg, see~\cite{Schillebeecks:2012,Helgesson:2015,Helgesson:2017,Neudecker:2018,Neudecker_Template:2019}. This may lead to an evaluator choosing either to alter the uncertainties provided in the literature for certain data or to outright reject those data that, in their opinion, are of such poor quality (\eg, seriously discrepant) that their inclusion in the evaluation process would lead to unacceptable distortions in the evaluated results (\eg, see the case of Staples data~\cite{Staples:1995} in \PU(n,f) Prompt Fission Neutron Spectra (PFNS) evaluation~\cite{Neudecker:2014}).

In evaluating the IAEA Standards (2009)~\cite{Badikov:2007,Standards:2009}, a reasonable value for the evaluated solution chi square per degree of freedom was obtained by including outlying uncertainties with medium energy range correlation and increasing the uncertainties of apparent outlier data points based on the observed differences between the actual data points and the prior (assumed to be the previously evaluated mean value). These adjustments to the uncertainty values were made whenever the differences from the evaluation were more than two sigma for a single point or more than one sigma for two or more consecutive energy points. The changes in the cross-section mean values resulting from this procedure were small because the discrepant points usually exhibited large uncertainties and therefore had limited influence on the evaluated mean values.

\subsection{Consequences of Deficiencies}

A specific concern that emerged a while ago (for Neutron Standards this issue was raised at CSWEG in 1991) relates to the topic of unrealistically low evaluated uncertainties.
Strong reduction of uncertainties was observed for R-matrix model fits, especially for cross-section data  fitted with small numbers of parameters (uncertainties as small as 0.02\% were observed for $^6$Li(n,t) in the 1/v energy range). Expert assessment based on assumed normal probability population statistics (the so called \ql 2/3 rule\qrs) was used to increase uncertainties. This issue was addressed explicitly in papers by Gai~\cite{Gai:2007,Gai:2007a} and by Badikov and Gai~\cite{Badikov:2003}. Of particular concern is the impact of this phenomenon on the standards evaluations~\cite{Badikov:2007,Standards:2009,Standards:2018,ENDFB-VIII:2018}.
Pronyaev~\cite{Pronyaev:2003a} has explored the matter of too-low uncertainties that appears to arise when model fitting is used in evaluating data rather than basing evaluations solely on experimental data. The concern for too-low uncertainties has also been raised in several papers presented at four workshops during the past decade that were devoted to nuclear data covariances~\cite{CW2008,CW2011,CW2014,CW2017}, as well as being discussed in numerous additional reports, theses, conference proceedings, and published papers. While it remains an open area of research, an emerging consensus view in the nuclear data community is that this effect is primarily related to deficiencies in the input data rather than to evaluation methods, provided that evaluators incorporate the available capabilities provided by these methods properly. This phenomenon can be attributed to under-estimation of experimental data uncertainties (for various reasons) and, in many instances, also to inadequate consideration of uncertainty data correlations~\cite{Neudecker_PUB:2018,Neudecker:2013}.

What harm can be done if evaluations generate small uncertainties? Of course there is no harm, only benefits, as long as these predicted uncertainties are realistic and truly reflect the quality of the underlying data being evaluated. Otherwise, if the estimated uncertainties are indeed too small, there is a danger that users of these evaluated data will be misled into thinking that the quality of the provided information is better than it really is. This can have serious consequences for analyses of cost, reliability, and safety when these data are employed in modeling nuclear systems and/or in making predictions. Of course, if the evaluated uncertainties are too large relative to the uncertainties of the underlying data, this too can have important implications. Then, it may happen that costly and time consuming effort might be devoted to improving the quality of the underlying data when it would not be warranted if the evaluation had produced results that represented the data properly. Clearly, the goal of proper data evaluation is to produce {\textit{realistic}} results that represent the available underlying data well. As mentioned above, we should be reminded of Aristotle's observation that was expressed long ago~\cite{Aristotle}.

\section{UNRECOGNIZED SOURCES OF UNCERTAINTIES (\USUn)}\label{sect:USU}

It can happen that no matter how hard evaluators attempt to apply rigorous evaluation procedures faithfully, and to track down, adjust, and include all the different sources of uncertainty known to them (including their correlations) for the various experiments considered, the resulting evaluated uncertainties may still appear to be inconsistent with the input data, and thus will be perceived as unacceptable by data users. Most often, these evaluated uncertainties are perceived as being too small, as mentioned above.

The present investigation focuses on scenarios where concerns for too-small uncertainties arise. Then, additional, non-specific sources of uncertainty may need to be postulated, and values estimated and introduced into the evaluation process, in order to address such perceived deficiencies in these evaluations. In this paper they are referred to by the term \underline{U}nrecognized \underline{S}ources of \underline{U}ncertainties (\USUn), \eg, see~\cite{ENDFB-VIII:2018}. It is important to understand that when the term \USU is mentioned by itself in this paper, it refers to the concept rather than to a specific uncertainty value. If it is the value that is being discussed, then it should be stated that it is the "magnitude of the \USU contribution". Also, note that we prefer to use ``\underline(S)ources'' rather than ``\underline(S)ystematic'' in the acronym \USUn, since both correlated or uncorrelated unrecognized sources of uncertainty have been identified as elaborated below, although the reader needs to be aware that the term \USU frequently has been used by other authors solely in the context of systematic uncertainties. This issue will arise if an evaluator can think of no known sources of uncertainty that could be included to remedy the problem of too-small evaluated mean-value uncertainties, or when an evaluator concludes that other possible known sources of uncertainty are simply too difficult to estimate objectively in the given circumstances. Here, we should mention another insightful observation from Aristotle~\cite{Aristotle}: {\textit{... Only the man on the spot can decide whether the bread is properly cooked ...}} . So, \USU are those which must be added to the uncertainty budget of an evaluation when no other practical options seem to be available to the evaluator to permit reasonable evaluated uncertainties to be generated based on consideration of known uncertainty sources. Of course, the challenge is to determine how this should be accomplished in a reasonable and consistent manner.

\subsection{\USU Historical Background}

As mentioned before, there has been a long history of criticism of nuclear data evaluated uncertainties, in particular of the Neutron Standards' uncertainties as being underestimated. The main problem in the case of evaluations based on primarily experimental data (as is the case of Neutron Standards) remains in the realm of properly estimating correlations between the input experimental data. %Evaluators are challenged by two problems: 1- a disagreement between the uncertainty distribution derived from the declared uncertainty estimates, and expected statistical uncertainty distribution~\cite{Gai:2007}. Attempts to improve the distribution by rejection of (discrepant) data introduce badly controlled uncertainties, and does not lead to a proper estimation of  the systematic uncertainties which are crucial for a proper UQ; 2- essential differences in the covariance matrices obtained with various statistical models. The integral data uncertainty (data averaged over a broad energy spectrum) depends on the local uncertainties in a rather complex way, and is very sensitive to evaluations of systematic uncertainties (that lead to long-rage correlations).
The proposed unrecognized uncertainty-estimation method~\cite{Badikov:2003,Gai:2007} addresses this problem, and it is equivalent to the determination of hidden uncertainties. The method has been applied to cross-section evaluations in the Neutron Cross Section Standards project~\cite{Badikov:2007,Standards:2009,Standards:2018}, and it has been also used to evaluate the uncertainties of the latest release of the Russian evaluated nuclear data library BROND--3~\cite{Gai:2008,Blokhin:2016}.

In Ref.~\cite{Hermanne:2018} it was written: \textit{This method allows the determination of experimental systematic uncertainties that have been underestimated by their authors, and some implicit correlations of the data can also be obtained. Additional systematic and statistical uncertainties of each experimental study are determined in accordance with the observed distribution of the data around the calculated mean value~\cite{Gai:2007}. Application of the method is based on the iterative procedure of minimizing the mean squares deviations with the assessed statistical and systematic uncertainties}.

The unrecognized uncertainty-estimation method~\cite{Badikov:2003,Gai:2007} inspired the introduction of \USU in the latest Neutron Data Standards~\cite{Standards:2018}. %Those analyses of the uncertainties were based on the \ql a priori\qrs assumption of equal reliability of all available experimental data which of course excludes proven erroneous results~\cite{Gai:2007}.
In the Neutron Standards publication we have defined \textit{the unrecognized (or unknown) systematic uncertainty as a practical minimum uncertainty that can be achieved using a given measuring method (or measuring tool). No matter how many times the measurements are repeated, if we use the same method, we cannot get a result with lower uncertainty~\cite{Standards:2018}}.

For the Neutron Standards evaluation, each of the cross sections evaluated had the normalization for absolute measurements analyzed statistically (considering weights) to obtain the standard deviation of that distribution. Derived standard deviation was regarded as an additional component of the unrecognized systematic uncertainty. The assumption was made~\cite{Standards:2018} that the unrecognized systematic uncertainty is not energy dependent. While this approximation was valid for data partially evaluated by the R-matrix theory\footnote{Eventually all data are combined in the GLS fit.}, where the normalization of experimental data relied on the unitarity of the formalism, the situation is different for other cross sections of neutron-induced reactions on heavy elements. In the latter case the approximation of \USU as an energy independent quantity may be questionable, but it was introduced in the standards evaluation as a first approximation toward deriving more realistic uncertainty estimates.

Even though it has been demonstrated that inclusion of energy dependent \USU values results in an added need to iterate the generalized least-squares (GLS) procedure in performing an evaluation~\cite{Capote-Neudecker:2018}, it is worth noting that the increased uncertainties derived for quantities like the \PU~fission cross sections generally have been justified by using independent methods (\eg, the PUBs estimate~\cite{Neudecker_PUB:2018}).

%Note that the increased uncertainties due to \USU for quantities like the \PU~fission cross sections have been generally justified by using independent methods (\eg, the PUB estimate~\cite{Neudecker_PUB:2018}).

\subsection{\USU Definition and Characteristics}\label{ssec:definition}

It is important that a clear definition of \USU be provided as the basis for further discussions in this paper:

\underline{Definition}: {\textit{\USU are those uncertainties for which one has no or only limited idea of their cause (at present), and that limit the accuracy and precision of quantifying an observable.}}

%%%Georg insertion start%%%
We should keep in mind that it's meaningless to discuss a systematic \USU if only one experimental data point is available for a particular physical process. Furthermore, there is no possibility to identify a systematic \USU effect if all the available data correspond to just one experiment, no matter the number of data points within it.

Taking into account the existence of \USUn, the relationship between the measured values $y_i$ associated with potentially different observables and the underlying true value $\mu$ for one experiment, leads us to write down the following equation which is an extension of Eq.~\eqref{eq:model1}:
%\eqnarrow
\begin{equation}
y_i =
\mu +
\varepsilon_i +
\eta_i +
\delta_i \,.
\label{eq:USU}
\end{equation}
%\eqnormal
This formulation could be easily generalized to multiple measurements $K$ by introducing an additional index $(k)$. This will be omitted for clarity in the ensuing discussion.
The difference between the measured values $y_i$ and the true value $\mu$ is given as the sum of various errors.
The error due to a finite counting statistics is denoted as $\varepsilon_i$ and the error due to recognized systematic effects, such as the detector efficiency, as $\eta_i$.
The error due to \USU is denoted as $\delta_i$.
These quantities are all considered as random variables. The errors $\eta_i$ and $\delta_i$ of different measurements can be correlated, \eg, if the measurements were performed using the same detector or the same sample.
It is noteworthy that only the measured values $y_i$ are known and the values of all other variables are uncertain and therefore have to be estimated using statistical assumptions.
For instance, if the $y_i$ represent the measured values \textit{corrected} for recognized systematic errors, a reasonable and common assumption is that the most likely $\eta_i$ is zero.
Even though it is not necessary to do so from a statistical point of view, it is usually assumed that the random variables $\eta_i$, $\varepsilon_i$, and $\delta_i$ are governed by normal distributions, which leads to a simplified mathematical treatment.
Depending on the case, this choice may be motivated either by the central limit theorem, the principle of maximum entropy, or the fact that the normal distribution is the limit of several distributions, such as the Poisson distribution, \eg, see Refs.~\cite{Smith:1991,Smith-Otuka:2012}.
%The determination of uncertainties (or standard deviations) associated with the errors is one essential step in nuclear data evaluation.
%%%Georg insertion end%%%

Report JCGM 100:2008~\cite{JCGM100} from the \textit{Bureau International des Poids et Mesures} is a well-known guide to uncertainty quantification (UQ) for experimental measurements in the field of metrology and, while it is a useful reference for present purposes, its recommendations do not always apply in the case of nuclear data evaluation. As mentioned earlier in this paper, the latter often involves employing Bayesian prescriptions that combine available data, which may consist of both experimental and theoretically derived information, while in conventional metrology the data are assumed to be entirely observational in nature and sufficiently comprehensive to enable reliable probabilistic treatments of particular measurement processes without having to resort to prior assumptions, \eg, see~\cite{Smith:1991}.

Report JCGM 100:2008~\cite{JCGM100} categorizes uncertainty sources as belonging to one or the other of two basic types: One type consists of those uncertainties that potentially can be estimated objectively by analyses of experiments performed under controlled conditions (sometimes referred to as Type A). The other type consists of uncertainties that are more subjective and are NOT readily determined by objective procedures (sometimes referred to as Type B). Uncertainties belonging to both of these categories are characterized by underlying probability distributions~\cite{JCGM100}, according to the Bayesian interpretation of probability~\cite{Smith:1991}. It should be understood that these two categories refer to the methods by which these uncertainties are determined and NOT by how they are actually involved in specific UQ situations.

% Commented by RC
% \USU are experimental uncertainties that cannot be deduced simply by repetition of a specific method or other objective procedures used in a laboratory. They clearly belong in the Type B category.  Their estimation must rely on the experience of experimenters and evaluators. It should be noted that Report JCGM 100:2008~\cite{JCGM100} clearly indicates that what it
\USU are experimental uncertainty sources whose estimated magnitudes must be determined based on the experience of experimenters and evaluators. It should be noted that Report JCGM 100:2008~\cite{JCGM100} clearly indicates that what it considers to be \USU are completely indeterminable (unknowable). As mentioned earlier, nuclear data evaluators, motivated by practical necessities, must proceed, albeit cautiously, beyond this STOP sign! This observation is fully consistent with the discussion in Sect.~\ref{sect:metrology} that points out differences in how uncertainties are approached in traditional metrology and in nuclear data evaluation work.

With respect to their influence, \USU can be either correlated or uncorrelated, depending on the circumstances. Fully correlated \USU will determine the lowest possible accuracy to which it is possible to determine a certain physical quantity by a particular experimental technique due to unknown bias-inducing effects. Uncorrelated \USU may contribute in a quasi-random way to unexplained scatter in measured data when it is assumed by the experimenter that the measurement procedures are well understood and under control. This may occur because of unrecognized instabilities in the experimental apparatus. Uncorrelated \USU will certainly limit achievable precision, and possibly the attainable accuracy as well.

\subsection{Clues of \USU in Experimental Data}\label{ssec:clues}

Let us assume that after performing the evaluation procedures discussed in the preceding paragraphs of this section, an evaluator arrives subjectively at the conclusion that evaluated uncertainties generated without consideration of \USU are unrealistically small, and that additional \USU contributions should be included. Clues that will support objectively an evaluator's intuitive view regarding the need to consider including \USU contributions are discussed below.

Before examining these clues, it should be pointed out that if there exists a correlated bias (referred to traditionally as a systematic bias) that extends in the same manner across all available and included physics input data sets relevant to a particular evaluation (\eg, as a multiplicative shift of all these data), it will be impossible for an evaluator to suspect the presence of this bias (and, thus, a need to consider sources of \USUn) solely on an examination of these physics input data. Only when the evaluated data are employed in particular application simulations, where calculated and measured derived system observables are compared (C/E), might the existence of such a bias in the input data likely become evident, \eg, see~\cite{Pronyaev:2017}. An additional indication of bias could emerge from the comparison of a completely different type of experiments that measure the same physical quantity (\eg, capture cross section measured by TOF vs AMS, see Sect.~\ref{subsect:AMS}).

For example, suppose hypothetically that the evaluated physics data applicable to calculation of the criticality parameter $k_{\mathrm{eff}}$ for a particular nuclear reactor are very well known, with a single exception being the neutron multiplicity $\overline{\nu}$, and that the computational model and simulation procedures also are assumed to be precise and accurate, \eg, see~\cite{Divadeenam:1984,Axton:1986,Capote-Neudecker:2018}. Furthermore, let's postulate that a significant discrepancy does exist between the currently accepted value of $\overline{\nu}$ and its true value. Then, the calculated criticality parameter $k_{\mathrm{eff}}$ will differ from the measured one, and existence of the $\overline{\nu}$ data discrepancy may be revealed. Unfortunately, such sources of uncertainty that produce biases in application simulations rarely manifest themselves with such transparency in realistic situations.

Additionally, we should mention that ideally before looking for \USU related effects an evaluator is expected to:
\begin{enumerate}
\item remove outliers,
\item undertake a full UQ estimation exercise using available experimental information, and including the guidance of a template to estimate potentially missing uncertainties.
\end{enumerate}

Here are some well-defined clues that point to the need for an evaluator to consider \USU contributions in an evaluation:

\vspace{3mm}
\underline{Clue 1}:

If the spread of input data values is such that limited overlap of the uncertainty bars is observed in the plotted data points, this is a \textit{qualitative} clue that additional unaccounted-for uncertainty sources, \ie, \USUn, may need to be included in the evaluation process.

\vspace{3mm}
\underline{Clue 2}:

A \textit{quantitative} clue that \USU need to be considered is manifested when a calculated global chi-square-per-degree-of-freedom parameter ($\chi^2/df$) that exceeds 1 significantly is generated in an evaluation without consideration of an \USU contribution. This is the case especially if care has been taken by an evaluator to eliminate clearly discrepant experimental data, and to consider all known sources of uncertainty (as well as their correlations) in a realistic manner. Unfortunately, it is common practice for evaluators to simply multiply all the input data uncertainties by the factor $\sqrt{\chi^2/df}$ when this happens, and thus artificially force $\chi^2/df$ to be exactly 1, \eg, see~\cite{Smith:1991,Smith:1993,Smith-Otuka:2012}. This is not considered to be good evaluation practice because it treats all input data points equally, \eg, see~\cite{JCGM100}. Doing so may lead to biases owing to the excessive influence of those data points that contribute to the large chi square per degree of freedom. So, it is a poor substitute for actually uncovering the specific origin of the problem (\eg, discrepant data points, overlooked known sources of uncertainty, failure to consider correlations, \textit{etc}.), or for performing a detailed examination of the individual terms that contribute to the calculated global value of $\chi^2/df$, and only then introducing a targeted \USU contribution when no other options for eliminating the problem are envisioned.

\vspace{3mm}
\underline{Clue 3}:

It should be investigated whether the input data points tend to scatter such that the number of these data points seen to lie significantly outside the uncertainty bands defined by the population standard deviation (irrespective of the uncertainties assigned by the original experimenters) is larger (or smaller) than should be expected from applicable statistical criteria. For example, if the data are assumed to be normally distributed, as is generally the default assumption, then (as mentioned earlier) approximately two-thirds of the points should lie within the above-mentioned uncertainty bands. Excessive numbers of data points (or too few of them) seen to lie outside these limits would suggest that a problem exists with the assumption that these data are normally distributed. Then, evaluation by conventional methods, \eg, the least squares method, that assume normally distributed data could generate misleading results. Some of the rogue data points actually may be discrepant for physical reasons that transcend statistics, and this could lead to inappropriate reliance on statistical interpretation. Furthermore, most statistical rules become meaningful only when fairly large numbers of data points are involved.

\vspace{3mm}
\underline{Clue 4}:

If the population standard deviation of the input data points (without regard to assigned uncertainties) is noticeably different from the average of the magnitudes of uncertainties assigned to these data, then either the assigned uncertainties are over-estimated or under-estimated. In the latter case, the difference may be attributable to \USUn. This clue is related to Clue 1.

\vspace{3mm}
\underline{Clue 5}:

If a collection of experimental data to be evaluated can be organized into two or more groups, based on distinct experimental techniques employed in the measurements, and if evaluation of these data groups separately, according to the various group characteristic experimental techniques, produces noticeable differences in both the evaluated mean values and uncertainties for the distinct groups, this is strong evidence for the existence of a method-related bias attributable to \USU that must be taken into consideration by an increase in the evaluated uncertainties.

While there are statistical algorithms for identifying the tendency of collections of data to exhibit grouping, without regard to considering experimental technique, grouping data points based on measurement techniques, when this information is known by the evaluator, should be preferred to grouping them based solely on statistical analysis for purposes of establishing the need to consider introducing \USU contributions.

\vspace{3mm}
\underline{Clue 6}:

If the preceding clues indicate that evaluated uncertainties are likely under- or over-estimated, the ``Physical Uncertainty Bounds'' (PUBs) method developed by Vaughan \etal\cite{Vaughan_PUB:2014} can be applied to investigate quantitatively whether the evaluated uncertainties are realistic given the input quantities and pertinent known physics information.
One of the first nuclear-data-specific examples is shown in Ref.~\cite{Neudecker_PUB:2018}. It investigates whether the $^{239}$Pu(n,f) cross-section uncertainties that were produced in the Neutron Data Standards evaluation increased with \USU or without consideration of \USU are more realistic.
To this end, the evaluated quantity was separated into its physics sub-processes which are related to independent sources of uncertainties of the evaluated quantity\footnote{{\color{black}{In the present case it is assumed that these sources of uncertainties correspond to experimental sources of uncertainties.}}}.
Physics-based conservative and optimistic uncertainties and functional forms were assessed for each of these sub-processes.
A total minimum realistic and PUBs conservative uncertainty is obtained by adding up uncertainties of all independent sub-processes in quadrature.
If the original evaluated uncertainties are below the minimum realistic bound, they are likely underestimated {\it given all known uncertainties}.
If they are above the conservative bound, they are likely over-estimated given all known uncertainty sources.
For the case of $^{239}$Pu fission, this two-fold PUBs procedure clearly indicated that the experimental uncertainties were inadequately estimated, and that a more detailed UQ effort should be undertaken before estimating an \USU contribution.
If the minimum realistic and PUBs conservative uncertainties bracket the evaluated uncertainties, then the latter uncertainties are realistic {\it given all known uncertainties}.
However, if the evaluated uncertainties do not fall within the PUBs boundary values and, in particular, if the evaluated results fall below the minimum realistic PUBs bound, and no further objective uncertainty source can be identified, then an \USU component needs to be estimated and included in the evaluation process to remedy the situation.
It should be stressed here that the PUBs methodology assesses whether uncertainties are realistic given all {\it known} uncertainties rather than looking for \USUn.
However, consideration of the clues of above allows us to quantify if an \USU contribution is really needed or whether more realistic uncertainties might be obtained by a further detailed uncertainty analysis.
An example to illustrate this approach and the application of PUBs is given in Appendix~\ref{app:PUB} for the \CF~\nub.

\vspace{5mm}
The possibilities of being able to quantify \USU contributions in a reasonable way by pursuing these clues will depend on the specific conditions encountered by the evaluators.

Statistically speaking, the larger the databases to be evaluated for particular observable quantities, the better the chances are that reasonable estimates of \USU uncertainty contributions can be generated to include in the evaluations. If a particular input database is sparse, the possibility of estimating \USU contributions reliably is likely to be marginal.

It may be surmised from the preceding discussion that uncovering the existence of \USUn, and developing procedures by which these uncertainties can be taken into consideration, requires evaluators to examine features of the data points themselves, including both their mean values and assigned uncertainties, and not simply to accept the author-provided mean values and uncertainties as a matter of faith. Failure to consider the possible existence of \USUn, if the contributions are significant, will result in the very deficiencies that manifest themselves in the aforementioned clues.

\USU are additional uncertainties that need to be included in the input database in order to be able to perform a reliable evaluation. Introduction of \USU contributions after an evaluation has been performed is not comparable to introducing them in the input data prior to an evaluation, with the sole exception being the evaluation of a single physical quantity~\cite{Capote-Neudecker:2018}. When \USU contributions are included in the input data, their impact will be reflected in the output evaluated results. This is the proper way that \USU should be treated in an evaluation.

Finally, it should be stressed that the more robust the clues are, the better the chance of generating reasonably reliable estimates of \USU contributions by means of one or more of the methods described in the following section.

\subsection{\USU Estimation Techniques}\label{ssec:methods}

Acknowledging existence of a need to identify \USU involves considerable subjectivity, since it must be established whether adequate steps have been taken to avoid the necessity of introducing \USU by seeking to uncover and incorporate any objectively understood sources of uncertainty that might have been overlooked or improperly quantified in an evaluation conducted without consideration of possible \USU contributions. Consequently, there will be differences of opinion among evaluators on how to approach the matter of dealing with the issue of \USU in specific evaluation scenarios. Evaluated outcomes may differ noticeably depending on choices made by individual evaluators. As a general principle, it should be understood by evaluators that simply enhancing uncertainties to cover data discrepancies that may %simply
reflect failure to apply needed corrections is not a viable option unless these corrections are judged to be relatively small and so difficult to estimate objectively that common sense suggests that attempts to determine them would be impractical.

Regardless of which methods are applied, the following basic approach ought to be followed in dealing with \USU in nuclear data evaluations: As few arbitrary assumptions as possible should be involved in estimating \USU contributions, in accordance with the principle of Occam's Razor~\cite{OccamRazor}. Well-defined, transparent mathematical algorithms should be employed so that they can be implemented in a straightforward manner and adequately documented as part of the evaluation process. The assumptions and procedures involved should deliver results that could be replicated by a future evaluator who might wish to apply them to the same input data set.

The following are some suggested methods that may be considered by evaluators in estimating  the magnitudes of \USU contributions for inclusion in an evaluation. An evaluator may choose to use just one of these methods, or more than one (if feasible) and then compare the outcomes in reaching a decision of how to incorporate \USU estimates in a particular evaluation.

\subsubsection{Grouping Datasets by Experimental Method}\label{ssect:group}
This method may be used when the available experimental data can be grouped into two or more distinct sets according to the experimental methods employed in the measurements, as mentioned above in discussing Clue 5. The \USU contribution, when evaluating a single physical quantity, can be determined from the standard deviation of the various evaluated mean values obtained by analyzing the several data sets separately, according to the distinct measurement techniques (without regard for \USUn). However, if there are only two groups of data to consider, based on the measurement techniques, the best estimate of the \USU contribution probably should be the actual difference in the evaluated mean values for the two groups (see, \eg, Sect.~\ref{subsect:AMS}). If more than one physical quantity is evaluated, similar procedures can be applied separately when considering each of the individual quantities being evaluated.

\subsubsection{Population Standard Deviation $\sigma_P$~in the Normalization Factors}\label{ssect:popul}
When ranges of data are involved, \eg, as characterized by particle energy, then \USU contributions can be estimated as population standard deviations $\sigma_P$ in the average normalization factors of absolute data sets that span the whole energy range. This method was used in estimating \USU contributions for R-matrix data in performing the ENDF/B-VIII standards evaluation~\cite{Standards:2018}.
\subsubsection{Difference Between $\sigma_P$~and Average of Reported Uncertainties}
\USU can be determined from the difference between the population standard deviation of the experimental data and the average value of their reported uncertainties. This method might be appropriate if there is no clear evidence of data-set grouping based on experimental methodology.

\subsubsection{Statistical Estimation}
%A statistical test might be applied to establish whether or not grouping effects are present in the collective body of experimental data employed in an evaluation, regardless of any extraneous knowledge about experimental methodologies used to generate these data (see Appendix).
%\r \textbf{DENISE note: This section is difficult to understand}
If the mathematical relationship in Eq.~\eqref{eq:USU} between the measurements, the truth and the various errors and biases
is complemented with additional distribution assumptions, the maximum likelihood approach~\cite{Smith:1991} (see the Appendix \ref{app:MLE}) can be used to infer the most likely uncertainty and potential correlations connected to \USUn.
For the following, it is pertinent to write Eq.~\eqref{eq:USU} in vector form
$
\vec{y} =
\vec{\mu} +
\vec{\varepsilon} +
\vec{\eta} +
\vec{\delta}
$.
%Note that we have grouped the known statistical $\vec{\varepsilon}$ and systematic $\vec{\eta}$ uncertainties from Eq.~(\ref{eq:USU}) into the known experimental uncertainty $\vec{\varepsilon}_\textrm{exp}$. This combined experimental covariance matrix~$\vec{C}_\textrm{exp}$ is ideally provided by the experimentalists, but could be also determined by the evaluator using the available information from considered experiments.
%\r
The vector $\vec{\mu}$ is the result of a model function.
In the case of a single quantity it may be just a constant.
If $\vec{y}$ contains measurements of the same reaction cross section at different incident energies, the model function can be given by a Pad\'e approximant.
As a last example, in the case of a linear model with a parameter vector~$\vec{\theta}$, we would have $\vec{\mu} = \vec{S} \vec{\theta}$, being $\vec{S}$ a constant scaling factor.
The dependence on the model parameters can be made explicit by writing $\vec{\mu}(\vec{\theta})$.

We assume that the covariance matrices $\vec{C}_{\vec{\varepsilon}}$ and $\vec{C}_{\vec{\eta}}$ are known.
The former covariance matrix reflects the uncertainty in the errors $\vec{\varepsilon}$ due to counting statistics and the latter the uncertainty in the systematic errors $\vec{\eta}$.
The covariance matrix $\vec{C}_{\vec{\delta}}$ associated with \USU contribution is unknown.

In order to estimate both $\vec{\mu}(\vec{\theta})$ and $\vec{C}_{\vec{\delta}}$, we need to make assumptions about the distributions.
Assuming multivariate normal distributions and using the abbreviation $\vec{C}_\textrm{known} = \vec{C}_{\vec{\varepsilon}} + \vec{C}_{\vec{\eta}}$, the likelihood $\mathcal{L}$ for $\vec{y}$ is given by
\eqnarrow
\begin{multline}
\label{eq:mvn_mle_method}
\mathcal{L}(\vec{y} \,|\, \vec{\mu}, \vec{C}_\textrm{known}, \vec{C}_{\vec{\delta}}) =
\frac{1}{
\sqrt{ (2\pi)^N
  \det \left(
    \vec{C}_\textrm{known}
    + \vec{C}_{\vec{\delta}}
  \right)
}
} \times \\
\exp\left\{-\frac{1}{2}
  (\vec{y} - \vec{\mu})^T
  \left(
    \vec{C}_\textrm{known}
    + \vec{C}_{\vec{\delta}}
  \right)^{-1}
  (\vec{y} - \vec{\mu})
\right\}
\end{multline}
\eqnormal
The maximum likelihood principle states that $\vec{\mu}$ and $\vec{C}_{\vec{\delta}}$ should be chosen to maximize the likelihood.
This is an optimization problem.

There are usually not enough measured data available to estimate~$\vec{C}_{\vec{\delta}}$ without additional structural assumptions.
Therefore known distinguishing factors, such as the measurement method, can be used to define the structure of the covariance matrix. This is the same underlying idea as employed in \textit{Method 1} (see \ref{ssect:group}) and \textit{Method 2} (see \ref{ssect:popul}).

As an example for a structural assumption, we may assume that two \USU errors $\delta_i$ and $\delta_j$ have the same value (are fully correlated) if the i$^{th}$ and j$^{th}$ experimental data points have been measured with the same technique. In mathematical terms, elements in the USU covariance matrix are given as $\vec{C}_{\delta,ij} = \Delta$ if both $y_i$ and $y_{\textrm{exp},j}$ have been measured with the same technique and $\vec{C}_{\delta,ij} = 0$ otherwise.
In effect, the covariance matrix $\vec{C}_{\vec{\delta}}$ can be considered as a function of $\Delta$.

Given all the mentioned assumptions above, it means to choose $\Delta$ and $\vec{\theta}$ so that the value of
$
\det \vec{C} + \vec{z}^T \vec{C}^{-1} \vec{z}
$
with
$
\vec{z} = \vec{y} - \vec{\mu}(\vec{\theta})
$
and
$
\vec{C} = \vec{C}_\textrm{known} + \vec{C}_{\vec{\delta}}
$
is minimized.
This expression is obtained by taking the logarithm of Eq.~\eqref{eq:mvn_mle_method} and dropping constants as they do not change the location of the maximum.
This task can be solved by a numerical optimization algorithm.
See the Appendix \ref{app:MLE} for further details on the maximum likelihood method.

\section{EXAMPLES RELATED TO NUCLEAR DATA EVALUATION WITH \USUn}\label{sec:examples}
%\r Toni, Allan, Don, Roberto \b
In the following discussion, it is assumed that the evaluation methods employed are based on the least-squares approach, e.g., as embodied in code \GMA\textit{P}~\cite{Pronyaev:2003}. The estimated \USU contributions, when significant, need to be introduced as components of augmented covariance matrices associated with all the input experimental data. It has been shown by Capote and Neudecker~\cite{Capote-Neudecker:2018} that when more than a one-dimensional quantity is to be evaluated (\eg, energy-dependent cross sections), the evaluated mean values, as well as the derived covariances, will be altered by the use of augmented covariance matrices that include the \USU contributions.
This is intuitively evident because a change in the inclusive data covariance matrix changes the effective weighings of the input data points and this will alter the evaluated mean values. Therefore, a-priori determined \USU contribution (\eg, as done for a one-dimensional quantity \CF~\nub, see Sect.~\ref{subsect:cf-nubar} below) cannot be added a-posteriori to evaluated uncertainties while keeping the originally evaluated mean values. Instead, the evaluations need to be repeated including these augmented covariance matrices that contain an \USU component in the experimental covariance matrix.

This section presents several specific examples where the possibility of identifying the need for and an estimate of \USU contributions is explored. These examples are all based on situations that incorporate actual nuclear data rather than artificial data. However, they sometimes treat isolated and/or truncated data sets for reactions that normally would be linked to several other reactions through measured ratios in an actual comprehensive, simultaneous evaluation. By this means, the examples introduced here have been simplified and reduced in scope in such a way as to make discussions of \USU more transparent than generally would be the case for actual evaluations. In no instance should the results generated in these examples be interpreted as substituting for the results provided in existing evaluations. %This point is further clarified as each specific example is discussed.

In addition to presenting examples that illustrate how the need for \USU might be established, and quantitative estimates of these uncertainties produced, a few additional cases are mentioned as representative of situations where it would be either impossible, or at best inconclusive, to try to identify and quantify \USUn, due simply to a lack of adequate information. We choose to label such situations as \ql unmanageable\qr ones.

Clearly, an investigator needs to assess whether making an effort to identify and quantify \USU contributions would be a waste of time for very practical reasons.

\subsection{\USU in \CF~\nub~Evaluation}\label{subsect:cf-nubar}
%\r provided by Frank, edited by Roberto, to be modified by Denise and Georg \b

The 15-point data set of the neutron multiplicity \nub~experimental values for \CF~as
compiled by Axton~\cite{Axton:1986} had been used for the Thermal
Neutron Constants (TNC) fit~\cite{Pronyaev:2017} that involved more
{\color{black}{than}} 100 experimental points of various thermal quantities
\cite{Otuka:2017} within the IAEA Neutron Data Standards 2017~\cite{Standards:2018}.
The IAEA Neutron Data Standards were adopted for the ENDF/B-VIII.0
library~\cite{ENDFB-VIII:2018}.

Due to the small uncertainty of measured \CF~\nub, it is well established that this quantity is practically
independent of other TNC datasets and its evaluation can be undertaken
independently. A simple least-square fit assuming all 15
experiments to be independent gives an uncertainty of 0.13\%
\cite{Capote-Neudecker:2018}, which is practically identical to the one
quoted by Axton as well as what was derived in a recent comprehensive evaluation
\cite{Pronyaev:2017}. This uncertainty was also obtained in
the previous Neutron Standards evaluation~\cite{Badikov:2007,Standards:2009}. A more conservative uncertainty was
adopted in the IAEA Standards 2017 using previously discussed \USU
concept~\cite{Standards:2018}, which raised the estimated uncertainty to
0.42\% (see Ref.~\cite{Capote-Neudecker:2018} for discussion).

Note that the selected fifteen experimental values do not include any
outliers and their evaluation generated a $\chi^2/df$ that is very close to one. Dividing
the data points into three groups, according to the measurement technique,
gives statistically coherent results in spite of the fact that large differences between
mean values derived using different methods were observed~\cite{Divadeenam:1984,Axton:1986}.

Under these circumstances, and
without further knowledge, the evaluated weighted mean and its variance
are expected to be acceptable. However, the estimated uncertainty of
0.13\% is practically equal to the value that would be obtained if we
neglect all correlations between experiments, which could be an
indicator of missing {\color{black}{inter}}-experiment correlations. In
addition an uncertainty boundary quantification of the same experiments
using the PUBs methodology~\cite{Neudecker_PUB:2018,Vaughan_PUB:2014} summarized in
Appendix~\ref{app:PUB} clearly shows that the 0.13\% uncertainty is
significantly lower than the estimated {\color{black}{minimum realistic
uncertainty bound of 0.23}}\%. Such {\color{black}{a}} difference also
indicates that it is very likely that there are missing {\color{black}{uncertainties of single
experiments and}} correlations between several measurements. At the same
time the estimated {\color{black}{conservative PUBs uncertainty of
0.38}}\% was smaller than the Neutron Standards value of 0.42\%.

Therefore, a reevaluation of the \CF~\nub with a better uncertainty quantification is
needed, as this quantity is critically important in neutron applications.
A careful study of this dataset in a comparison with EXFOR revealed
sometimes sizeable differences~\cite{Otuka:2017} in the values of their uncertainties.

\begin{figure}[!thbp]
\vspace{-2mm}
\centering
\includegraphics[width=\columnwidth]{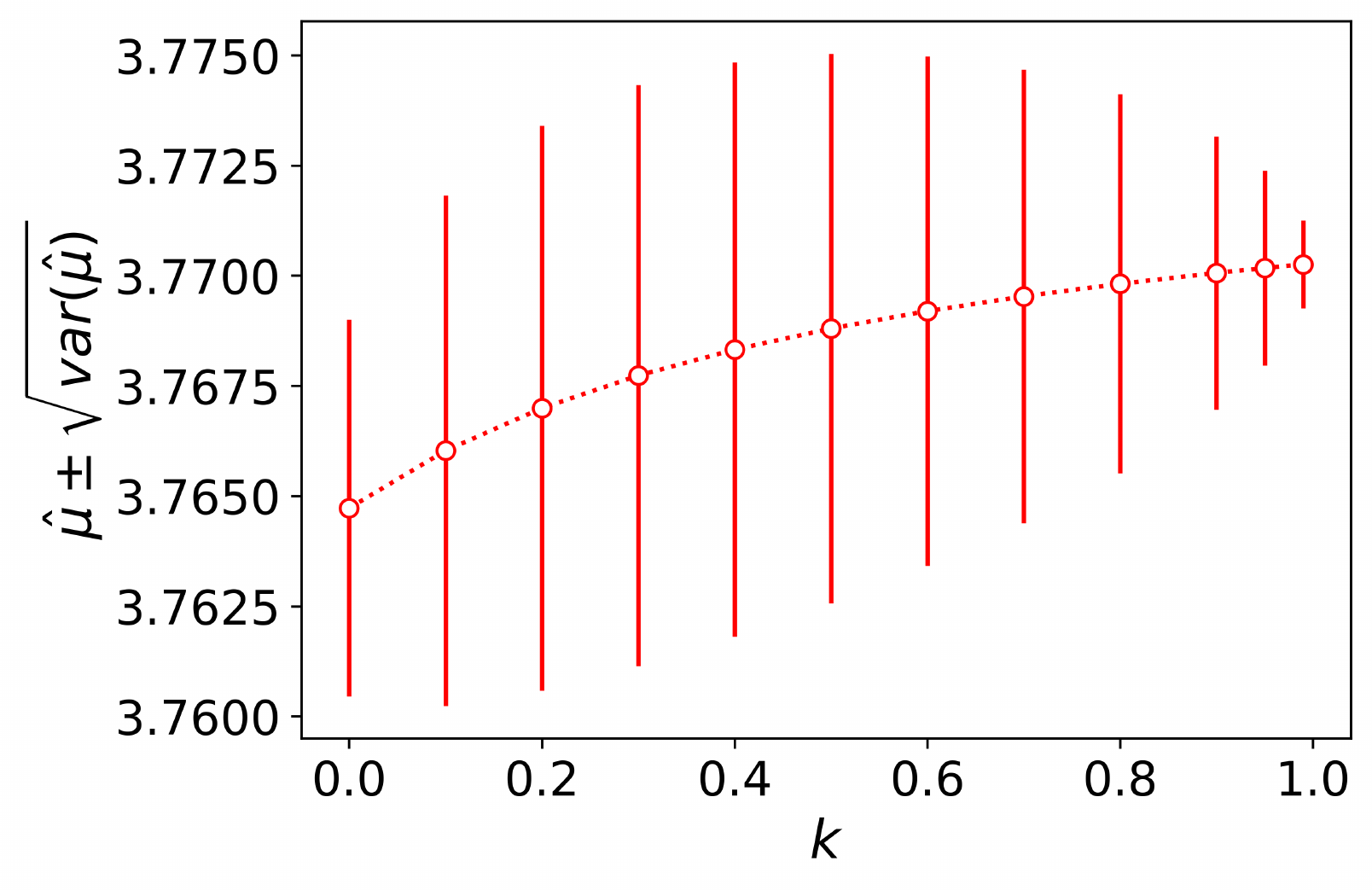}
\vspace{-6mm}
\caption{(Color online) The weighted mean value and its uncertainty of the \CF~\nub- using a
correlation matrix with all off-diagonal elements set to the same constant value
$k$, as a function of~$k$. Note that this is simple test case; a constant correlation
coefficient $k$ is physically implausible due to the very different measurement techniques and length
of correlations.}
\label{fig:percentage}
\vspace{-2mm}
\end{figure}

While we will not attempt here to re-evaluate the \CF~\nub, we will use
this dataset as an example and observe the influence of using the
generalized least-squares approach of Eq.~\eqref{eq:chi2} with
correlations. We employed the values and uncertainties as stated in
Ref.~\cite{Capote-Neudecker:2018}. When introducing correlations, the
total uncertainties are considered preserved. From these data we then
constructed the covariance matrix and estimated the weighted mean value $\hat{\mu}\equiv \bar{\nu}_\text{tot}$ and its variance, without introducing prior information. The act of constructing the covariance
matrix artificially from a correlation matrix is carried out here for
illustration purposes. Of course, we realize that this is not the recommended way to undertake
uncertainty propagation. As an example we set all off-diagonal elements
of the correlation matrix to the same value $k$. For $k=0$ we get a
weighted mean value of $\hat{\mu} = 3.765$ with an an uncertainty of 0.11\%.
If $k$ increases, the weighted mean value slowly increases, and its
uncertainty increases to a maximum of 0.18\% for $k \approx 0.3$, and
then decreases to a smaller value of 0.03\% when $k\rightarrow 1$, as
illustrated in Fig.~\ref{fig:percentage}. The covariance matrix remained
positive definite as $k$ increased, except for $k=1$ (as expected). This is a straightforward
demonstration of the PPP effect resulting from employing this artificially
constructed covariance matrix in the present least-squares evaluation process.
A more realistic case would include an
improved UQ leading to a realistic estimation of the covariance matrix,
possibly using correlations, for example, based on the PUBs approach
\cite{Neudecker_PUB:2018}. It is also clear that for an actual evaluation
of \CF~\nub, not only the covariances must be carefully analyzed and
quantified, but also the original mean values, and in particular the quoted
uncertainties, must be reassessed. It is beyond the scope of this paper to undertake this work.
However, the analysis performed in the present example clearly suggests that consideration of correlations between the experimental data used in the evaluation process, as well as introduction of more realistic uncertainties for the individual data points, and perhaps inclusion of an \USU component, will certainly lead to a larger, and probably more realistic, evaluated uncertainty for this physical quantity than the one recently obtained in Ref.~\cite{Croft:2019}.

%\subsection{Unrecognized Effect in a Counting Experiment} \label{sec:USUstat}
%\subsection{Uncorrelated \USU in a Counting Experiment}\label{subsect:USUstat}
\subsection{\USU in a Counting Experiment}\label{subsect:USUstat}
\USU contributions typically have been associated with correlated uncertainties that may
originate from common detection methods with unknown uncertainties/biases \cite{Standards:2018}.
However, the \USU concept could be extended to include unrecognized (hidden)
experimental uncertainties of unknown origin which may be partially correlated
or even uncorrelated. In this section we illustrate how certain uncertainty components, including
unrecognized (hidden) uncertainties which we are referring to as \USUn, can be estimated by dedicated
measurements and analysis procedures. These could then be viewed as Type A uncertainties, as described in Sect.~\ref{sect:USU} of this paper.

Experimental results  that are used to evaluate a physics quantity
include ideally an assessment of the main uncertainty components
involved in the measurement process. A detailed evaluation of these
components should reduce the need to consider \USUn. However, doing this requires
dedicated experiments and analysis procedures which rely on statistical
inference and variance hypothesis testing as well as a detailed UQ by the
evaluator across all measurements used as input data for the evaluation.
For example, Borella \etal\cite{BORE07} assessed the impact of the sample
properties on capture cross section data that are derived from total energy detection measurements combined
with the pulse height weighting technique. From a series of measurements
using samples with different characteristics, they concluded that the
minimum uncertainty is about 1.7\% in case an external normalization
based on a saturated resonance is applied~\cite{BORE07,SCHI12}. Massimi
\etal\cite{MASS14} demonstrated experimentally that this uncertainty can
be reduced to less than 1.0\% when an internal normalization is
applied. These conclusions are valid provided that the procedures described
by Borella \etal\cite{BORE07} and Massimi \etal\cite{MASS14} are applied, as
discussed in~\cite{SCHI12}.

The uncertainty of a radioactivity measurement using a low-geometry
alpha counting system is investigated in Ref.~\cite{SCHI19}.
%To derive the $\alpha$-activity of a thin layer sample, the sample is placed in a
%vacuum chamber and the emitted $\alpha$-particles are detected by a
%Si-detector placed in a well defined geometry. Both the sample and
%detector are covered by a diaphragm such that the solid angle is defined
%by the distance between the sample and detector and the diameters of the
%two diaphragms.
An experimental procedure was defined to verify if the
uncertainty due to counting statistics has to be combined with an \USU component, and
to estimate the variance of its distribution. The procedure is based on
a one-way ANOVA (ANalysis Of VAriance) model
combined with hypothesis testing~\cite{BENN54}. Therefore, a series of
repetitive (replicate) measurements was undertaken. The measurement
time was set to yield a total number of counts such that the Poisson
distribution approaches a Gaussian distribution.
Before the start of  a series of repetitive measurements the vacuum of
the chamber was eliminated, the sample was placed at its appointed position,
the sample-detector distance was measured, the diaphragms were placed
and the vacuum pump was again activated.

\begin{table*}[!thb]
\begin{center}
\caption{One-way ANOVA table for the homoscedastic assumption applied to a group of replicate measurements. The last column gives the formulae for an equal number $n$ of replicate measurements }
\label{tab:SUMSQ}
\begin{tabular}{l|c|c|c}
\hline\hline\noalign{\smallskip}
                          & ~~Sum of squares                                                                                                             &~~Degrees of freedom~~&~~Expectation value            \\
\noalign{\smallskip}\hline\noalign{\smallskip}
Between groups~~& $S^2_{g} =\sum\limits_{i=1}^{K} n_i (\hat{\mu}_{i} - \hat{\mu})^2 $                       &  $K - 1$  &       $  \hat{\sigma}^2_\epsilon + n \hat{\sigma}^2_\delta = \frac{S^2_g}{K-1} $           \\
Within groups      &  $S^2_r = \sum\limits_{i=1}^{K}\sum\limits_{j=1}^{n_i} (x_{ij} - \hat{\mu}_{i})^2$           &  $N - K$  &       $  \hat{\sigma}^2_\epsilon = \frac{S^2_r}{N-K} $                                               \\
Total                   & $S^2 =  \sum\limits_{i=1}^{K}\sum\limits_{j=1}^{n_i} (x_{ij} - \hat{\mu})^2$     &              &                                                                                                                                \\
\noalign{\smallskip}\hline\hline
\end{tabular}
\end{center}
\vspace{-0.2cm}
\end{table*}

A data set for one sample and sample-detector distance consists of a total of $N= \sum\limits_{i=1}^{K} n_i$ counts consisting of $K$  groups ($i= 1, ...,K$) of $n$ replicate measurement results ($j = 1,...,n_i$).
Each measurement  result $y_{ij}$ can be represented by an error structure (or model):
\begin{equation}\label{eq:ERST}
y_{ij}=\mu + \epsilon_{ij} +  \delta_i
\end{equation}
It is the sum of the true value $\mu$, an unrecognized error $\delta_i$ due to the group in which the observation occurs, and a random error $\epsilon_{ij}$ representing the variation from the mean value of the $i^{th}$ group. It is supposed that the measurements in the same group have error components $\epsilon_{ij}$ from the same distribution with zero mean and standard deviation $\sigma_{\epsilon}$ and the error components $\delta_i$ are identically distributed coming from a distribution with zero mean and standard deviation $\sigma_\delta$ (\USUn).
A set of sum of squares reported in Table \ref{tab:SUMSQ} are calculated from the experimental data. These values are used to quantify the degree to which the $\delta_i$ values of different groups are different or alternatively what is the variance $\sigma^2_\delta$ of the values across different groups.
For each group $i$  the variance $s^2_i$  is calculated:
\begin{equation}\label{eq:si2}
s^2_{i}=\frac{1}{n_i - 1} \sum\limits_{j=1}^{n_i} (x_{ij}-\hat{\mu}_{i})^2\,,
\end{equation}
with $\hat{\mu}_{i}$ the mean value for group $i$:
\begin{equation}\label{eq:Sg}
\hat{\mu}_{i} =\frac{\sum\limits_{j=1}^{n_i} x_{ij}}{n_i}\,,
\end{equation}
where $x_{ij}$ are the raw measured data.
In the homoscedastic case\footnote{In statistics, a vector of random variables is homoscedastic if all its random variables have the same finite variance. This is also known as homogeneity of variance.}, these values  are pooled to provide an estimator $\hat{\sigma}^2_{\epsilon}$   of  the variance  of the random error,
\begin{equation}\label{eq:esteta}
\hat{\sigma}^2_{\epsilon} =\frac{S^2_r}{N - K} = \frac{\sum\limits_{i=1}^{K} (n_i - 1)s^2_{i} }{N - K}\,,
\end{equation}
where relevant definitions are given in Table~\ref{tab:SUMSQ}.
Using this estimator of $\sigma^2_{\epsilon}$, the \USU variance $\sigma^2_{\delta}$ is calculated as:
\begin{equation}\label{eq:vareta}
\hat{\sigma}^2_{\delta} =\frac{N}{N^2 - \sum\limits_{i=1}^{K} n^2_i} \left[S^2_g -(K-1) \hat{\sigma}^2_{\epsilon} \right]\,,
\end{equation}
with the between group sum of squares $S^2_g$ defined in Table~\ref{tab:SUMSQ}, % by:
%\begin{equation}\label{eq:Sg}
%S^2_{g} =\sum\limits_{i=1}^{K} n_i (\hat{\mu}_i - \hat{\mu})^2\,,
%\end{equation}
and the overall mean  $\hat{\mu}$  or estimate of the true value by:
\begin{equation}\label{eq:Sg}
\hat{\mu} =\frac{\sum\limits_{i=1}^{K}\sum\limits_{j=1}^{n_i} x_{ij}}{N}\,.
\end{equation}

\begin{figure}[t]
\vspace{-2mm}
\begin{center}
\includegraphics[width=\columnwidth,trim=2mm 2mm 8mm 2mm]{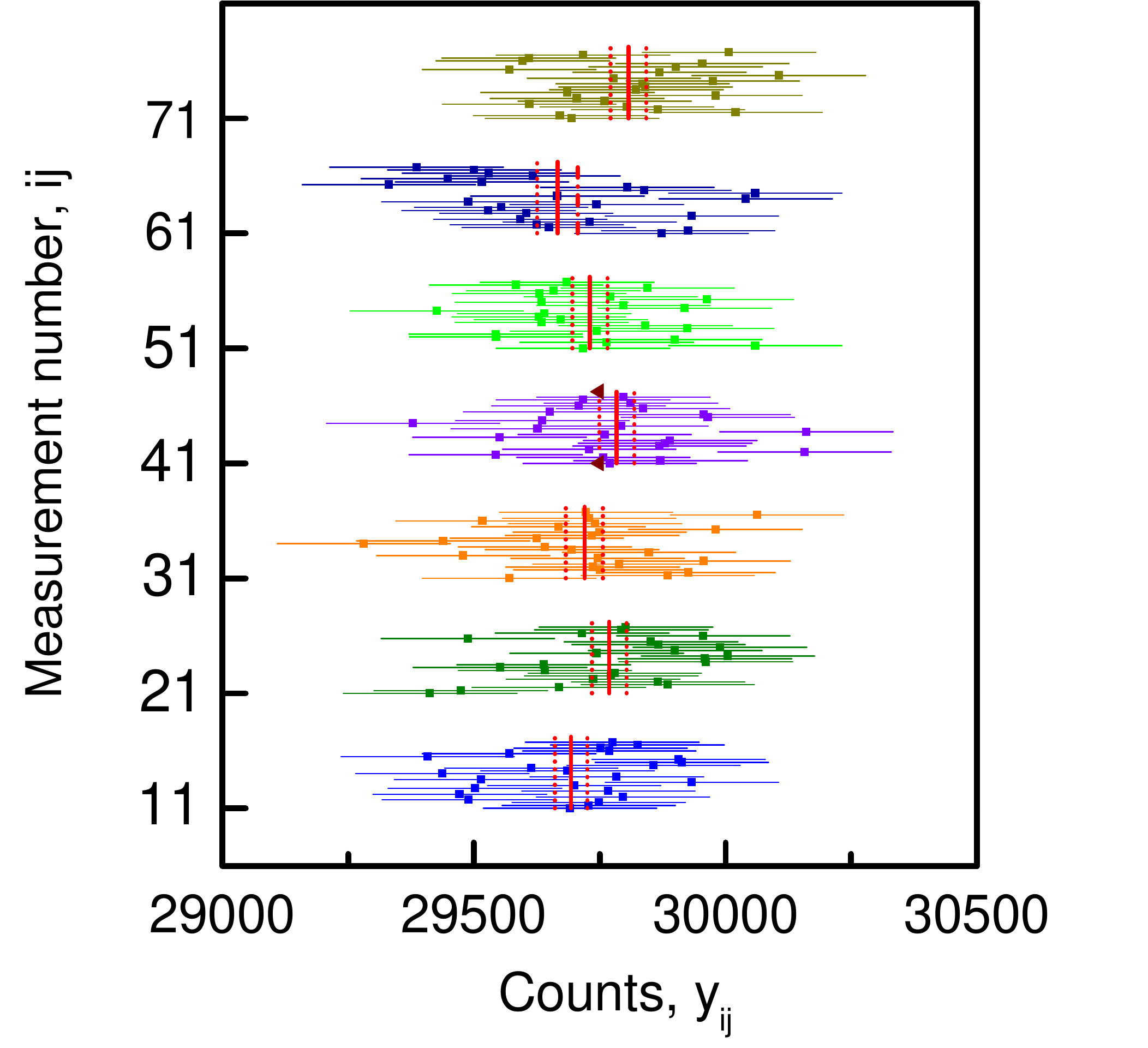}
\vspace{-4mm}
\caption{(Color online) Results of repetitive measurements to determine the $\alpha$-activity of a thin layer sample.}
 \label{fig:COUNTS}
\end{center}
\vspace{-4mm}
\end{figure}

\begin{figure}[t]
\vspace{-3mm}
\begin{center}
\includegraphics[width=1.02\columnwidth,trim=4mm 4mm 12mm 12mm]{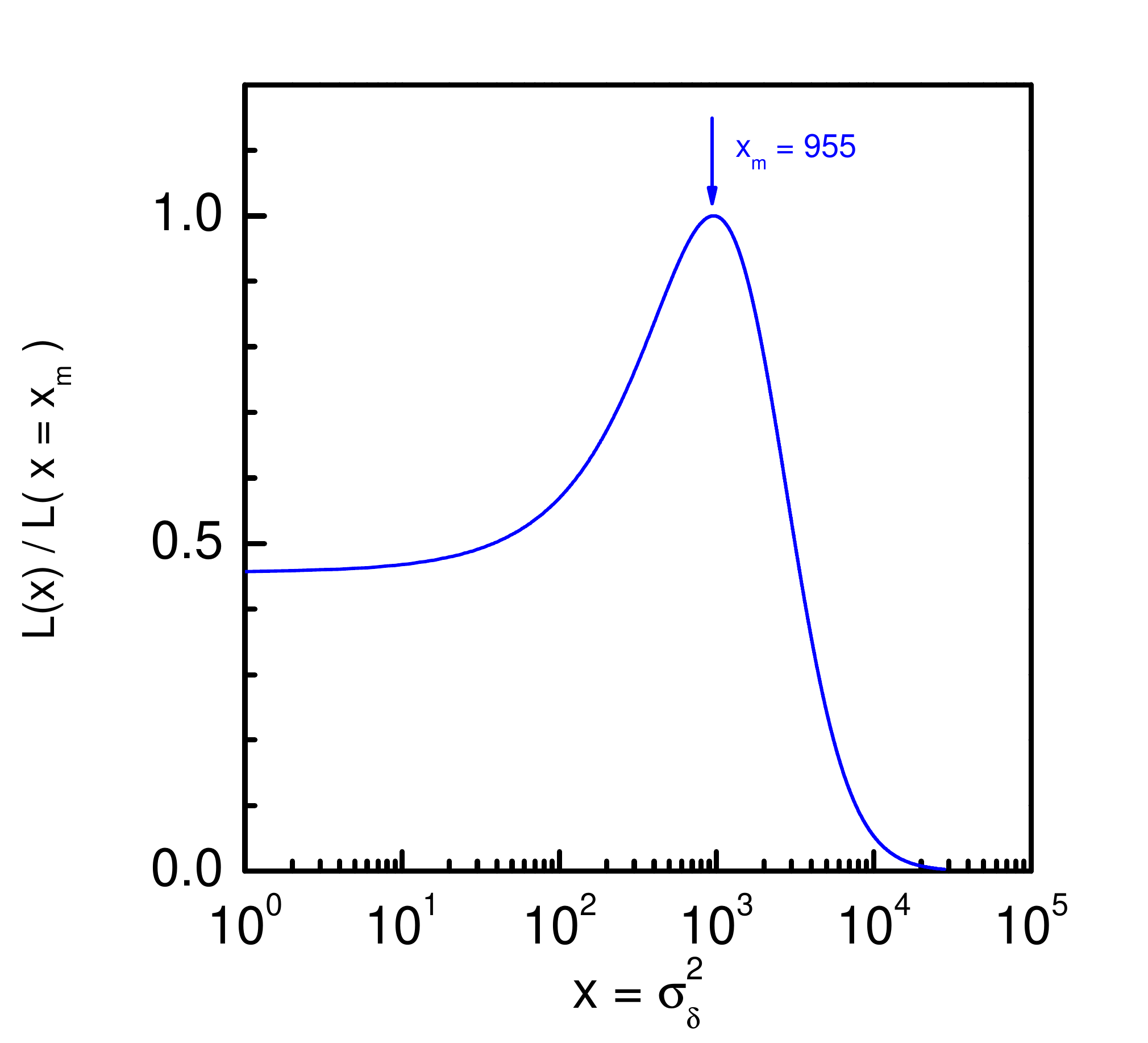}
\vspace{-4mm}
\caption{(Color online) Likelihood of the variance $\sigma^2_\delta$ for an  $\alpha$-activity measurement in a well-defined low-geometry condition. The likelihood $\mathcal{L}$ is normalised to unity at the maximum $\sigma^2_\delta$ = 955.}
\label{fig:PROBWIL}
\end{center}
\vspace{-4mm}
\end{figure}

\begin{table*}[!bht]
\begin{center}
\vspace{-4mm}
\caption{Hypothesis testing of  repetitive measurement data analysed by ANOVA. The last column reports the probability that values are larger than the test-value $T_t$.}
\label{tab:ANOVA_DATA}
\begin{tabular}{l|c|c|c}
\hline\hline\noalign{\smallskip}
 Hypothesis                                                                                                                  & ~~Test-value, $T_t$~~                 &~~Degrees of freedom~~& ~~$P(T \geq T_{t})$             \\
\noalign{\smallskip}\hline\noalign{\smallskip}
  $\hat{\sigma}^2_{\epsilon} = \frac{\sum\limits_{i=1}^{K} (n_i - 1)s^2_{i} }{N - K}$       & $\chi^2_{B} = 3.38$                                &             6                     &   0.76  \\
& & &\\
  $  \hat{\sigma}^2_{\epsilon}  = \hat{\mu}$                                                        &     $\chi^2_{\sigma^2_\epsilon} = 158$      &            161                    &  0.55     \\
& & & \\
  $  \sigma^2_{\delta}  = 0 $                                                                                            &     $F_{\sigma^2_\delta} = 2.13 $                                       &            (6,161)                   & 0.05        \\
\noalign{\smallskip}\hline\hline
\end{tabular}
\end{center}
\vspace{-4mm}
\end{table*}

The homoscedascity of the $K$ variances $s^2_i$ is verified by the Bartlett test~\cite{BENN54}. The quantity $\chi^2_{B}$ :
\begin{equation}\label{eq:bartlett}
\chi^2_{B}  =\frac{ (N-K) \ln{\hat{\sigma}^2_{\epsilon}} -\sum\limits_{i=1}^K  (n_i - 1) \ln{s^2_i}}  {1+ \frac{1}{3(K-1)} \left[\sum\limits_{i=1}^K \frac{1}{n_i - 1}- \frac{1}{(N-K)}  \right]   }
\end{equation}
is computed, which follows a $\chi^2$-distribution with $(K-1)$ degrees of freedom.
The hypothesis that the variance of the random error is due only to counting statistics, \ie, based on a Poisson distribution, is tested by the quantity:
\begin{equation}\label{eq:esteta}
\chi^2_{\sigma^2_\epsilon} =\frac{(N-K) \hat{\sigma}^2_{\epsilon}}{\hat{\mu}} ,
\end{equation}
with the best estimate of the true value  $\hat{\mu}$  used as an estimator of the variance.
The variable $\chi^2_{\sigma^2_\epsilon}$  follows the $\chi^2$-distribution with $(N-K)$ degrees of freedom.
Finally, the hypothesis of the absence of \USUn, or the hypothesis that $\sigma^2_{\delta} = 0$,  is tested by the ratio:
\begin{equation}\label{eq:F}
F_{\sigma^2_\delta} = \frac{\frac{S^2_g}{K-1}} {\frac{S^2_r}{N-K} } ,
\end{equation}
which follows the $F$--distribution with $(K-1)$ degrees of freedom in the numerator and $(N-K)$ degrees of freedom in the denominator.

The one-way ANOVA model  together with the Bartlett and $\chi^2$-tests
were applied to identify and quantify the \USU using the results of the $\alpha$-counting
experiments for different samples and different sample-detector
distances.  A detailed description of the exercise is reported in
Ref.~\cite{SCHI19}.  The data of the measurement for one sample and
sample-detector distance are shown in Fig.~\ref{fig:COUNTS}. The
results of a statistical analysis of these repetitive measurements are
summarised in Table~\ref{tab:ANOVA_DATA}.  The data in
Table~\ref{tab:ANOVA_DATA} reveal that the variance of the random
component  can be derived from the estimated mean, provided that  the
counting statistics is enough to approximate the Poisson distribution by
a normal distribution. The $F$-test suggests the presence of a \USU with a
variance $\hat{\sigma}^2_{\delta}$~=~1365, estimated from
the sum of squares   $S^2_g$ and  $S^2_r$.

The \USU variance ${\sigma}^2_{\delta}$ can also be estimated starting from grouped data $(y_i, u_i)$ based on a maximum likelihood estimation as in \eg, Ref.~\cite{Willink:2002}.  The mixed error-component model is similar to the one in Eq.~\eqref{eq:USU}, however, without any correlated error component:
\begin{equation}\label{eq:ERST2}
y_{i}=\mu + \epsilon_{i} +  \delta_i\,.
\end{equation}

\begin{figure*}[!bht]
\vspace{-4mm}
\centering
\includegraphics[width=0.99\textwidth]{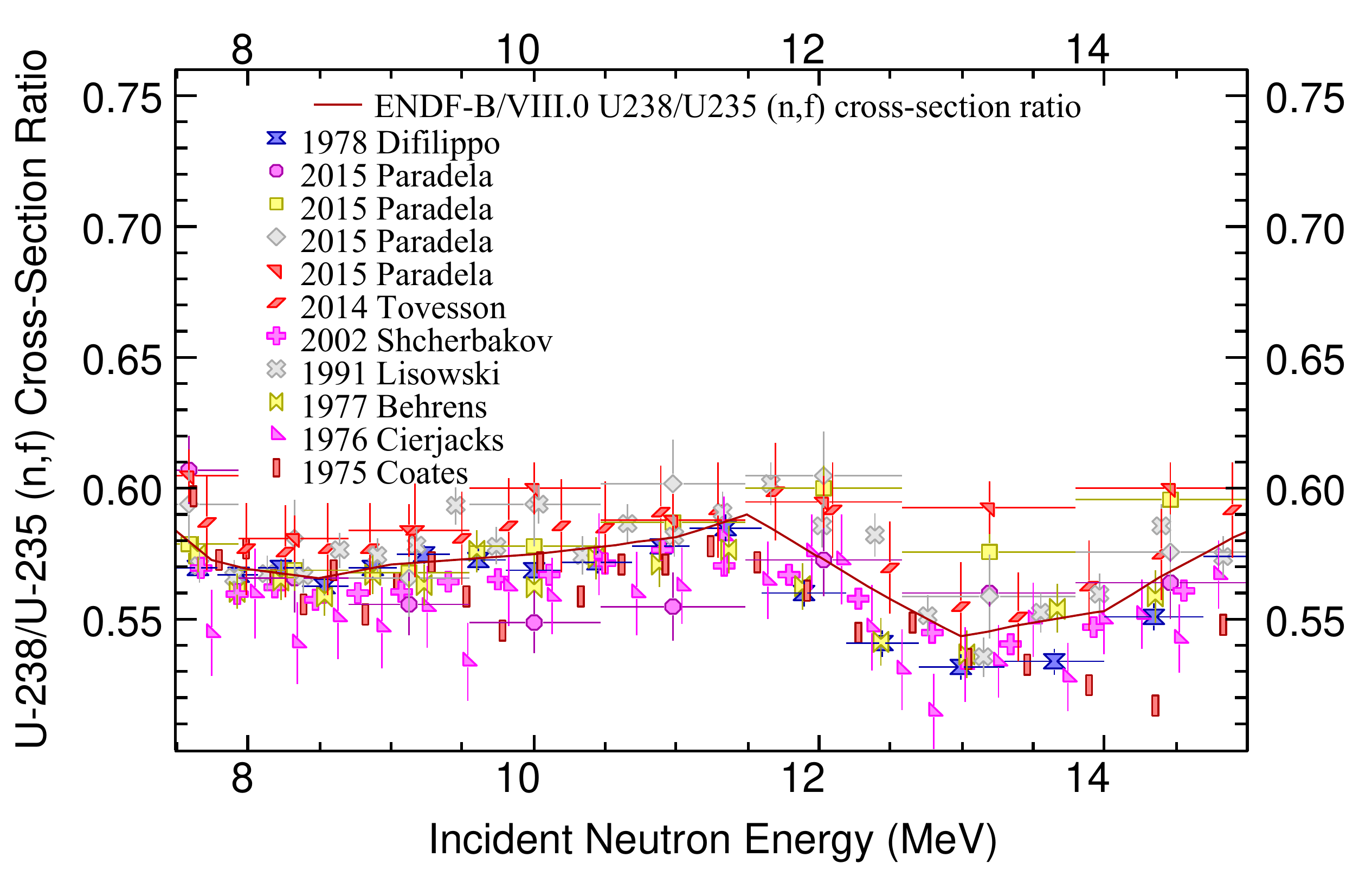}
\vspace{-2mm}
\caption{(Color online) Overview of the \UF/\UT~neutron-induced fission cross section ratio measurements from 7.5 up to 15~MeV. Note the weak energy dependence.}
\label{fig:U8toU5datasets}
\vspace{-3mm}
\end{figure*}

It is assumed that the  error components $\epsilon_{i}$ and $\delta_{i}$ have a normal statistical distribution with mean zero, \ie, $\mathcal{N}(0,u^2_i)$ and $\mathcal{N}(0,{\sigma}^2_{\delta})$ $\mathcal{N}(0$,${\sigma}^2_{\delta}$),  that they are mutually independent, and that the unrecognized effect is  independently sampled from a single distribution. The \USU variance  ${\sigma}^2_{\delta}$ of this distribution is estimated by maximizing the likelihood:
\begin{equation}\label{eq:LIKE_WILL}
\mathcal{L}(\mu, \sigma^2_\delta) = \prod_{i=1}^K \frac{1}{\sqrt{2\pi(u^2_i +\sigma^2_\delta )}} e^{-\frac{(y_i-\hat{\mu})^2}{2(u^2_i +\sigma^2_\delta )}}.
\end{equation}
The likelihood $\mathcal{L}$, which is represented in Fig.~\ref{fig:PROBWIL}, results in
Maximum Likelihood Estimates (MLE) $\hat{\sigma}^2_{\delta}$ = 955 and 970 using  $s^2_i$  and
$\hat{\mu}$, respectively, for the uncertainties $u^2_i$. These
estimates are fully consistent with the one derived from  $S^2_g$ and
$S^2_r$. To evaluate the full measurement uncertainty of the result of
an activity measurement the \USU variance needs to be
determined as a function of count rate and the impact of the correction
for the solid angle needs to be investigated. The result of these
studies will be reported in~\cite{SCHI19}.

\subsection{\USU in \UF/\UT~Fission Cross-section Ratio Measurements}\label{subsect:u238-u235-ratio}
%\r Nacho, Frank, Peter, Vladimir, Denise {\color{teal}{to add refs}, Georg, Sergei \b
The IAEA Neutron Standards evaluations~\cite{Badikov:2007,Standards:2009,Standards:2018} offer an example when energy-dependent cross sections, their ratios and other energy-dependent combinations are used in a non-model combined evaluation of data on a common energy grid. The data reduction to the \GMA\textit{P}~energy nodes (grouping) is undertaken using a \GMA\textit{P}~auxiliary code prior to the least-squares fit, and this is not discussed in the present paper\footnote{The grouping preserves the fully correlated uncertainty over the group and it is aimed at reducing the uncorrelated uncertainty, \ie, at reducing data fluctuations due to, \eg, limited counting statistics in the spirit of the Neutron Standards.}.
%Both \UT(n,f)~and \UF(n,f)~cross sections are Neutron Standards starting at 2~MeV up to 200~MeV~\footnote{The \UT(n,f)~is an standard cross section at thermal, from 7.8--11~eV (the integral), and above 150~keV up to 200~MeV.}. One of the commonly measured quantities is the ratio of neutron-induced fission cross sections of \UF and \UT. Such ratio data, which will be denoted by $R_{8/5}$, are neutron energy dependent quantities. Their uncertainties, including possible components due to \USU, are expected to depend on energy. In this section the presence of \USU for  fission cross-section ratio data is investigated.

Both \UT(n,f)~and \UF(n,f)~cross sections are Neutron Standards in a broad energy region that includes neutron energies from 7.5~MeV up to 15~MeV. One of the most-often measured quantities is the ratio of the neutron-induced fission cross sections of \UF~to \UT. Such ratio data in the region of interest, which will be denoted by $R_{8/5}$, are weakly energy-dependent quantities. Their uncertainties, including possible components due to \USUn, are also expected to depend on energy. In this section the presence of \USU for  fission cross-section ratio data is investigated for neutron energies from 7.5~MeV up to 15~MeV. The data were taken from the Neutron Standards database, {\textit{i.e.}}, from the \GMA\textit{P}~input database, covering the energy range of interest.
The eleven data sets\footnote{reduced to the \GMA\textit{P}~energy nodes.} that were considered are listed in Table \ref{tab:U5U8-input} of Appendix~\ref{app:nf-ratio}, and are shown in  Fig.~\ref{fig:U8toU5datasets}. Note that for the exercise presented in this work all data sets were assumed to result from absolute ratio measurements.

The data shown in Fig.~\ref{fig:U8toU5datasets} were used  to evaluate the cross section ratio $R_{8/5}$ at 9~MeV and 10~MeV from
an independent analysis of the data in the energy regions (8.5~MeV $\le E_n < $ 9.5~MeV)  and  (9.5~MeV $\le E_n < $ 10.5~MeV), respectively.
It is supposed that the reported uncertainties consist of an uncorrelated component
and one fully correlated energy independent component in each of the energy regions. The former is labelled $u_\epsilon$ in Table \ref{tab:U5U8-input}.
The latter is derived by averaging the combined variance of all the correlated components listed in Table~\ref{tab:U5U8-input}, assuming full correlation.
The fully correlated contribution is equivalent to applying for each data set a correction (or normalization) $N=1$ with an uncertainty $u_{N,i}$.
The uncertainties  $u_{N,i}$ for the different data sets are specified in Table~\ref{tab:nf_estim_01}.
It is assumed that there are no correlations between different datasets over the whole energy region.

The analysis is done in two different ways. \textbf{In a first analysis} a weighted average, considering only the uncorrelated uncertainty component, was calculated for each data set within each energy interval.
This weighted average $\hat{\mu_i}$, combined with the total correlated uncertainty $u_{N,i}$,  provides an estimate of the ratio $R_{8/5}$ together with its uncertainty  $u_{\hat{\mu}_i}$ for each data set $i$. These estimates are shown in Figs.~\ref{fig:U8toU5_GLSresult_noUSU_9} and \ref{fig:U8toU5_GLSresult_noUSU_10} and listed in Table~\ref{tab:nf_estim_01}.
A weighted average of these data results in evaluated values for the ratio $R_{8/5}$ at 9 MeV and 10 MeV, respectively, which are also listed in Table~\ref{tab:nf_estim_01}.
%= $0.5731 (28)$  and $\hat{\mu}$ =$ 0.5783 (29)$  for the ratio $R_{8/5}$ at 9 MeV and 10 MeV, respectively.
The correction $(N, u_N)$ is applied by iteration based on the overall estimate. This procedure avoids a bias (underestimation) due to the PPP effect when the uncertainties $u_{N,i}$ are propagated, as discussed in Refs.~\cite{D_Agostini:1994,Froehner:2003}.

\begin{table*}[!thb]
\vspace{-4mm}
\caption{Estimates $\hat{\mu_i}$ of the cross-section ratio $R_{8/5}$ and corresponding uncertainty $u_{\hat{\mu_i}}$ at 9~MeV and 10~MeV derived from the data specified in the first column. Both absolute and relative uncertainties (in \%) are listed. The uncertainties are the total uncertainties resulting from a propagation of the reported correlated and uncorrelated components. The correlated uncertainty is specified in the column $u_{N,i}$. The individual \USU contribution of each dataset $i$, derived by Eq.~\eqref{eq:USU_GMA}, is reported in the column $\hat{\sigma}_{\delta,i}$ and as a percent in the last column. The group \USU $\hat{\sigma}_{\delta}$ averaged over all datasets for each energy bin (derived by Eq.~\eqref{eq:group-USU}) is reported in the last row of the table.}
\label{tab:nf_estim_01}
\begin{tabular}{l|c|rlcrcc|rlcrcc}
\hline \hline
                                     &            & \multicolumn{6}{c|}{ 9 MeV}  &       \multicolumn{6}{c}{ 10 MeV}   \\
          Dataset                    & $u_{N,i}$(\%)  & $\hat{\mu_i}$~~ &($u_{\hat{\mu_i}}$) & $u_{\hat{\mu_i}}$(\%) & $z_i$~ & $\hat{\sigma}_{\delta,i}$ & $\hat{\sigma}_{\delta,i}$(\%) &
                                                    $\hat{\mu_i}$~~ & ($u_{\hat{\mu_i}}$) & $u_{\hat{\mu_i}}$(\%) & $z_i$~ & $\hat{\sigma}_{\delta,i}$ & $\hat{\sigma}_{\delta,i}$(\%) \\
 \hline
01 -- Tovesson \etal\cite{Tovesson:2015}       &~0.84~& 0.5792   &(49)  &0.85~& ~1.25 &  0.00371 &0.64~& 0.5842 & (50)  & 0.86 & ~1.17 & 0.0031 & ~0.53 \\
02 -- n\_TOF (1)~\cite{Paradela:2015}          &~1.73~& 0.5843   &(140) &2.40~& ~0.80 &  0.0     &0.00~& 0.6005 & (143) & 2.38 & ~1.55 & 0.0169 & ~2.82 \\
03 -- n\_TOF (2)~\cite{Paradela:2015}          &~2.49~& 0.5659   &(209) &3.69~& -0.34 &  0.0     &0.00~& 0.5938 & (216) & 3.64 & ~0.72 & 0.0    & ~0.00 \\
04 -- n\_TOF (3)~\cite{Paradela:2015}          &~3.35~& 0.5683   &(208) &3.66~& -0.23 &  0.0     &0.00~& 0.5782 & (212) & 3.67 & -0.01 & 0.0    & ~0.00 \\
05 -- n\_TOF (4)~\cite{Paradela:2015}          &~3.74~& 0.5565   &(241) &4.33~& -0.67 &  0.0     &0.00~& 0.5496 & (253) & 4.60 & -1.13 & 0.0136 & ~2.47 \\
06 -- Behrens \etal\cite{Behrens:1977}         &~0.81~& 0.5630   &(60)  &1.07~& -1.66 &  0.00805 &1.43~& 0.5709 & (63)  & 1.10 & -1.18 & 0.0039 & ~0.69 \\
07 -- Difilippo \etal\cite{Difilippo:1978}     &~2.39~& 0.5692   &(138) &2.42~& -0.28 &  0.0     &0.00~& 0.5713 & (140) & 2.45 & -0.50 & 0.0    & ~0.00 \\
08 -- Cierjacks \etal\cite{Cierjacks:1976}     &~1.09~& 0.5604   &(110) &1.96~& -1.14 &  0.00614 &1.10~& 0.5670 & (99)  & 1.75 & -1.15 & 0.0056 & ~0.98 \\
09 -- Coates \etal\cite{Coates:1975}           &~2.92~& 0.5775   &(176) &3.05~& ~0.25 &  0.0     &0.00~& 0.5720 & (177) & 3.09 & -0.36 & 0.0    & ~0.00 \\
10 -- Shcherbakov \etal\cite{Shcherbakov:2001} &~2.45~& 0.5618   &(142) &2.53~& -0.80 &  0.0     &0.00~& 0.5662 & (144) & 2.54 & -0.84 & 0.0    & ~0.00 \\
11 -- Lisowski \etal\cite{Lisowski:1991}       &~0.94~& 0.5806   &(64)  &1.10~& ~1.18 &  0.00396 &0.68~& 0.5824 & (68)  & 1.17 & ~0.60 & 0.0    & ~0.00 \\
\hline
\multicolumn{2}{c|}{ \textbf{Evaluated values $\mu$ ($u_{\hat{\mu}}$)}}  & \textbf{0.5731} &\textbf{(28)}  & \textbf{0.49\%}  & $\hat{\sigma}_{\delta}=$ & \textbf{0.0035} & (\textbf{0.6\%}) &
                                                                           \textbf{0.5783} &\textbf{(29)}  & \textbf{0.50\%}  & $\hat{\sigma}_{\delta}=$ & \textbf{0.0069} & (\textbf{1.2\%}) \\

 \hline \hline
\end{tabular}
%\vspace{-2mm}
\end{table*}

\begin{figure}[bt]
\vspace{-4mm}
\centering
\includegraphics[width=1.01\columnwidth,trim=15mm 9mm 13mm 10mm]{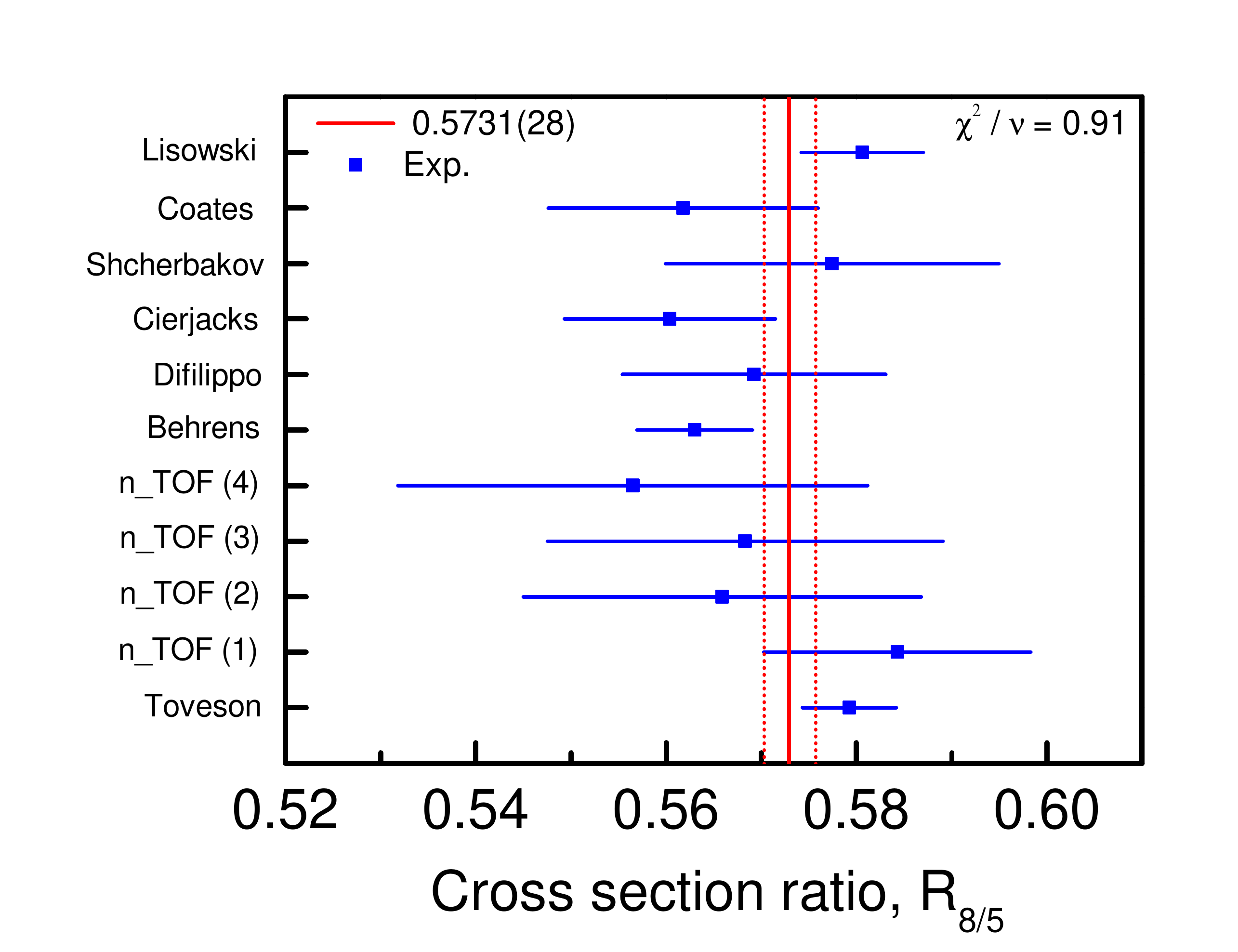}
\vspace{-3mm}
\caption{(Color online) Reduced datasets for the energy group $8.5 \leq E_n < 9.5$ and the evaluated ratio with its associated uncertainty based on $\sigma_\delta = 0$ (\ie, no \USU component).}
\label{fig:U8toU5_GLSresult_noUSU_9}
\vspace{-3mm}
\end{figure}
\begin{figure}[!bthp]
\vspace{-4mm}
\centering
\includegraphics[width=1.01\columnwidth,trim=15mm 9mm 13mm 9mm]{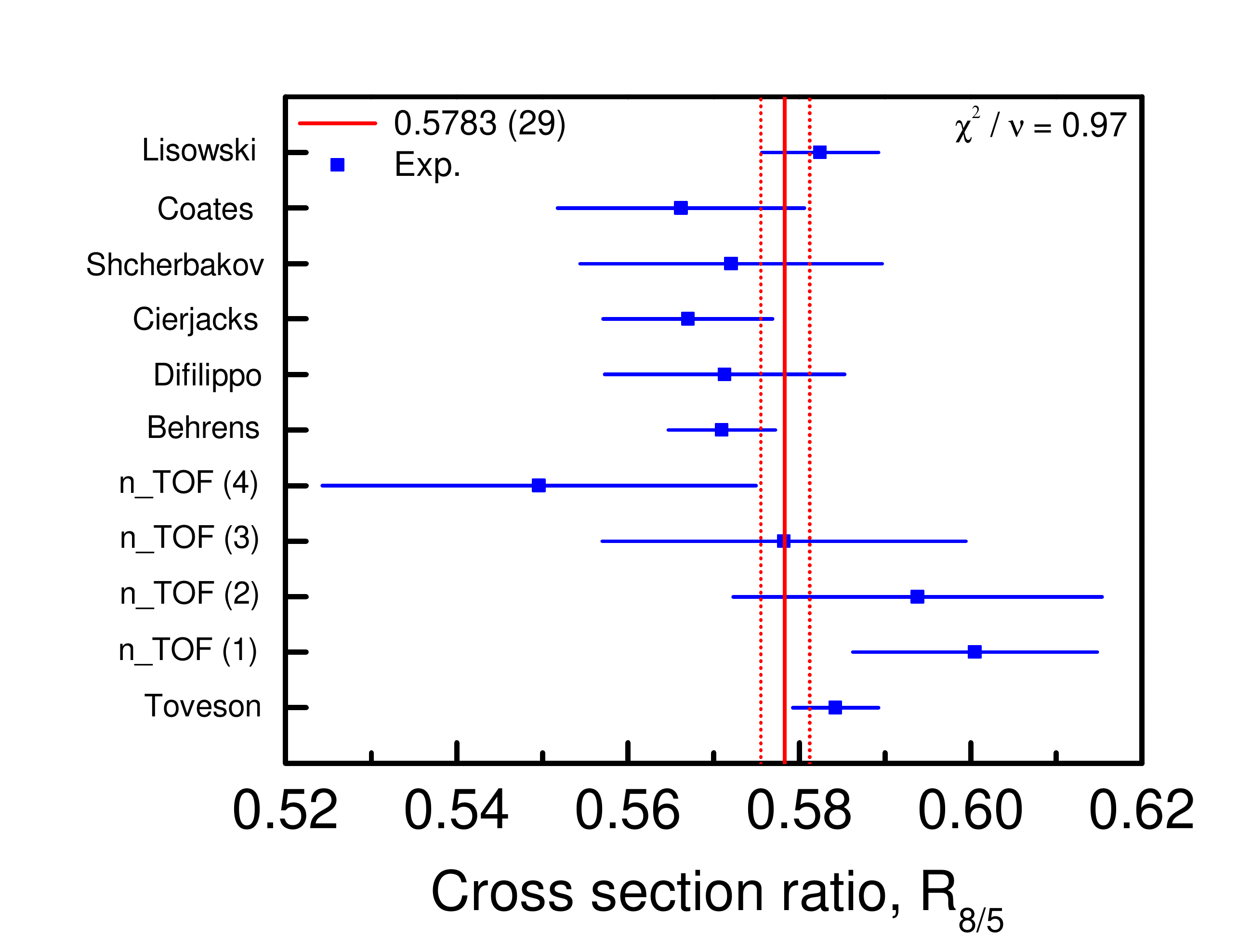}
\vspace{-3mm}
\caption{(Color online) Reduced datasets for the energy group $9.5 \leq E_n < 10.5$ and the evaluated ratio with its associated uncertainty based on $\sigma_\delta = 0$ (\ie, no \USU component).}
\label{fig:U8toU5_GLSresult_noUSU_10}
\vspace{-3mm}
\end{figure}

\begin{table*}[!thb]
\vspace{-2mm}
\caption{Estimates $\hat{\mu}$ of the cross section ratios and uncertainties at 9~MeV and 10~MeV derived from the data listed in Table \ref{tab:U5U8-input} of Appendix~\ref{app:nf-ratio}. The results derived by different order polynomials ($M$) to model the energy dependence of the cross section ratio are compared.}
\label{tab:nf_estim_02}
\begin{tabular}{c|cccccc|cccccc}
\hline \hline
    & \multicolumn{6}{c|}{ 9 MeV}                     &   \multicolumn{6}{c}{ 10 MeV}   \\
 $M$--order &\multicolumn{2}{c}{$\sigma_\delta = 0$}& \multicolumn{2}{c}{$\sigma_\delta = 0.0031$} & \multicolumn{2}{c|}{$\sigma_\delta = 0.0042$} & \multicolumn{2}{c}{$\sigma_\delta = 0$}& \multicolumn{2}{c}{$\sigma_\delta = 0.0018$} & \multicolumn{2}{c}{$\sigma_\delta = 0.0031$} \\
 polynomial & $\hat{\mu}$    & $\chi^2/\nu$ &  $\hat{\mu}$    & $\chi^2/\nu$ & $\hat{\mu}$    & $\chi^2/\nu$ & $\hat{\mu}$   &$\chi^2/\nu$ & $\hat{\mu}$    & $\chi^2/\nu$& $\hat{\mu}$    & $\chi^2/\nu$  \\
 \hline
0  & 0.5731  (28)&1.27&0.5727 (31)&1.22&0.5724 (33)&1.18&0.5783 (29)&1.11&0.5782 (30)&1.10&0.5780 (32)&1.07\\
1  & 0.5738  (28)&0.69&0.5733 (31)&0.63&0.5730 (33)&0.60&0.5786 (29)&1.07&0.5785 (30)&1.05&0.5783 (32)&1.02 \\
2  & 0.5730  (29)&0.68&0.5725 (32)&0.62&0.5723 (34)&0.59&0.5797 (30)&1.06&0.5795 (31)&1.04&0.5793 (33)&1.01 \\
3  & 0.5730  (30)&0.71& 0.5725(32)&0.65&0.5722 (34)&0.61&0.5795 (31)&1.10&0.5794 (32)&1.08&0.5791 (34)&1.05 \\
 \hline \hline
\end{tabular}
\vspace{-3mm}
\end{table*}

\textbf{In the second approach} the different data sets are not treated separately in each energy interval but analyzed simultaneously. One covariance matrix for the whole data set is constructed by adding the correlated and uncorrelated components. A GLS method is applied, assuming an energy dependence of the cross-section ratio as a function of incident neutron energy in a given energy region.  The energy dependence of the cross-section ratio is approximated  by a polynomial function of order $m$ with a maximum order of $M$:
\begin{equation} \label{eq:polyn}
f(E_n, \vec{a},M) =  \sum\limits_{m=0}^{M} a_m  E_n^m,
\end{equation}
and $E_n$ is the incident neutron energy.

In each energy region the parameters $a_m$  and  covariance matrix $\vec{C}_{\vec{a}}$ are derived from a fit to the whole data set within that region.
To avoid any PPP effect, the correlated contribution of the covariance matrix of the input data is derived iteratively based on the model function $f(E_n, \vec{a},M)$.
From the parameters $a_m$ and covariance matrix $\vec{C}_{\vec{a}}$, the best estimate at the central energy of each interval is calculated. The results, which are summarized in Table~\ref{tab:nf_estim_02} as a function of $M$, suggest a weak energy-dependence of the fission cross-section ratio. Since the data do not cover a broad energy region, the best estimate will not depend strongly on the model to express the energy dependence. This is confirmed by the results listed in Table~\ref{tab:nf_estim_02} for different order polynomials up to $M=3$.

The $\chi^2/df$ corresponding to the different results  do not suggest the presence of \USU components or outliers as already observed for the \CF~\nub evaluation that is discussed earlier in this section (see Sect.~\ref{subsect:cf-nubar}). This can also be concluded from the $z$-scores for each data set that are defined by:
\begin{equation} \label{eq:zscore}
z_i = \frac{\hat{\mu}_i - \hat{\mu}}{u_{\hat{\mu}_i}}\,,
\end{equation}
and whose values are listed in Table~\ref{tab:nf_estim_01}. They are all smaller in magnitude than 2 with a maximum absolute value of 1.66.

However, the existence of \USU cannot be excluded by our analysis. The low probability for the presence of \USU in this situation is confirmed by a maximum likelihood analysis using Eq.~\eqref{eq:LIKE_WILL} similar to the analysis described in Sect.~\ref{subsect:USUstat}. The likelihoods as a function of \USU variance ($=\sigma^2_\delta$) for the data at 9~MeV and 10~MeV are presented in Fig.~\ref{fig:U8toU5deltausu}. They are relatively flat, ranging in value between  $\sigma^2_\delta = 0$  and $\sigma^2_\delta \approx 10^{-5}$,  and then declining rapidly as seen in Fig.~\ref{fig:U8toU5deltausu}. The maxima at  $\sigma^2_\delta=0.000018\approx 0.0042^2$ (\USUn$\approx$~0.7\%)  and  $\sigma^2_\delta=0.0000036 \approx 0.0018^2$ (\USUn$\approx$ 0.3\%)  for the data at 9~MeV and 10~MeV, respectively, are not pronounced, especially the maximum corresponding to the 10~MeV data.

\begin{figure}[!thbp]
\vspace{-4mm}
\centering
\includegraphics[width=1.06\columnwidth,trim=19mm 9mm 13mm 9mm]{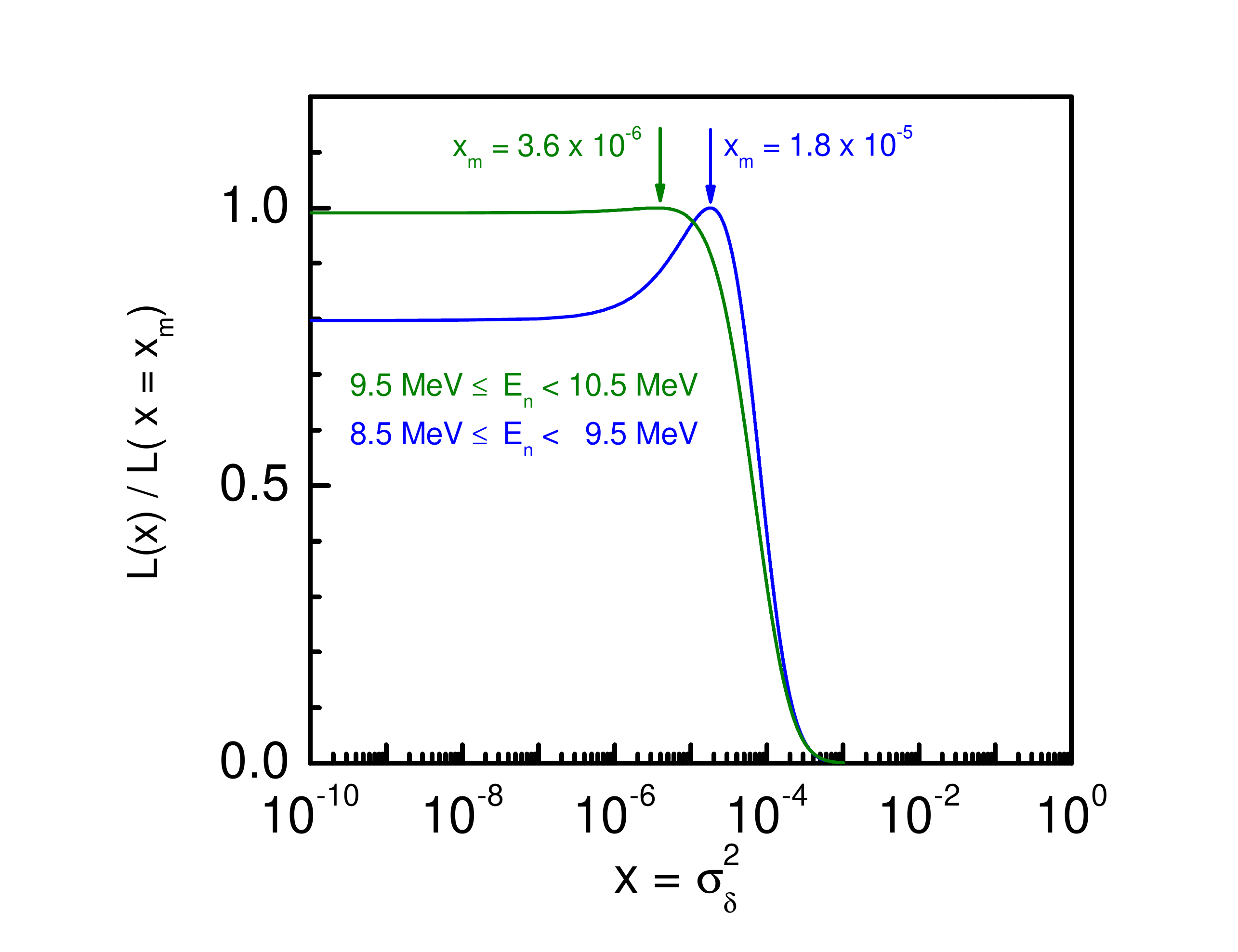}
\vspace{-6mm}
\caption{(Color online) Likelihood of the \USU variance $\sigma^2_{\delta}$ derived from the data listed in Table~\ref{tab:nf_estim_01}. The likelihoods are normalized to unity at the maxima located at $\sigma^2_\delta \approx 0.002^{2}$ (\USUn$\approx 0.2$\%) and at $\sigma^2_\delta \approx 0.004^{2}$ (\USUn$\approx 0.3$\%) for 9~MeV and 10~MeV, respectively.}
\label{fig:U8toU5deltausu}
\vspace{-4mm}
\end{figure}

\begin{figure}[!thbp]
\vspace{-3mm}
\centering
\includegraphics[width=1.08\columnwidth,trim=12mm 9mm 12mm 9mm]{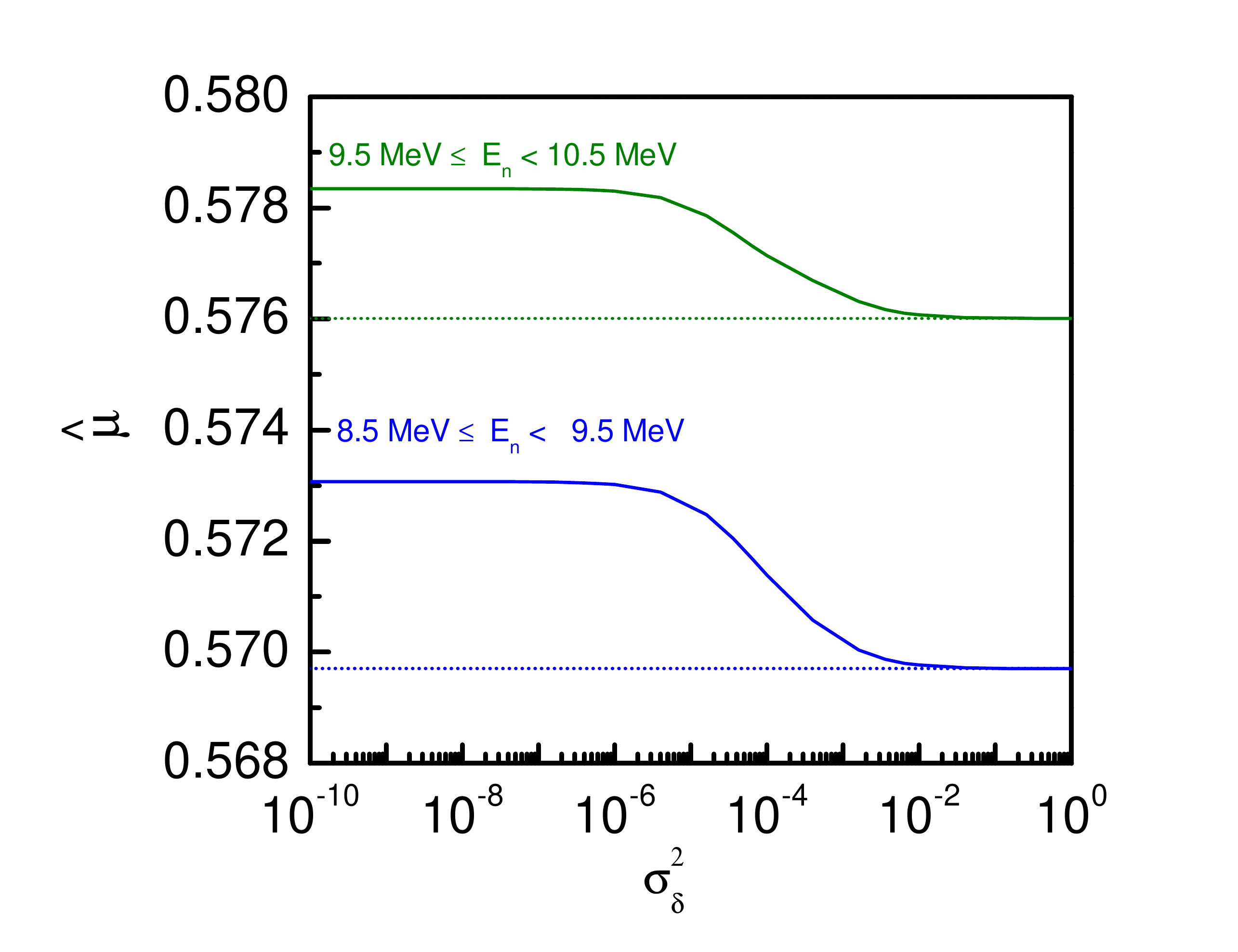}
\vspace{-3mm}
\caption{(Color online) Estimated cross-section ratio as a function of the \USU variance $\sigma^2_\delta$. The dotted lines represent the ratio based on an unweighed average of the data in Table~\ref{tab:nf_estim_01}.}
\label{fig:U8toU5mudeponusu}
\vspace{-2mm}
\end{figure}

By introducing an \USU component, a GLS estimated mean value will change, as discussed in Ref.~\cite{Capote-Neudecker:2018}. This is illustrated in Fig.~\ref{fig:U8toU5mudeponusu} where the estimated values of $\hat{\mu}$ for 9~MeV and 10~MeV are plotted as a function of $\sigma^{2}_{\delta}$. For $\sigma^{2}_{\delta}= 0$ we obtain for the energy group $[8.5, 9.5)$ the evaluated ratio $\hat{\mu} = 0.5731 (28)$ (0.5\%) and for the energy group $[9.5, 10.5)$ the evaluated ratio $\hat{\mu} = 0.5783 (29)$ (0.5\%). The estimates practically do not change for the most probable \USU values, and become $\hat{\mu} = 0.5724 (34)$ (0.6\%)  and $\hat{\mu} = 0.5782 (30)$ (0.52\%)  for the ratio at 9~MeV and 10~MeV, respectively. Only for unrealistically large \USU values, when the \USU value is much larger than the reported uncertainties, the evaluated ratios converge to the unweighed average of the data $\hat{\mu}_i$ and becomes $\hat{\mu} = 0.5705(59)$ (1\%) and $\hat{\mu} = 0.5765(58)$ (1\%) for 9~MeV and 10~MeV, correspondingly. The probability of the largest \USU value can be considered to be very small, therefore the latter estimate would be the most conservative one.

For completeness of the present analysis, we have also studied how the estimated mean values, using the polynomial model given by Eq.~\eqref{eq:polyn}, depends on the \USU variance for two different typical cases compared to the case of no assumed \USUn:
$\sigma_\delta = 0$ (no \USUn), $\sigma_\delta = 0.0031$ (\USU $\approx 0.3$\%), and  $\sigma_\delta = 0.0042$ (\USU $\approx 0.7$\%) at 9~MeV, and
similarly $\sigma_\delta = 0$ (no \USUn), $\sigma_\delta = 0.0018$ (\USU $\approx 0.3$\%), and  $\sigma_\delta = 0.0031$ (\USU $\approx 0.5$\%) at 10~MeV.
Estimates $\hat{\mu}$ depend very weakly on the \ql reasonable\qr \USU variances, as shown in Table~\ref{tab:nf_estim_02}, and practically do not depend on the assumed polynomial model.

Inconsistencies between data sets are often due to an underestimation of the (uncertainties of) normalization factors, in  the present example of the uncertainty  $(u_N)$.
An alternative method to identify an underestimation of the uncertainty is to compare the  magnitude of the estimated absolute uncertainty component $u_{\hat{\mu_i}}$
and the difference between the observed value $\hat{\mu}_i$ for each set $i$ and the best estimate $\hat{\mu}$. Applying this method the \USU variance $\hat{\sigma}^2_{\delta,i}$, is calculated using Eq.~\eqref{eq:USU_GMA} for each data set $i$:
\begin{equation}\label{eq:USU_GMA}
\hat{\sigma}^2_{\delta,i}= \max{[0, ((\hat{\mu}_{i} - \hat{\mu})^2 - (u_{\hat{\mu_i}})^2)  ]}.
\end{equation}
This quantity, which can be interpreted as being an individual \USU value for a specific data set, is listed in Table~\ref{tab:nf_estim_01}. A similarly defined additional uncertainty\footnote{The additional uncertainty in the Neutron Standards is numerically different due to different criteria used in its definition.} has been applied to outliers in Neutron Standards  evaluations \cite{Standards:2009,Standards:2018} before carrying out the \GMA\textit{P}~GLS fit.

The group \USU contribution for a given energy can be defined as the averaged \USU variance over all evaluated datasets at that energy bin, as given by:
\begin{equation}\label{eq:group-USU}
\hat{\sigma}^2_{\delta} \equiv  \frac{1}{K}\sum\limits_{i=1}^K \hat{\sigma}^2_{\delta,i}
\end{equation}

The data in Table~\ref{tab:nf_estim_01} suggest that the uncertainties reported by Toveson \etal\cite{Tovesson:2015}, Behrens \etal\cite{Behrens:1977}, Cierjack \etal\cite{Cierjacks:1976} and Lisowski \etal\cite{Lisowski:1991} are all underestimated.

From the data in Table~\ref{tab:nf_estim_01} we estimate $\hat\sigma^{2}_{\delta} \approx 0.0035^2$ (0.6\%)   and $\hat{\sigma}^2_{\delta} \approx 0.0069^2$ (1.2\%) for the \USU contribution at 9~MeV and 10~MeV, respectively. These values are only in the first approximation consistent with those derived from a Maximum Likelihood Estimate (MLE) approach, being the agreement better for the 9 MeV group.
Note that both the MLE approach, \eg, see Eq.~\eqref{eq:MLE} in Appendix~\ref{app:MLE} and Eqs.~\eqref{eq:USU_GMA} and \eqref{eq:group-USU} are founded on the basic assumption that the \USU contribution for each data set is sampled independently from a distribution with the same variance.

\begin{table*}[t]
\vspace{-1mm}
\caption{Energy nodes $\hat{E}^m_n$ and corresponding evaluated cross-section ratio values $\hat{\mu}^m$ ($\hat{\mu}^m_+$) with uncertainties $u^m_{\hat{\mu}}$ ($u^m_{\hat{\mu}+}$) obtained without (with) individual \USU components calculated by Eq.~\eqref{eq:USU_GMA} at each GLS iteration. Evaluated values were derived by iterative \GMA\textit{P}~fit for 11 experimental data sets of the \UF/\UT(n,f) cross-section ratio values (see Table~\ref{tab:U5U8-input} of Appendix~\ref{app:nf-ratio}). The lower-triangle correlation matrix obtained at the final GLS iteration (including \USU components) is also shown.
The group \USU $\hat{\sigma}^m_{\delta}$ at a given energy group $m$ was derived after the final iteration by Eq.~\eqref{eq:group-USU} averaging over all datasets and contributed to the total uncertainty $u^{m,tot}_{\hat{\mu}+} \equiv \sqrt{(u^m_{\hat{\mu}+})^2 + (\hat{\sigma}^m_{\delta})^2}$ at each energy node and to the diagonal of the covariance matrix. Values of uncertainties and the correlation matrix were rounded to 3 digits to fit in the table, so this is an approximate covariance matrix which may not be positive definite due to round-off errors.}
\label{tab:T1}
\vspace{1mm}
\begin{tabular}{c|c|c|c|c|c|c|ccccccccccc}
\hline \hline
\T Node $m$ & $\hat{\mu}^m$ & $u^m_{\hat{\mu}}$ & $\hat{\mu}^m_+$ & $u^m_{\hat{\mu}+}$ & $\hat{\sigma}^m_{\delta}$ & $u^{m,tot}_{\hat{\mu}+}$\\
$\hat{E}^m_n$ (MeV)        & (no \USUn)  & (\%)  & (with \USUn) & (\%)  & (\%)  & (\%) &\multicolumn{11}{c}{Lower triangle of the correlation matrix, $\vec{\rho}$}  \\ \hline
 8.0         &  ~0.5699~ & ~0.51  & ~0.5698~ & ~0.51  &  ~0.29    & 0.59     & ~1.00  &        &       &       &       &       &       &        &        &        &     \\
 8.5         &  ~0.5684~ & ~0.44  & ~0.5679~ & ~0.45  &  ~0.45    & 0.64     & ~0.74  &  1.00  &       &       &       &       &       &        &        &        &     \\
\textbf{9.0} & \textbf{~0.5746~} & \textbf{~0.42}  & \textbf{~0.5745~} & \textbf{~0.43} & \textbf{~0.52} & \textbf{0.68}
                                                                                         & ~0.67  &  0.81  &  \textbf{1.00} &       &       &       &       &        &        &        &     \\
\textbf{10.0}& \textbf{~0.5772~} & \textbf{~0.43}  & \textbf{~0.5765~} & \textbf{~0.45} & \textbf{~1.16} & \textbf{1.25}
                                                                                         & ~0.63  &  0.76  &  \textbf{0.79} & \textbf{1.00}  &       &       &       &        &        &        &     \\
 11.0        &  ~0.5828~ & ~0.46  & ~0.5826~ & ~0.48  &  ~1.00    & 1.11     & ~0.55  &  0.66  &  0.70 & 0.75  &  1.00 &       &       &        &        &        &     \\
 11.5        &  ~0.5914~ & ~0.49  & ~0.5912~ & ~0.52  &  ~0.92    & 1.06     & ~0.53  &  0.62  &  0.64 & 0.70  &  0.71 &  1.00 &       &        &        &        &     \\
 12.0        &  ~0.5784~ & ~0.52  & ~0.5782~ & ~0.69  &  ~2.11    & 2.22     & ~0.45  &  0.55  &  0.53 & 0.61  &  0.61 &  0.70 &  1.00 &        &        &        &     \\
 13.0        &  ~0.5461~ & ~0.43  & ~0.5456~ & ~0.45  &  ~2.88    & 2.92     & ~0.59  &  0.69  &  0.72 & 0.70  &  0.70 &  0.69 &  0.62 &   1.00 &        &        &     \\
 14.0        &  ~0.5567~ & ~0.53  & ~0.5559~ & ~0.55  &  ~0.75    & 0.93     & ~0.50  &  0.56  &  0.56 & 0.56  &  0.54 &  0.56 &  0.54 &   0.68 &   1.00 &        &     \\
 14.5        &  ~0.5706~ & ~0.53  & ~0.5670~ & ~0.58  &  ~2.82    & 2.88     & ~0.49  &  0.57  &  0.57 & 0.58  &  0.53 &  0.57 &  0.61 &   0.65 &   0.66 &   1.00 &     \\
 15.0        &  ~0.5863~ & ~0.49  & ~0.5860~ & ~0.50  &  ~1.33    & 1.42     & ~0.54  &  0.60  &  0.61 & 0.60  &  0.56 &  0.54 &  0.46 &   0.68 &   0.65 &   0.66 & 1.00\\
\hline\hline
\end{tabular}
\end{table*}

\subsubsection{\GMA\textit{P}~Least-squares Estimation}\label{sect:GMA}
% Pronyaev, Roberto
The application of the \GMA\textit{P}~code to generate the Neutron Standards~\cite{Standards:2009,Standards:2018} creates a need to estimate \USU contributions and to include those additional unrecognized uncertainties within the least-squares fit procedure performed by \GMA\textit{P}. Those uncertainties may include correlated and uncorrelated \USU contributions. In the Neutron Standards' fit it is assumed that the first-pass posterior result $\hat{\mu}$ is a very good approximation of the searched \textit{true} value, therefore the \USU component can be determined relative to this value. %e posterior evaluated mean value.%, similarly to the formulation given by Eqs.~\eqref{eq:USU_GMA} and \eqref{eq:group-USU} using the estimated uncertainty $u_{\mu}$ at each iteration for the comparison.

The \GMA\textit{P}~code used  as input the 11 selected experimental data sets listed in Table \ref{tab:U5U8-input} of Appendix~\ref{app:nf-ratio}. The starting individual \USU component $\hat{\sigma}^{m}_{\delta,i}$ is defined as an additional uncertainty for a dataset $i$ at each energy node with index $m$, and it is similar to the quantity given by Eq.~\eqref{eq:USU_GMA} with an additional energy index $m$. It is also assumed that this additional \USU uncertainty is fully correlated over all energies for a given dataset $i$ in the GLS fit. For the following GLS iterations the individual \USU components for each dataset $i$ are recalculated by Eq.~\eqref{eq:USU_GMA} using the newly estimated posterior evaluation $\hat{\mu}^m$. Note that the dataset mean value $\hat{\mu}^m_i$ and the corresponding uncertainty $u^m_{\hat{\mu_i}}$ do not change in the GLS iterations. Very few iterations are needed for convergence.

The total uncertainty $u^{m,tot}_{\hat{\mu}+}$ taking into account the group \USU component, is obtained after convergence from the square root of the sum of variances, evaluated without accounting for the individual \USU component $(u^m_{\hat{\mu}})^2$ and the evaluated group \USU component of uncertainty $(\hat{\sigma}^m_{\delta})^2$.

The evaluated values, with and without considering the individual \USUn, the total uncertainties including the group \USU calculated by Eq.~\eqref{eq:group-USU}, and the lower triangle of the correlation matrix obtained with \GMA\textit{P}~fit considering the individual \USUn, are listed in the Table~\ref{tab:T1}.

The estimated cross-section ratio values $R_{8/5}$ for 9 and 10 MeV are in very good agreement with those obtained in the previous section (within quoted uncertainties). Estimated \USU contributions of 0.52\% and 1.16\% at 9~MeV and 10~MeV (highlighted in bold) are also in fairly good agreement with those listed in Table~\ref{tab:nf_estim_01}, derived by a similar method (0.6\% and 1.2\% at 9~MeV and 10~MeV, respectively), but the value at 10~MeV is larger than the MLE estimate. The introduction of individual \USU components was found to increase only slightly the uncertainty of the evaluation (see small differences between $u^m_{\hat{\mu}}$ and $u^m_{\hat{\mu}+}$ in Table~\ref{tab:T1}), but the group \USU component $\hat{\sigma}^m_{\delta}$ makes a larger contribution.

\begin{table*}[!thb]
%\vspace{-2mm}
\caption{The average relative uncertainties for each data set $i$ resulting from the variance analysis proposed by Badikov~\etal\cite{Badikov:1992}  ($<\sigma_i>, <\sigma_{\eta,i}>, <\sigma_{\epsilon,i}>$) are compared with the ones derived from the reported uncertainties ($<u_i>, <u_{\eta,i}>, <u_{\epsilon,i}>$). The total uncertainties are indicated without a sub-index. For the uncertainties due to systematic and random effects the sub-indices $\eta$ and $\epsilon$, respectively, are used. They are derived by taking the average of the relative values for the three energy groups  multiplied by 100. The last three columns give the ratio between the values from the variance analysis and the reported ones for each component and for the total uncertainty. Uncertainty ratios that differ significantly from one indicate overestimated ($>1$ in \r red\b) or underestimated ($<1$ in \a blue\b) reported uncertainties.}
\label{tab:nf_pade_01}
\vspace{1mm}
\begin{tabular}{l|ccc|ccc|ccc}
\hline \hline
{Dataset}           & \multicolumn{3}{c|}{Reported values} & \multicolumn{3}{c|}{Variance analysis} &  \multicolumn{3}{c}{Ratio}  \\
                         &$<\sigma_i>$& $<\sigma_{\eta,i}>$ & $<\sigma_{\epsilon,i}>$  & $<u_i>$  & $<u_{\eta,i}>$ &$<u_{\epsilon,i}>$  & $total$ & $\eta$  & $\epsilon$\\
 \hline
Tovesson            &0.91 & 0.84 &0.34   &2.38&2.31&0.55  &2.63&\r 2.76\b&\r 1.63\b\\
n\_TOF (1)          &2.40 & 1.73 &1.77   &2.60&2.08&1.56  &1.08&1.20&0.94\\
n\_TOF (2)          &3.65 & 2.49 &2.67   &2.66&2.34&1.27  &0.73&0.94&\a 0.48\b\\
n\_TOF (3)          &3.65 & 3.35 &1.44   &2.43&2.12&1.18  &0.67&\a 0.63\b&0.82\\
n\_TOF (4)          &4.39 & 3.74 &2.30   &3.04&2.29&2.00  &0.69&\a 0.61\b&0.87\\
Behrens             &1.46 & 0.81 &1.21   &2.57&2.26&1.22  &1.76&\r 2.79\b&1.01\\
Difilippo           &2.48 & 2.39 &0.66   &2.77&2.20&1.68  &1.12&0.92&\r 2.54\b\\
Cierjacks           &3.01 & 1.09 &2.81   &2.77&2.36&1.45  &0.92&\r 2.16\b&\a 0.52\b\\
Coates              &3.44 & 2.92 &1.82   &3.27&2.31&2.32  &0.95&0.79&1.27\\
Shcherbakov         &2.53 & 2.45 &0.64   &2.47&2.31&0.88  &0.98&0.94&1.37\\
Lisowski            &1.54 & 0.94 &1.22   &2.59&2.28&1.24  &1.68&\r 2.42\b&1.01\\
 \hline \hline
\end{tabular}
%\vspace{-2mm}
\end{table*}

\subsubsection{Estimation of Average \USU by Pad\'e Approximants}\label{sect:pade-res}
A statistical model to identify averaged (over energy groups) contributions of unrecognized uncertainty components, both correlated and uncorrelated, was proposed by Badikov~\etal\cite{Badikov:1992}. The method is based on a variance analysis and it can be applied when experimental data are reported without uncertainties or with unreliable uncertainties. At present, reported uncertainties are ignored, but estimated uncertainties can be compared with reported ones to assess their quality. A Bayesian procedure that takes into account reported uncertainties was discussed in Ref.~\cite{Schnabel:2017}. A procedure that takes into account reported uncertainties in the GLS adjustment procedure using Pad\'e approximants is in preparation.

\begin{table*}[!tbh]
\vspace{-4mm}
\caption{Estimates of the fission cross-section ratio resulting from an analysis of the experimental data sets specified in Table~\ref{tab:U5U8-input} of Appendix~\ref{app:nf-ratio}. The results are obtained by applying the variance analysis method proposed by Badikov~\etal\cite{Badikov:1992,Badikov:2003}, using Pad\'e approximants as a model function and considering three different energy groups. The best estimates together with the final covariance matrix are given for 11 incident neutron energies. The covariance matrix is represented by the tabulated uncertainties and a correlation matrix.}
\label{tab:nf_pade_03}
\vspace{1mm}
\begin{tabular}{c|c|ccccccccccc}
\hline \hline
%\centering
 Energy  &     Ratio, $\hat{\mu}$    &       \multicolumn{11}{c}{Lower-triangle of the correlation matrix, $\vec{\rho}$}    \\
\hline
8.0  & 0.5670 (28)  &1.000 \\
8.5  & 0.5705 (28)  &0.995 & 1.000 \\
9.0  & 0.5740 (28)  &0.980 & 0.995 & 1.000 \\
10.0 & 0.5811 (29)  &0.916 & 0.949 & 0.976 & 1.000 \\
\hline
11.0 & 0.5872 (31)  &0.798 & 0.845 & 0.888 & 0.961 & 1.000 \\
11.5 & 0.5884 (34)  &0.707 & 0.751 & 0.794 & 0.880 & 0.971 & 1.000 \\
12.0 & 0.5853 (42)  &0.576 & 0.604 & 0.632 & 0.703 & 0.836 & 0.938 & 1.000 \\
\hline
13.0 & 0.5469 (46)  &0.525 & 0.541 & 0.551 & 0.549 & 0.535 & 0.543 & 0.585 & 1.000 \\
14.0 & 0.5583 (48)  &0.515 & 0.533 & 0.545 & 0.551 & 0.542 & 0.548 & 0.572 & 0.956 & 1.000 \\
14.5 & 0.5744 (49)  &0.528 & 0.541 & 0.548 & 0.542 & 0.531 & 0.548 & 0.603 & 0.967 & 0.978 & 1.000 \\
15.0 & 0.5863 (49)  &0.532 & 0.550 & 0.562 & 0.561 & 0.541 & 0.543 & 0.576 & 0.961 & 0.952 & 0.990 & 1.000\\
\hline \hline
\end{tabular}
\vspace{-2mm}
\end{table*}

The error model in Eq.~\eqref{eq:USU} can be applied using any model function.
Pad\'e polynomials~\cite{Pade:1892,Graves-Morris:1973,Baker:1975} have been used as a model function in nuclear data evaluations \cite{Vinogradov:1987,Badikov:1992,Badikov:2003,Standards:2009, Hermanne:2018}, and it can be expressed as a pole expansion~\cite{Badikov:1992}
\eqnarrow
\begin{equation}
f^{[L_1,L_2]}(E,\vec{\theta})=c+\sum\limits_{j=1}^I \frac{a_j}{E-r_j} + \sum\limits_{j=1}^J \frac{\alpha_j(E-\zeta_j)+\beta_j}{(E-\zeta_j)^2+\gamma^2_j}\, , \label{eqpade_1}
\end{equation}
\eqnormal
with the following definitions for the parameters: %$K$ -- the number of experiments, $M$ -- the number of energy groups.
$I$ -- the number of real poles of the Pad\'e approximant. Diagonal $L_1=L_2$, or near diagonal $L_1=L_2-1$, Pad\'e approximants are used, and $J$ can be determined from the equation $L_1+L_2+1=2I+4J+1$. The algorithm for the construction of Pad\'e polynomials was described in Ref.~\cite{Badikov:1992}.

The error model assumes that data within a given data set (given experiment) and energy group share the same unknown uncorrelated uncertainty. Each data set suffers from a systematic error with a variance that can be energy dependent and with covariance terms between energy groups. This unknown systematic error is not common between data sets, however, it is supposed that they have the same covariance matrix for each energy grouped data set.

The parameters of the Pad\'e approximants together with the estimated covariance matrix, including correlated and uncorrelated uncertainties, are derived from the experimental data using an iterative GLS method.

The data base consists of a total of $N$ data points resulting from $K$ different experiments or data sets ($i = 1, ... , K$). Each data set $i$ contains  $n_i$ results of different measurements ($j = 1, ..., n_i$). In the present example the results within a data set are obtained from measurements at a different incident neutron energy.

For each data set $i$ and energy group $m$ a systematic error $\hat{\eta}_{i,m}$ is estimated by:
 \begin{equation}\label{eqpade_2}
 \hat{ \eta}_{i,m} = \frac{1}{n_i}\sum\limits_{j=1}^{n_i}(y_{ij,m} - f(E_{j,m},\vec{\theta})),
  \end{equation}
where $y_{ij,m}$ are the data belonging to the data set $i$ with energies $E_j$ belonging to the energy group $m$.
These values are used to provide estimates of the systematic error $\hat{ \eta}_{m}$ for each energy group  $m$:
 \begin{equation}\label{eqpade_3}
 \hat{ \eta}_{m} = \frac{1}{K}\sum\limits_{i=1}^{K}\hat{ \eta}_{i,m},
  \end{equation}
together with its covariance matrix $\vec{C}_{\eta}$. The diagonal and off-diagonal terms of this matrix, which are denoted by $C_{\eta,{mm}}$ and $C_{\eta,{m_1m_2}}$, respectively, are given by:
\begin{equation}\label{eqpade_4}
  C_{\eta,{mm}} = \frac{1}{K-1}\sum\limits_{i=1}^{K}  (\hat{ \eta}_{i,m} -\hat{ \eta}_{m})^2
  \end{equation}
and
 \begin{equation}\label{eqpade_5}
  C_{\eta,{m_1m_2}} = \frac{1}{K}\sum\limits_{i=1}^{K}  \hat{ \eta}_{i,m_1} \hat{ \eta}_{i,m_2} -  \hat{ \eta}_{m_1}  \hat{ \eta}_{m_2} ,
  \end{equation}
with a constraint that:
 \begin{equation}\label{eqpade_5}
  C_{\eta,{m_1m_2}} \le \min{[C_{\eta,m_1m_1}, C_{\eta,m_2m_2}]}.
  \end{equation}

The variance of the distribution of the random error for a given data set $i$ and energy group $m$ is estimated from the formula
 \begin{equation}\label{eqpade_6}
 \hat{\sigma}^2_{{\epsilon}_{i,m}} = \frac{1}{n_{i} -1}\sum\limits_{j=1}^{n_i}  (y_{ij,m} - f(E_{j,m},\vec{\theta}) - \hat{\eta}_{i,m})^2.
  \end{equation}
These values form a covariance matrix  $\vec{D}_\epsilon$ with only diagonal terms reflecting the contribution of random effects due to, \eg, counting statistics.

The total covariance matrix, \ie, the sum of the covariance matrices $\vec{C}_\eta$ and $\vec{D}_\epsilon$ due to systematic and random effects, respectively,
is used in a GLS analysis to derive new estimates of the parameters $\vec{\theta}$  and their covariance matrix $\vec{C}_{\vec{\theta}}$ of the Pad\'e approximants. These new estimates are then introduced to start a new iteration sequence. The process is repeated until convergence is reached. From the final values ($\vec{\theta}$,$\vec{C}_{\vec{\theta}}$) the estimates of the ratio $R_{8/5}$  together with its covariance matrix are derived.

This method was applied using the 11 fission cross section ratio data sets specified in Table \ref{tab:nf_estim_01}.
The results between 8 MeV and 15 MeV were analyzed considering three energy groups:  $[8,10.33]$ MeV,  $[10.33,12.66]$ MeV and $[12.66, 15]$ MeV.
The results are summarized in Tables \ref{tab:nf_pade_01}, \ref{tab:nf_pade_02} and \ref{tab:nf_pade_03}.

Table \ref{tab:nf_pade_01} reports for each data set $i$ the average relative uncertainties derived from the ratio of the uncertainty  and the best estimate for each of the three energy groups. The average values derived from the reported total uncertainties, and the uncertainties due to systematic and random effects are compared with the values derived from the statistical analysis method proposed by Badikov \etal\cite{Badikov:1992}. The results in Table \ref{tab:nf_pade_01} confirm the conclusions based on Eqs.~\eqref{eq:USU_GMA} and \eqref{eq:group-USU} that the  uncertainties due to systematic effects reported by Toveson \etal\cite{Tovesson:2015},  Behrens \etal\cite{Behrens:1977}, Cierjacks \etal\cite{Cierjacks:1976} and Lisowski \etal\cite{Lisowski:1991} might be underestimated. In addition, there is an indication that the uncertainties due to random effects reported for the n\_TOF data (2)~\cite{Paradela:2015} and the data of Cierjacks \etal\cite{Cierjacks:1976} are overestimated, while those of Difilippo \etal\cite{Difilippo:1978} and Tovesson \etal\cite{Tovesson:2015} are underestimated.

The covariance matrix due to systematic effects for the three energy groups is given in  Table \ref{tab:nf_pade_02}. The data in  Table \ref{tab:nf_pade_02} reveal a relatively strong energy dependence for the uncertainty due to systematic effects. Such an energy dependence is difficult to explain from an experimental point of view, considering the relatively small energy region covered by the data points included in the analysis. The values in Table \ref{tab:nf_pade_02} are not consistent with those derived from a MLE analysis.

The final  estimate of the cross section ratio, together with its covariance matrix, is reported in  Table~\ref{tab:nf_pade_03}. The uncertainties for the ratio at 9 MeV and 10 MeV are very close to those derived from a GLS analysis using  only the reported uncertainties without consideration of contributions from \USUn. This is most probably due to the fact that the additional systematic effect is not treated as a common source of uncertainty between the data sets. Therefore, its impact will be reduced by increasing the number of data points.

\begin{table}[!thb]
\vspace{-2mm}
\caption{Results of the variance analysis proposed by Badikov~\etal\cite{Badikov:1992} applied to cross-section ratio data specified in the first column. The estimation of a systematic effect for three energy groups is given together with the corresponding covariance matrix. The covariance matrix is represented by the square root of the diagonal terms and the correlation matrix.}
\label{tab:nf_pade_02}
\begin{tabular}{c|c|ccc}
\hline \hline
\T Energy group  &  &  & & \B\\
$m$ (MeV)    & $\frac{\sqrt{C_{\eta,{m,m}}}}{\hat{\mu}_{m}}$ (\%) &  & Correlation matrix $\vec{\rho}$ & \B\\
\hline
8--10.33     & 1.50 & 1.00 &      &      \\
10.33--12.66 & 2.35 & 0.64 & 1.00 &      \\
12.66--15    & 3.08 & 0.49 & 0.76 & 1.00 \\
 \hline \hline
\end{tabular}
\vspace{-2mm}
\end{table}

\subsubsection{Discussion of Results Using Different Evaluation Methods}\label{subsect:eval-res}
\vspace{3mm}
We have offered results to demonstrate that if data are analyzed in a consistent way there is almost no difference in the results that can be attributed to: the interpolation
scheme (proxy function); treatment of correlated uncertainty component; almost no PPP effect, \textit{etc}.
So one can easily avoid such a source of discrepancies. Indeed, all our analyses show similarly underestimated correlated uncertainties for four datasets, independently of the method used.
However, there is a difference in treating the regions individually or together in accounting for the correlated uncertainty component.

With the assumption that the error model contains an \USU component, a rather consistent value of 0.6\% at 9~MeV was obtained by the various methods. The estimated \USU value at 10~MeV increases from 0.3\% up to 1.6\%, showing a larger variability within the methods. However this must be interpreted with caution, as several datasets are consistent with an error model without \USUn.
%An estimated \USU component of $\approx 0.6$\% at 9 MeV was obtained rather consistently by the various methods. The estimated \USU value at 10 MeV increases from 0.3\% up to 1.6\%, showing a larger variability within the methods.

Evidently, uncorrelated (not systematic) uncertainty components (\USUn) do not have a strong impact when a large data base is available. However, it is difficult to identify if data suffer from a common systematic \USU component, and even more difficult to quantify the magnitude of that uncertainty if it exists. Undoubtedly, such a hidden systematic uncertainty component will lead to \USU correlations that will have an impact and increase the evaluated uncertainties.

\subsection{\USU in Highly-Enriched Uranium (HEU) Criticality Experiments}\label{subsect:HEU}
% Roberto
Since the earliest years of applied nuclear technology, integral systems of less complexity than full-scale systems like nuclear power reactors, and commonly referred to as benchmarks, have been used to validate or adjust differential nuclear data and to test simulation computational procedures. This integral approach was the only practical one decades ago\footnote{Probably it is still the only one today as well.} when many fundamental nuclear data needed for simulation computations often were either unknown or insufficiently reliable. It was also perceived that experiments on these integral systems, especially those measuring criticality, were capable of generating far more accurate and precise results than those involved in measuring fundamental physics observables. Furthermore, due to existing computational limitations, the attainable accuracy of numerical simulations was lacking. Many such integral experiments were performed and their results documented in the literature over a period of many decades. Since most of these experiments cannot be repeated for a variety of reasons, the importance of preserving these data and accompanying information about the experiments cannot be overemphasized.

Several years ago the OECD Nuclear Energy Agency (NEA) organized, and it continues to coordinate, an international project aimed at collecting and making readily available detailed and verified information about benchmark experiments performed in different laboratories worldwide since the earliest days of nuclear technology. The main focus of this effort is on criticality safety matters. This information resource continues to be used for validation and/or adjustment of nuclear data, among other applications. The Handbook of International Criticality Safety Benchmarks Evaluation Project (ICSBEP)~\cite{ICSBEP:2016} contains a very large number of evaluations\footnote{This Handbook documents in excess of 5000 different integral benchmark configurations.} including critical, near-critical, and subcritical configurations.

More recently, the improved quality and scope of evaluated differential data, such as cross sections, as well as the burgeoning power of computers and use of highly sophisticated nuclear system models and simulation software, have led to a narrowing of the agreement and accuracy gaps between measured integral data (denoted in this section as $E$) and corresponding calculation results (denoted here as $C$). In fact, achievement of system $C/E$ ratio values approaching unity to within better than 1\% (= 1000 pcm) is now the expected norm for criticality parameters such as neutron multiplication factor $k_\mathrm{eff}$.	

The large international effort devoted to the project of evaluating, documenting, and cataloging integral data in readily accessible, machine-usable formats is usually justified on the grounds that system simulations are likely to continue to play an ever increasing role in future nuclear technology development. Additionally, it is assumed that integral data, which still tend to be considered as more accurate, on average, than differential data, will continue to be used to guide differential data development or to adjust differential data to specific applications.

\textit{Is the broad general assumption that integral criticality benchmark data are inherently more accurate than differential data universally valid?}
Since quoted accuracies that approach 0.1\% for certain experimental criticality integral data are fairly common, this would appear to be a rather bold assumption. Could it be that instances of \USU may exist in some data from integral measurements, especially since all criticality measurements involve using almost exactly the same technique? If so, this would cause one to doubt that integral data are \textit{always} the most accurate. This is a valid question that merits further exploration. In this section, we examine this possibility of \USU existing in integral benchmark data by exploring a relatively simple situation.

\begin{table}[t]
\vspace{-2mm}
\caption{List of HEU metallic bare (and fast spectrum) assemblies from the ICSBEP Handbook~\cite{ICSBEP:2016}. The first five configurations are spherical. The remaining thirteen are cylindrical assemblies with different height to radius ratios.}
\begin{tabular}{r|lll}
\hline\hline
\T No.&      ICSBEP Label & Short name &  Common name         \\
\hline
 1 &  \tt \small HEU-MET-FAST-001 &  hmf001    &   Godiva             \\
 2 &  \tt \small HEU-MET-FAST-008 &  hmf008    &   VNIIEF\_bare       \\
 3 &  \tt \small HEU-MET-FAST-018 &  hmf018    &   VNIIEF\_Sphere     \\
 4 &  \tt \small HEU-MET-FAST-100 &  hmf100-1  &   ORSphere-1         \\
 5 &  \tt \small HEU-MET-FAST-100 &  hmf100-2  &   ORSphere-2         \\ \hline
 6 &  \tt \small HEU-MET-FAST-015 &  hmf015    &   VNIIEF\_UnrCy1     \\
 7 &  \tt \small HEU-MET-FAST-065 &  hmf065    &   VNIIEF\_UnrCy2     \\
 8 &  \tt \small HEU-MET-FAST-051 &  hmf051-01 &   ORCEF-01           \\
 9 &  \tt \small HEU-MET-FAST-051 &  hmf051-02 &   ORCEF-02           \\
10 &  \tt \small HEU-MET-FAST-051 &  hmf051-03 &   ORCEF-03           \\
11 &  \tt \small HEU-MET-FAST-051 &  hmf051-04 &   ORCEF-04           \\
12 &  \tt \small HEU-MET-FAST-051 &  hmf051-09 &   ORCEF-09           \\
13 &  \tt \small HEU-MET-FAST-051 &  hmf051-14 &   ORCEF-14           \\
14 &  \tt \small HEU-MET-FAST-051 &  hmf051-15 &   ORCEF-15           \\
15 &  \tt \small HEU-MET-FAST-051 &  hmf051-16 &   ORCEF-16           \\
16 &  \tt \small HEU-MET-FAST-051 &  hmf051-17 &   ORCEF-17           \\
17 &  \tt \small HEU-MET-FAST-051 &  hmf051-18 &   ORCEF-18           \\
18 &  \tt \small HEU-MET-FAST-080 &  hmf080    &   Caliban\footnote{discarded, see text} \\
\hline\hline
\end{tabular}
\label{Table:U5bare}
%\vspace{-2mm}
\end{table}

It is well understood that it is difficult to identify sources of discrepancies and estimate uncertainties in $\Delta k_{\mathrm{eff}}\equiv C-E$ comparisons for complicated integral benchmarks that incorporate many types of materials and reaction processes. Therefore, we have selected eighteen bare, HEU criticality benchmarks from the ICSBEP catalog that were found to be suitable for this exercise, owing to their relative simplicity and very limited number of materials involved.
Being ICSBEP benchmarks for critical assemblies, the experimental values are usually given as a $k_\mathrm{eff}$ close to unity with an uncertainty $u$. The given benchmark uncertainty includes both experimental uncertainty as well approximations of the computational model of the benchmark.
The selected list, using identification codes from the ICSBEP catalog, is given in Table~\ref{Table:U5bare}, and  $\Delta k_{\mathrm{eff}}$ comparison results are shown in Fig.~\ref{fig:benchm-bare}.

Of these benchmarks, one of them (Caliban -- HMF--080) is excluded in Fig.~\ref{fig:benchm-bare}. The model of the French Caliban cylindrical assembly corresponds by far to the highest calculated $k_{\mathrm{eff}}$ of all the benchmarks considered here, differing by nearly +1\% (+1000 pcm $\equiv$ parts per 100,000) from the measured benchmark value of 1. A quick review of the benchmark model and benchmark input data undertaken by Oscar Cabellos\footnote{Private communication, December 2016.} revealed that the fissile material mass calculated from the volumes and the densities in the computational model is 0.5\% higher than specified in the benchmark documentation. This would account for about 400 pcm of the discrepancy, which is therefore not to be treated as an \USUn. However, this would still leave an additional discrepancy of about 500 pcm which could possibly originate from an \USU contribution. Further investigation of this benchmark and correction of its input data are certainly merited before undertaking an \USU analysis.

Differences $\Delta k_{\mathrm{eff}}$ between the calculated $k_{\mathrm{eff}}$ and the 17 remaining benchmark values are shown in Fig.~\ref{fig:benchm-bare} for IAEA CIELO~\cite{IAEA-CIELO,Capote:2018-U} (=ENDF/B-VIII.0~\cite{ENDFB-VIII:2018}), JEFF-3.2~\cite{JEFF32} and ENDF/B-VII.1~\cite{ENDFB71} nuclear data evaluations.

\begin{figure*}[!bht]
\centering
\vspace{-2mm}
\includegraphics[width=0.99\textwidth]{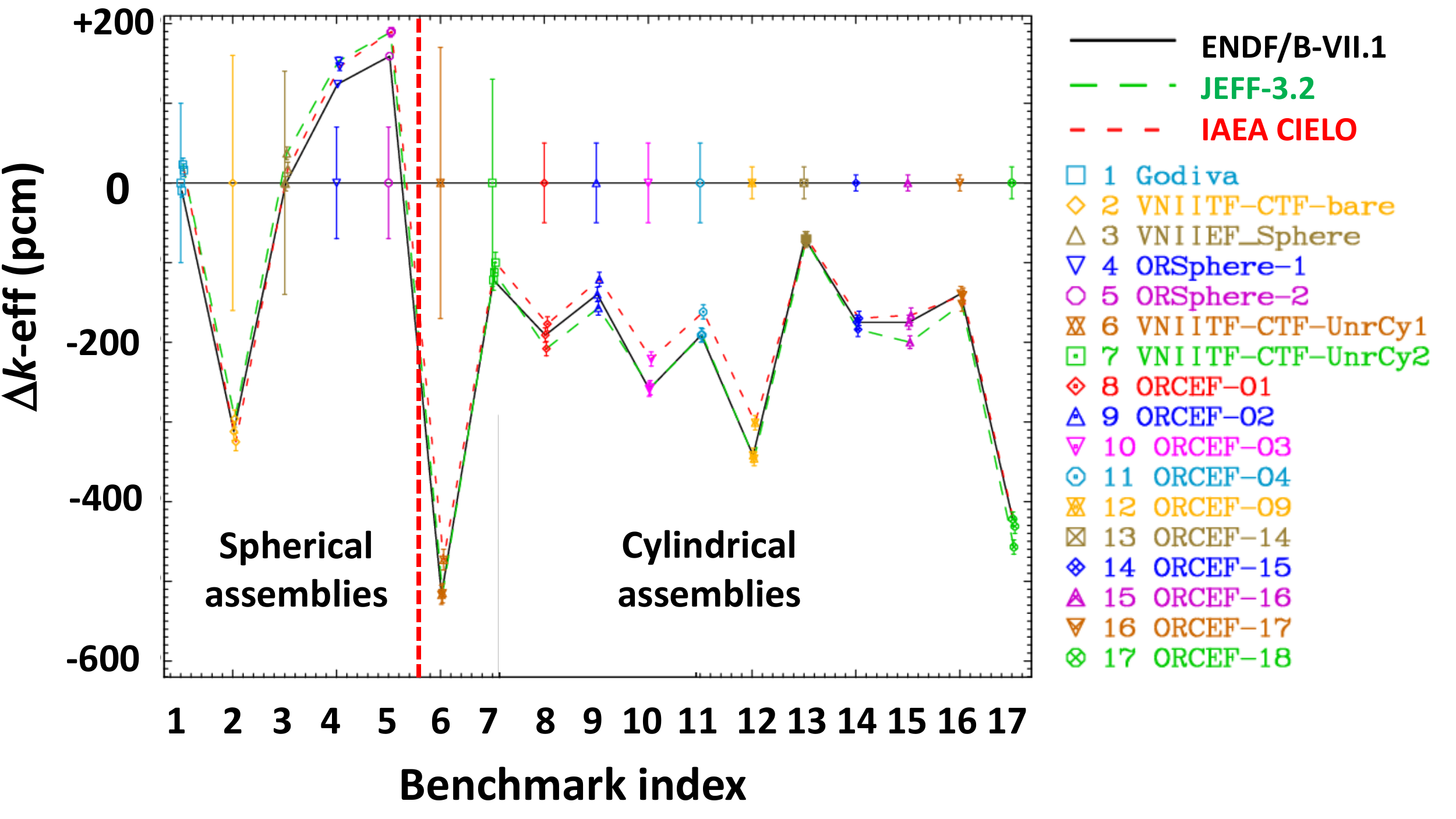}
\vspace{-4mm}
\caption{(Color online) Benchmark results for HEU metallic bare fast assemblies from ICSBEP benchmarks~\cite{ICSBEP:2016} are shown for 17 of the 18 assemblies from Table~\ref{Table:U5bare} (Caliban is excluded). The symbols are linked by lines to show the calculated $\Delta k_{\mathrm{eff}}\equiv C-E$ values (expressed in pcm) obtained from using the three different nuclear-data libraries. Clearly, there are significant discrepancies in $\Delta k_{\mathrm{eff}}$ that exceed the assigned uncertainties.}
\label{fig:benchm-bare}
\vspace{-4mm}
\end{figure*}

%\begin{itemize}
%\item The Russian bare cylinders and spheres HEU-MET-FAST-008, HEU-MET-FAST-015, HEU-MET-FAST-065, and HEU-MET-FAST-018 show a spread of more than 500 pcm with a bias to a lower reactivity,
%    obtained using exactly the same nuclear data as other mentioned benchmarks that show better agreement. Therefore, for benchmarks HEU-MET-FAST-008, HEU-MET-FAST-015, and HEU-MET-FAST-065 we need to look for other possible origins of the discrepancies other than nuclear data if we assume that benchmarks HEU-MET-FAST-001 and HEU-MET-FAST-018 are acceptable.
%\item The six Oak Ridge cylinders HEU-MET-FAST-051 exhibit a relatively small spread, but their C/E values are found to be lower compared to Godiva (HEU-MET-FAST-001) by about 250 pcm. The HEU-MET-FAST-051 suite actually includes 17 cases. Since they all show similar trends, only six of them are included in the present analysis. Some of these benchmarks exhibit unreasonably small uncertainties. On the other hand, the two Oak Ridge sphere HEU-MET-FAST-100 cases yield $C/E$ values that are high when compared to Godiva by about 200 pcm. Overall, the Oak Ridge benchmarks also have a spread of nearly 500 pcm. The origins of these ORNL benchmark discrepancies clearly need to be investigated.
%nvestigation of this benchmark is certainly merited.
%\end{itemize}
%
Specific comments on the individual benchmark experiments listed in Table~\ref{Table:U5bare} have been discussed in Ref.~\cite{Capote:2018-U}.
Some additional comments on sources of $\Delta k_{\mathrm{eff}}$ discrepancies in these benchmarks, as well as a more general discussion of discrepancies and the possibility of encountering \USUn, appear below:
The first five benchmarks listed in the Table (with indexes 1--5) are highly enriched \UT~spheres (HEU).
The benchmarks with indexes 6--17 are cylinders with different geometries (ratios of the heights to the diameters) consisting of two separated parts.

The $\Delta k_{\mathrm{eff}}$ results for spherical benchmarks obtained using different nuclear data libraries coincide practically within the limits of statistical uncertainty of the Monte Carlo simulation calculations, therefore discrepancies cannot be assigned to nuclear data; those discrepancies are clearly related to differences between the experimental measurements and the computational models.
We may conclude that uncertainties estimated for the group of spherical metal fast benchmarks 1--5 contain contributions from \USUn; being that the spread of
$\Delta k_{\mathrm{eff}}$ of about 450 pcm (from $\approx -300$~pcm for HMF-008 to $\approx +150$~pcm for HMF-100 benchmark). Note that the observed spread should be compared with the average estimated experimental uncertainty of about 100 pcm. The spread is a factor of 4 larger than the declared (recognized) uncertainty!

This is less obvious for a group of cylindrical benchmarks (indexes 6--17), which shows a smaller spread than the spherical ones.
%The $\Delta k_{\mathrm{eff}}$ results for cylindrical benchmarks are sensitive to the used nuclear data, especially angular distributions (different neutron leakage because of different geometry).
On average, all cylindrical benchmarks exhibit lower calculated values by about 250 pcm than the spherical benchmarks. % However, this bias could be potentially eliminated by adjusting the angular distributions.
At the same time, the estimated experimental uncertainties given in the specifications for the cylindrical benchmarks 12 to 17 are considered to be unusually small.

As discussed above, the observed biases in $\Delta k_{\mathrm{eff}}$ of spherical metallic fast benchmarks cannot be removed by any adjustments of the evaluated nuclear data used in the analyses. This conclusion also does not depend on the particular method or code used for the calculations of $k_{\mathrm{eff}}$ (Monte Carlo, deterministic, \textit{etc.}), provided that the same code is used for all the benchmark calculations. Therefore, additional factors that influence the $\Delta k_{\mathrm{eff}}$ outcomes need to be considered.

The most significant quantity in all these criticality experiments is the mass of the assembled fissile material in the benchmark assembly when it is close to the criticality point (critical mass). Measured physical parameters (other than nuclear data) that influence determination of the critical mass are the geometry of the fissile material, its isotopic composition and density, the geometry and material composition of the support structure, and characteristics of the surrounding environment. The neutron leakage from cylinders with various geometries is rather sensitive to the energy-angular distribution of neutrons scattered in the benchmark fissile material and less sensitive for geometrically similar spheres. Uncertainties of the physical parameters (\eg, dimensions) for manufactured cylinders of fissile material are much smaller than those for comparable spheres. Criticality measurements are very sensitive to fissile material geometries. Errors in determining benchmark physical parameters, underestimation of their uncertainties, failure to apply required corrections, and deficiencies or simplifications in the models used in simulations can affect the $\Delta k_{\mathrm{eff}}$ comparisons significantly. The excessive impact of these benchmarks on data validation and/or data adjustment could be mitigated by considering \USUn.

Numerical mistakes by an integral data evaluator in cataloging the benchmark facility physical parameters for the ICSBEP documentation could also occur. While this might affect one or more experiments randomly (outliers), it is unlikely to affect a collection of experiments of the same type systematically, as observed here, unless all of them were evaluated by the same individual.
It should be possible to resolve some of these sources of discrepancy provided that adequate records were generated and well maintained, and access to other important experimental information (say from the original experimenter or evaluator) could be obtained (see such an example below). Also, some discrepancies could be removed by repeating measurements or by rechecking some of the facility parameters (dimensions, \textit{etc}.). However, it would be difficult or even impossible to do this for many of these benchmarks since, for most of these experiments, the apparatus is no longer assembled or operational. Thus, discrepancies that might seem to be removable in principle might, for practical reasons, be not amenable to correction.
%This could lead to consideration of \USUn.

Determination as to whether $\Delta k_{\mathrm{eff}}$ discrepancies, such as those mentioned in this section, can be resolved, or whether \USU components should be introduced, depends on the individual situation. Owing to their importance, it is imperative to examine the matter of possible \USU contributions for all \UT~benchmarks with significant $\Delta k_{\mathrm{eff}}$ discrepancies. Doing so may lead to identifying additional benchmark uncertainties, and that could be important for those cases where unreasonably low-declared uncertainties are reported. An action should be taken to encourage the ICSBEP project curators to assign experienced benchmark evaluators to re-examine the afflicted benchmarks, with the objective being to identify possible reasons for such large discrepancies and, if located, to revise, as needed, the benchmark specifications, quoted benchmark uncertainties, or the corresponding MCNP computational models.

Finally, we offer an example where an existing integral benchmark discrepancy has been resolved. This followed from the re-examination of a particular set of ORNL sphere benchmarks that was performed in 2002 by Mihalczo \etal\cite{Mihalczo:2002}. The revised $k_{\mathrm{eff}}$ obtained from this work ($0.9994\pm0.0004$), after converting the \UT~abundance of these benchmarks to conform to the density of the Godiva I benchmark, is consistent with the Godiva I experimental value ($1.0000\pm0.001$).

\begin{figure*}[t]
\vspace{-3mm}
\centering
\includegraphics[width=\textwidth]{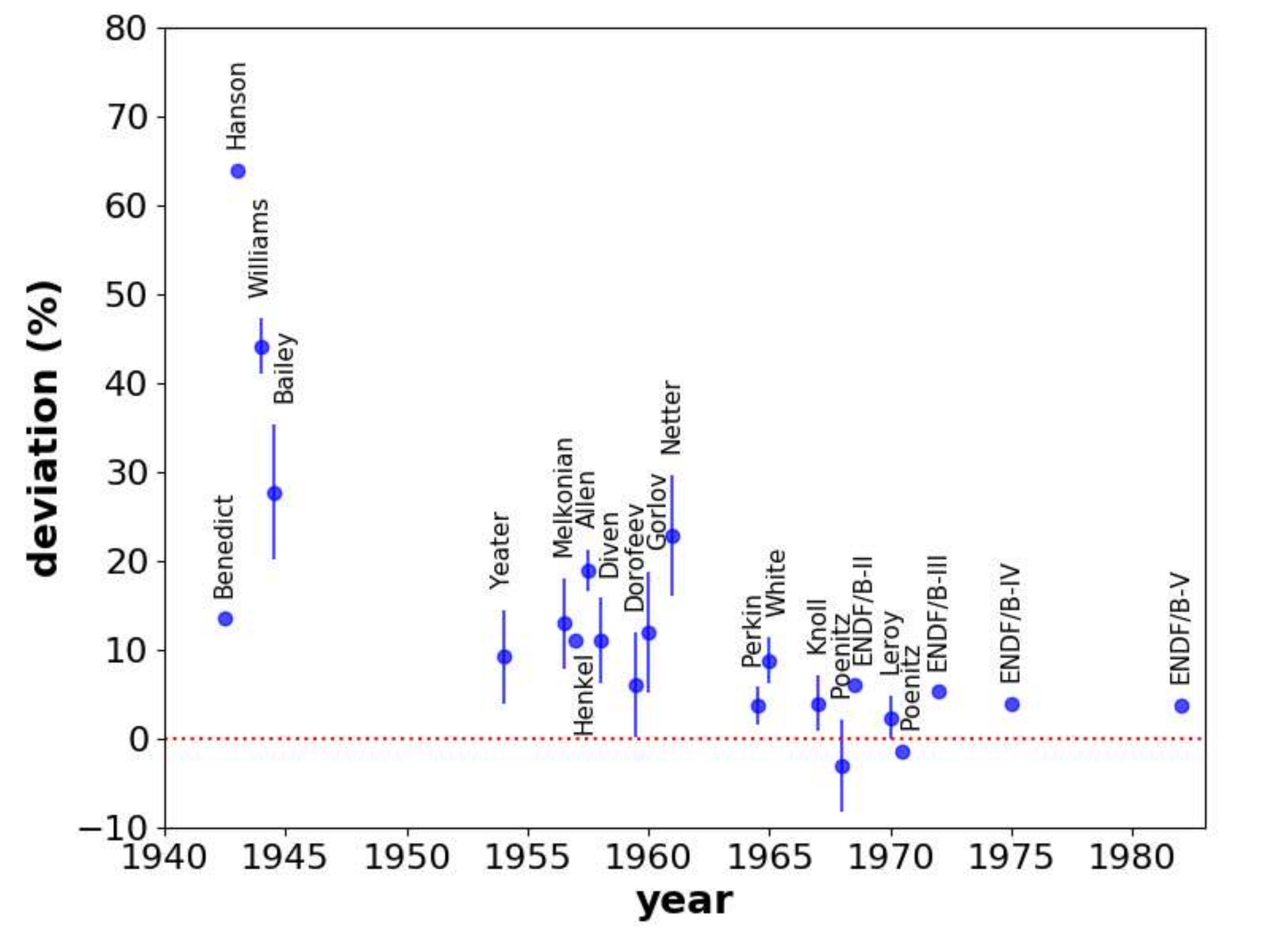}
\vspace{-8mm}
\caption{(Color online) Average deviation of various measurements of the \UT($n$,f) cross section in the 20~keV to 1~MeV energy range from an early least-squares evaluation by Poenitz (1970)~\cite{Poenitz:1970}. The more recent evaluations are not shown to improve visibility. They are generally slightly higher than the Poenitz evaluation.}
\label{fig:U235nf-time}
\vspace{-3mm}
\end{figure*}

\subsection{Resolving \USU and Unmanageable Situations}\label{ssect:unmanag}
%\subsection{Miscellaneous examples}

This section includes four examples that illustrate situations where it would be either impossible, or at best inconclusive or impractical, to try to identify and quantify \USUn, due simply to circumstances or a lack of adequate information. Often, such impossibility is linked to experimental data that have been improperly corrected (or not corrected at all) for a particular physical effect.

\subsubsection{The \UT(n,f) Cross Section as a Function of Time}\label{subsect:u235T}
The {\bf{first example}} that we discuss here involves the anomalous behavior of measured \UT(n,f) cross sections as a function of time over the period 1940 to mid 1980. Fig.~\ref{fig:U235nf-time} shows these results. The data in this plot are the average deviations from a 1970 evaluation reported by Poenitz~\cite{Poenitz:1970} of various measurements and evaluations of the \UT(n,f) cross section in the 20~keV to 1~MeV energy range. These \UT(n,f) cross section data were particularly high in value for the first measurements that were made compared to the Poenitz  evaluation~\cite{Poenitz:1970}. Even up to the early 1960s the cross-section values still appear to be abnormally high. Apparently, for these earliest works the experimental conditions were not the best for such measurements to be made, thus leading to a distinct bias which, if it had been suspected in those times, could have been considered as due to an \USU effect.

\begin{figure*}[!htb]
\vspace{-3mm}
\centering
\includegraphics[width=\textwidth]{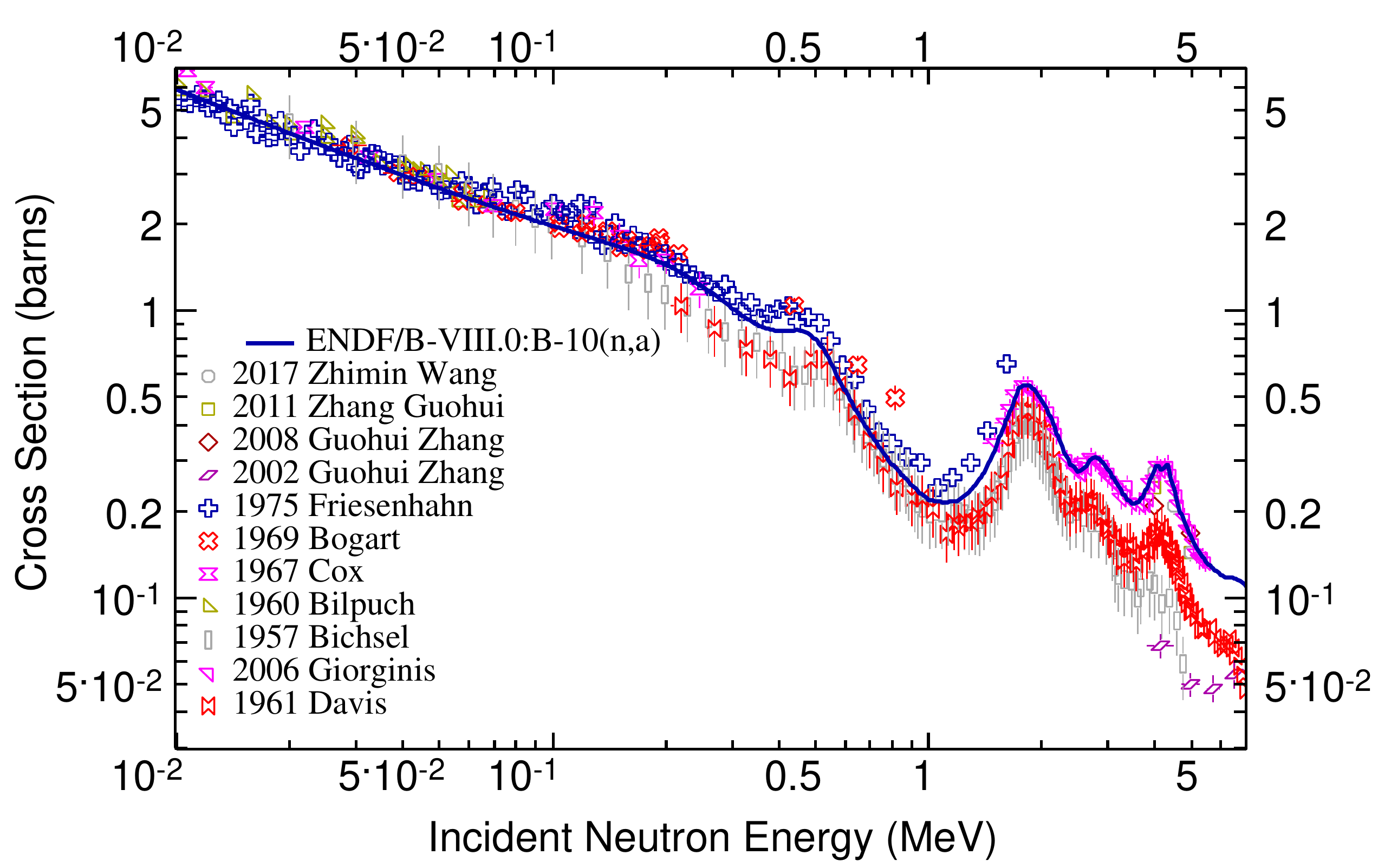}
\vspace{-2mm}
\caption{(Color online) Selected measurements of the $^{10}$B($n$,$\alpha$) cross section compared with the IAEA Neutron Standards evaluation adopted into the ENDF/B-VIII.0 library~\cite{Standards:2018,ENDFB-VIII:2018}.}
\label{fig:particle-leakage}
\vspace{-3mm}
\end{figure*}

It has been conjectured that difficulties in quantifying and correcting for backgrounds accurately may be a large factor in this discrepant trend. Since \UT(n,f) is a non-threshold reaction, with large cross sections at low energy, it is particularly vulnerable to the effects of scattered neutrons which are inevitably lower in energy than primary neutrons. Often a determination of the background effects will be underestimated. Thus the reaction-rate measurement will contain some background contributions that, if underestimated, will lead to a larger cross section. This argument seems reasonable since more modern experiments are designed with improved determination and reduction of backgrounds. Attention to this effect has led to lower cross-section value determinations in more recent times. So, in general, the later investigations appear to have learned from the problems and difficulties of the earlier work, thus leading to more reliable data. While it is not known for certain that above-mentioned background problems can explain all the differences noted for the oldest experiments, where the effect is quite large, or if the experimental conditions of these older measurements could be better established so that corrections could be made now and the possibility of quantifiable \USU sought, it is probably more practical to simply reject the older results, especially in those cases where the work is poorly documented, than to try and rescue these data, or to arbitrarily try to increase their uncertainties.

\subsubsection{Particle Leakage Effect with Ionization Chambers}\label{subsect:leaking}
The {\bf{second example}} we consider here illustrates that unforeseen problems can arise from the introduction of new and presumed more sophisticated measurement techniques if initially they are not completely understood.

Many measurements of the $^{10}$B(n,$\alpha$)$^7$Li and $^6$Li(n,t)$^4$He standard cross sections with charged-particle detection performed using ionization chambers have problems with separation of the reaction-product detector signals, especially at neutron energies in the high-MeV energy range. The use of multi-grid Frisch ionization chambers, with digitalization of the signals and full kinematic analysis, was introduced to enable better separation of events from the different reaction products, \eg, see~\cite{Giorginis:2006}. However, it was discovered that there exists a problem when both reaction products are emitted in the same (or nearly the same) direction in the chamber. An analysis of the resulting signal would determine that it does not possess the proper characteristics for one of the reaction products and it would be rejected. This effect has been called \textit{particle leaking}. It leads to a loss of events and thus a systematically lower cross section.

\begin{figure*}[!tbh]
\vspace{-3mm}
\centering
\includegraphics[width=\textwidth]{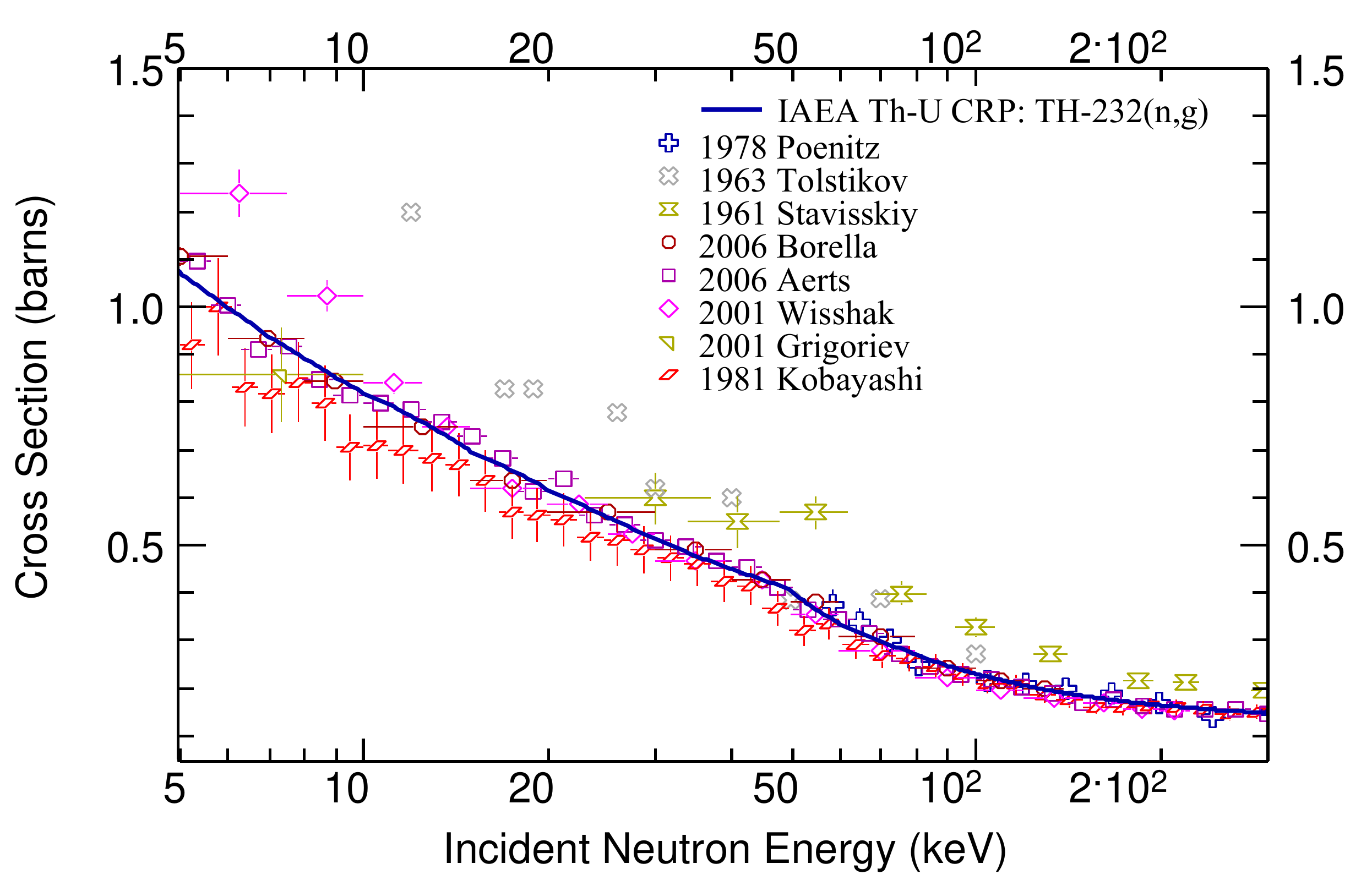}
%\caption{(Color online) Selected measurements of the \TH($n$,$\gamma$) cross section compared with the ENDF/B-VIII.0 evaluation~\cite{ENDFB-VIII:2018,Standards:2018}.}
\vspace{-3mm}
\caption{(Color online) Selected measurements of the \TH($n$,$\gamma$) cross section compared with the IAEA Th-U CRP evaluation~\cite{Sirakov:2008,IAEA-th-eval:2010}.}
%\footnote{This evaluation has been adopted by ENDF/B-VII.1~\cite{ENDFB71}, ENDF/B-VIII.0~\cite{ENDFB-VIII:2018}, and JEFF-3.3~\cite{JEFF33} libraries.}
\label{fig:water_in_sample_figure}
\vspace{-3mm}
\end{figure*}

An example of this effect is seen in the measurements of $^{10}$B(n,$\alpha$)$^7$Li performed on two separate occasions at the same laboratory using the same multi-grid fission chamber, by Zhang \etal in 2002~\cite{Zhang:2002} and by Zhang \etal in 2011~\cite{Zhang:2011}. For the earlier measurements, it was not known that \textit{particle leaking} had occurred. Thus, the measured cross sections were significantly lower than other measurements, as is seen in Fig.~\ref{fig:particle-leakage}. Initially these data could have been considered to have had an \USU component. For the later measurements it was understood that \textit{particle leaking} can occur in such experiments, and a more complete kinematic analysis was carried out to remove its effect. Those data agree with recent measurements, such as those of Giorginis and Khryachkov~\cite{Giorginis:2006} who understood the \textit{particle leaking} problem, as is shown in Fig.~\ref{fig:particle-leakage}.

Thus measurements with ionization chambers that do not take \textit{particle leakage} into account, and for which the \textit{particle leaking} effect cannot be corrected, should not be considered as containing a correctable \USU contribution.  A component of uncertainty should not be added to these data to compensate for a deficient correction of raw experimental data.
In most cases it would be best to discard such data when it is known or assumed that such a correction was not made.
% However, an additional component of uncertainty possibly could be added to these data to account for the oversight.

\subsubsection{\TH(n,$\gamma$) Measurement with Complications Due to Water in the Sample}\label{subsect:th232g}
The {\bf{third example}} is illustrative of the influence that a sample-related impurity that is difficult to identify and quantify, and that can lead to measured cross-section results that cannot be easily corrected, can have on the experimental results.

% Allan
Neutron capture cross-section measurements made using the activation technique are sometimes performed using rather thick samples, thereby requiring corrections for neutron multiple scattering and interfering  reactions. The scattering of incident neutrons in these samples reduces the average neutron energy. In some cases, when the sample is a hygroscopic chemical compound that can absorb a substantial amount of water impurity, the effect can be large. This results from scattering on hydrogen and oxygen in the sample that also will significantly reduce the average neutron energy (due to light hydrogen and oxygen atoms assuming considerable recoil energy). Furthermore, due to the increase of capture cross sections at lower neutron energy, this will effectively increase the sample activation and the cross section derived from the data if left uncorrected. However, while correction for this effect is possible in principle, it requires knowledge of the exact amount of water in each sample. This  is usually very difficult, if not impossible, to determine. Also, for obvious reasons it could be variable during the course of the experiment.

Early neutron capture activation measurements by Stavisskii and Tolstikov~\cite{Stavisskii:1961} and Tolstikov\etal\cite{Tolstikov:1963} with sealed \TH~oxide samples are shown in Fig.~\ref{fig:water_in_sample_figure}. As later measurements were made of this cross section with metal samples, it became clear that cross sections derived from these earlier data were significantly larger than those obtained with metal samples. Not all measured experimental data sets are shown in Fig.~\ref{fig:water_in_sample_figure}. Although the authors tried to avoid water absorption in their samples, it was found later that the increase of the capture cross section could be explained qualitatively by the presence of water in those samples. Initially this difference might have been considered to be due to an \USU effect. However, in this case, the physical origin of the discrepancy became known but no correction for the neutron multiple scattering due to hydrogen and subsequent capture in \TH, could be made. Therefore, a large uncertainty (not an \USU component) should be assigned to these data if they are to be considered for evaluation purposes. An equivalent (and perhaps best) alternative is to discard these uncorrected data and not use them for data evaluation.

Later measurements were made by Wisshak \etal\cite{Wisshak:2001} using the time-of-flight technique and direct detection of gammas with a total absorption detector. A thorough analysis of all corrections with different components of uncertainty was performed. Results from this work also show some increase of the capture cross section below 12~keV, as is shown in Fig.~\ref{fig:water_in_sample_figure}. This increase in the cross section cannot be explained by any physical effects considered in model calculations. Two new independent measurements by Borella \etal\cite{Borella:2006} and Aerts \etal\cite{Aerts:2006},  that also are shown in Fig.~\ref{fig:water_in_sample_figure}, are very consistent and do not exhibit this energy dependence. The origin of this deviation in the Wisshak \etal data is not clear, and an \USU component should be assigned to these data below 12~keV. Aerts \etal\cite{Aerts:2006} also came to the following conclusion: \textit{... that many of the discrepancies with previous experiments, when exceeding the reported systematic uncertainties, could be explained by unrecognized errors in the normalization procedures, affecting the whole data set in the same way, or by insufficient corrections to background estimations, applied weighting functions, or flux determinations ...}. This conclusion conforms well with the present conceptualization of \USUn.

\subsubsection{Accelerator Mass Spectrometry (AMS) vs Time-of-Flight (TOF) for (n,$\gamma$) Cross-section Measurements}\label{subsect:AMS}
The {\bf{fourth example}} illustrates large systematic differences between two methods of measurement of the same physical quantity that may suggest the need for yet unidentified corrections and additional uncertainty of an \USU nature.
%% Toni
AMS and TOF represent two independent and complementary techniques, \eg, for cross section measurements. Fig.~\ref{fig:AMS-vs-nTOF} shows results for four neutron capture reactions for the specific case of 30~keV Maxwellian-Averaged Cross Sections (MACS). For these reactions both AMS and TOF measurements had been performed recently as they are of particular interest to nuclear astrophysics.

\begin{figure}[!thbp]
\vspace{-3mm}
\centering
\includegraphics[width=0.98\columnwidth]{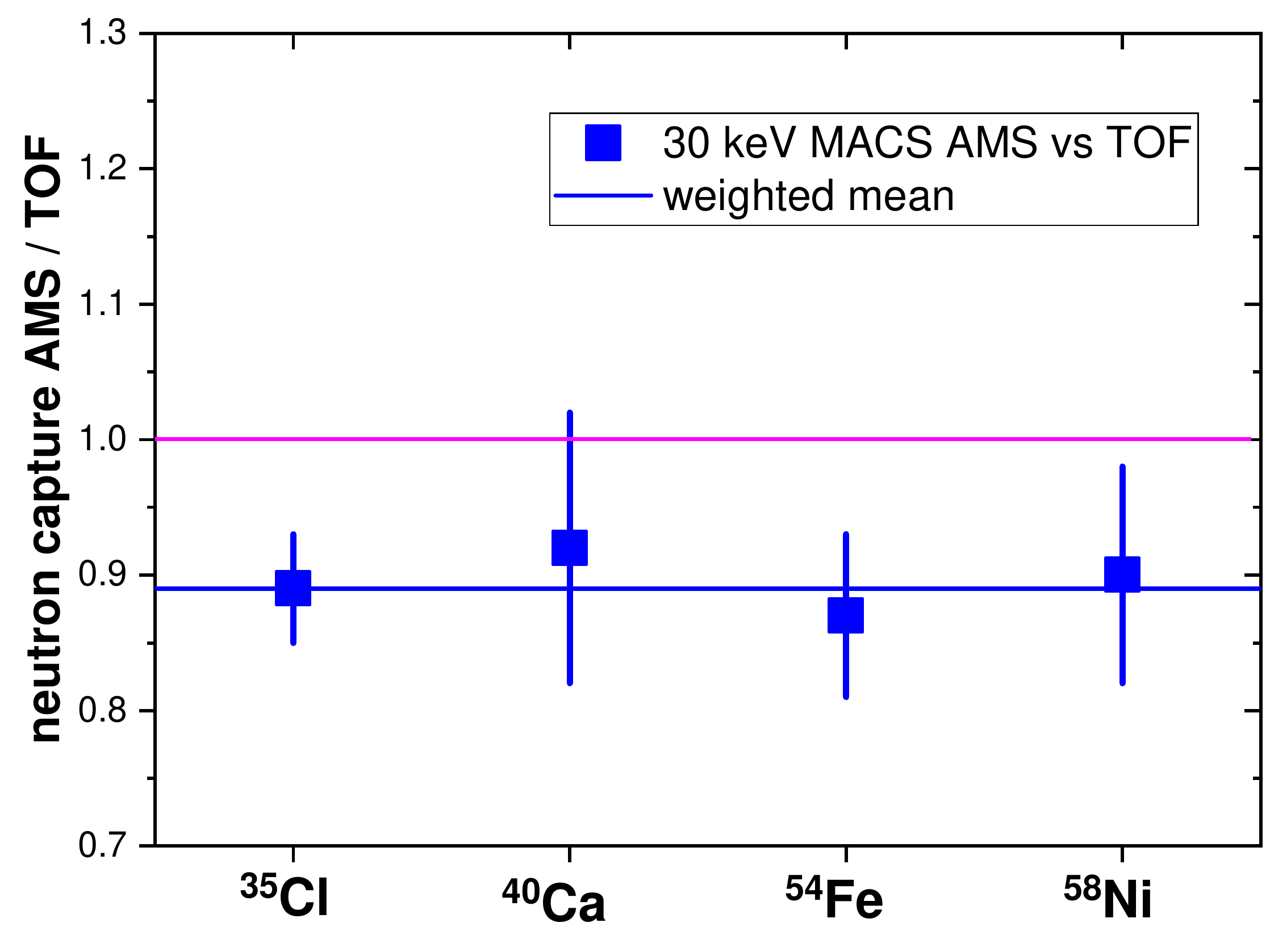}
\vspace{-3mm}
\caption{(Color online) Ratio of Maxwellian Averaged Cross Sections (MACS) for neutron induced capture reactions at 30~keV measured with Time-of-Flight (TOF) \textit{vs.} Accelerator Mass Spectrometry (AMS) measurements.}
\label{fig:AMS-vs-nTOF}
\vspace{-2mm}
\end{figure}

\begin{table}[!thb]
\vspace{-2mm}
\caption{Comparison of Maxwellian Averaged Cross Sections (MACS) for neutron induced capture reactions at $30$~keV measured with TOF vs AMS measurements. All cross sections are given in mb.}
\label{tab:AMS-vs-TOF}
\centering
\begin{tabular}{c|l|l|l}
\hline \hline
   Reaction             &     AMS data                   &  TOF data                                  & Ratio            \\
\hline
\T
$^{35}$Cl(n,$\gamma$) &  8.33(0.32)~\cite{PWD19}  & 9.39(0.29)~\cite{35Cl-TOF}            &  0.89~(0.04) \\
$^{40}$Ca(n,$\gamma$) &  6.18(0.37)~\cite{DHK09}  &  6.7(0.70)~\cite{40Ca-TOF}            &  0.92~(0.10) \\
$^{54}$Fe(n,$\gamma$) &  26.7(1.10)~\cite{WBB17}  & 30.8(1.60)~\cite{54Fe-TOF}            &  0.87~(0.06) \\
$^{58}$Ni(n,$\gamma$) &  30.4(2.3)~\cite{LRD17}  & 34.1(1.70)~\cite{58Ni-TOF1,58Ni-TOF2} &  0.89~(0.08) \\
\hline
Average                 &                                &                                            &    0.89          \\
\hline \hline
\end{tabular}
\vspace{-2mm}
\end{table}

AMS measurements are independent of the half-lives of the reaction products. The combination of activation and subsequent AMS measurement of the same quantities was applied for a range of measurements. The AMS method is a two-step process: the first step is the irradiation of a sample (based on the standard activation technique). The second step is the subsequent AMS detection of the reaction product. The experimental cross section is calculated from two quantities, the isotope ratio (conversion ratio), which is directly measured by AMS, and the neutron fluence. The fluence is usually determined independently, \eg, in case of the neutron irradiations from gold monitor foils simultaneously irradiated with the samples\footnote{Note that the revised Au cross section value~\cite{Standards:2018} was applied for all AMS data.}. The particular advantage of the AMS method is that the cross section is determined by the measured isotope ratio only, completely independent of the sample mass and the decay properties of the product nucleus.

In a series of irradiations at Karlsruhe Intitute of Technology (KIT), neutron capture reactions for a $25$--keV Maxwell-Boltzmann neutron energy distribution were studied, mainly for reactions that are relevant for s-process nucleosynthesis. AMS data obtained from such identical activations are used here, but they were converted into $30$-keV MACS values for a direct comparison with TOF data.
TOF measurement techniques have been well-established for several decades. In TOF, usually the prompt signature from the de-excitation of the reaction product is measured and used to generate a cross-section value. This approach involves different and also more sophisticated data processing procedures compared to AMS. The TOF data plotted here are based on measurements at ORELA ($^{35}$Cl, $^{40}$Ca and $^{58}$Ni) and n\_TOF/CERN ($^{54}$Fe and $^{58}$Ni).
%% Anton: $^{58}$Ni has been measured twice: the value in the final table, is it a mean value?
In the activation reaction studies for AMS, the irradiation setup was designed such that the integrated neutron energy distribution resembled closely a Maxwellian distribution. Thus the measured cross section approximated an energy-averaged value. In TOF the energy-dependent cross sections are folded with the respective neutron flux energy-distribution for the MACS.
There are no correlations between these two techniques. We find a systematic deviation between the two methods as seen in Table \ref{tab:AMS-vs-TOF}. AMS data are systematically lower by about 11\% compared to TOF data for these four reactions. In general, AMS measurements are normalized to reference materials which are independent from each other, and thus there is no correlation in AMS between different reactions. AMS and TOF data were both acquired at two different laboratories (Univ. of Vienna and TU Munich, and ORELA and n\_TOF, respectively). Currently, it is not known what the cause for this systematic deviation could be. The uncertainties of these ratios are generally lower than the observed deviation, leading to a potential estimate of an \USU value. For instance, a numerical value for an \USU contribution could be assumed to be the minimum deviation that make all measured data statistically consistent. If so, this could be estimated to be 5--7\%. %This \USU should be assigned to the data measured by one of the methods.

\section{SUMMARY AND RECOMMENDATIONS}

This paper explores the problem of identifying and estimating ``Unrecognized Sources of Uncertainties'' (\USUn).
These were formally defined in Sect.~\ref{ssec:definition} as those uncertainties for which one has at present no or only very limited understanding of their cause.
\USU contributions adversely impact the accuracy and precision of quantifying an observable.
When we speak here of an observable, it is understood as an evaluated quantity, specifically, in the field of nuclear data evaluation.
But the recommendations presented here might apply beyond this narrow field.

The \USU component is assumed here to become evident by comparing several experimental data sets used as input for evaluating the desired nuclear physics quantities, or derived from specially designed experiments and analysis procedures. The latter option is preferred, but it is often not available.
If \USU contributions are apparent for these data, the resulting evaluated data and uncertainties may be biased and the uncertainties incorrectly estimated. In many cases, the latter will be underestimated.
As nuclear data are the input for computational simulations of nuclear application systems, these simulations can be hampered by \USU contributions that affect evaluated mean values.
Moreover, if the evaluated uncertainties are distinctly underestimated, so will be the simulated safety, economic and operational margins of these systems computed from flawed estimates of the uncertainties.
Hence, it is of importance to identify and estimate the effect of these \USUn.

It should be emphasized that \USU refers to ``unrecognized'' sources of uncertainties as the term implies.
That means that before considering the possibility of \USU effects, all known sources of biases and uncertainties should be addressed, estimated, and corrected for to the best of our knowledge. A detailed analysis of uncertainty evaluation by the experimentalists may hopefully reduce the need for hidden errors or unrecognized sources of uncertainty (\USUn).
More specifically, clearly wrong data should be rejected as they have the potential to bias the resulting mean values as well as significantly affecting the evaluated uncertainties.
Known biases should be corrected, {\it e.g.}, by updating to the newest standard reference cross-section or half-life values.
A detailed uncertainty quantification (UQ) exercise should be undertaken using the literature that is pertinent to the respective data set.
Furthermore, it should be investigated whether known sources of uncertainties for this particular measurement type are missing and should be added.
Templates of expected uncertainties~\cite{Schillebeecks:2012,Helgesson:2015,Helgesson:2017,Neudecker:2018,Neudecker_Template:2019} for specific measurement types can help identify and estimate these missing covariances.
Once all these steps have been completed for all input data of an evaluation, and still an unexplained scatter in the data considering their standard deviations is observed, one should take recourse to identifying whether \USU components are present and estimate them.

In Sect.~\ref{ssec:clues}, methods are given concerning how to search for clues to identify that \USU effects need to be quantified.
Several approaches to estimating \USU effects are presented in Sect.~\ref{ssec:methods}.
These identification methods and estimation approaches were illustrated by several examples in Section~\ref{sec:examples}.

The \CF~\nub evaluation was chosen as a first example since \USU estimation techniques were applied for the recent Neutron Standards evaluation~\cite{Standards:2018}.
However, this example  highlights that what was perceived as a possible contribution from \USU for this evaluation could be traced back to incomplete UQ steps mentioned in the paragraph above before undertaking to quantify \USUn. Consideration of many of the \USU clues indicated that there is no need to quantify \USU for this specific example. However, one considered clue highlighted the point that uncertainties of single experiments and correlations between them might be missing. This emphasizes again the importance of an adequate uncertainty quantification based on known physics considerations for experimental data rather than using \USU contributions as a cover-up for avoiding the hard work of seeking to identify all possible objective (non-\USUn) sources of uncertainty.

An example of dealing with $\alpha$-counting experiments illustrated how the clues can be used to identifying non-correlated \USU effects and approaches to estimating them.
Another example explored the large scatter in calculated versus experimental criticality values of metal, fast neutron spectrum, highly enriched uranium criticality benchmarks.
It was discussed in detail that in all but one case no obvious physics reason for the spread in the data nor missing uncertainties of the size of the scatter could be identified.
Hence, it was recommended that \USU effects could account for the unknown spread in these data until it is resolved.

In the same vein, an example was shown where the (n,$\gamma$) cross sections of several isotopes experimentally determined by the TOF flight technique differed systematically from those obtained via the AMS approach.
While an uncorrected bias in one or both measurement techniques is evident from just looking at the data, no physics explanation exists to date to resolve this difference.
These two latter cases are typical ones where \USU is the last resort providing realistic uncertainties quantifying the limited present-day knowledge for a combination of data.

Also, examples of resolved \USU cases were shown in order to highlight how they were turned into a known sources of uncertainty in the past.
For instance, three examples were given ($^{235}$U(n,f) cross sections as a function of time, \textit{particle leakage} effects in ionization chambers, and $^{232}$Th(n,$\gamma$) measurements adversely impacted by water in the sample) where the need to introduce an \USU component could be avoided by understanding the physics effect causing the scatter in the data.
In these cases, it was recommended to reject data that cannot be corrected for this effect rather than increasing uncertainties of the data with an introduced uncertainty contribution attributed to \USUn.
This recommendation is based on the fact that the evaluated mean values might be biased by clearly wrong data from a physics point of view if the uncertainties are just increased rather than the mean values corrected.
These examples also showcase that \USU effects can be reduced or even eliminated by studying the physics cause for the discrepancy in the data by dedicated experiments that are designed to investigate the possible effects.

One example of \USU effects, formulated by considering measurements of neutron-induced fission cross-section ratios of $^{238}$U/$^{235}$U,  highlighted the point that different \USU estimation approaches result in slightly different but yet consistent evaluated mean values and uncertainties. Under/overestimation of reported uncertainties have been found. This example illustrated the subjective nature of estimating the effect of \USUn.
However, this subjectivity does not apply only to estimating \USU effects but also holds true for evaluating mean values and covariances in general.
After all, these derived values depend on the evaluation techniques utilized, input data selected and subjective choices made on the parameters of models, corrections of experimental data and on methods for estimating covariances of both, model and experiment.
Hence, our evaluated results are always subjective---with and without consideration of \USUn.
However, if one neglects adding obviously necessary contributions from \USU to uncertainties of input data, the evaluated mean values might be more biased and the evaluated uncertainties will be underestimated, in turn, adversely impacting application calculations.

Methods were suggested during the course of the present investigation on how to update the Neutron Data Standards evaluation with an improved formulation of contributions from \USUn.
These approaches will be applied in the future for this particular evaluation effort.

\section*{ACKNOWLEDGEMENTS}

The IAEA is grateful to all participating laboratories for their assistance in the work and for supporting the IAEA meetings and activities. Work described in this paper would not have been possible without contributions from the IAEA Member States.
%Work at Argonne National Laboratory and Los Alamos National Laboratory was supported by the US Department of Energy, Office of Science, Office of Nuclear Physics.
The IAEA staff acknowledges the valuable contributions made by V. Zerkin by improving his online plotting package.

%%%********** End of text entry ****************

\newpage
\section*{APPENDIX}
The Appendix contains three subsections: one where the numerical input data used in Sect.~\ref{subsect:u238-u235-ratio} are listed (see~\ref{app:nf-ratio}), a second one that includes a comprehensive explanation of the Maximum Likelihood estimation method (see~\ref{app:MLE}), and finally a third one, covering the Physical Uncertainty Bounds method (see~\ref{app:PUB}).

% \subsection{Statistical tests}
\subsection{\UF/\UT~Fission Cross-section Ratio Measurements}\label{app:nf-ratio}
A search for \USU components was undertaken for eleven selected experimental data sets of \UF/\UT~neutron-induced fission cross-section ratio. The datasets and corresponding recognized uncertainties, as estimated for the \GMA\textit{P}~ least-squares evaluation code input, are listed in Table~\ref{tab:U5U8-input}.

\newpage
\LTcapwidth=\textwidth
\begin{longtable*}{c|c|cc|ccccc|c|c}
\caption{Selected absolute \UF/\UT~cross-section ratio data from 8.0 up to 15~MeV. $E_n$ is the incident neutron energy. $R_{8/5}$ is the cross-section ratio.
$u_{\epsilon}$, $u_{\eta}$, $u_{norm}$, and $u_{tot}$ are the cross-section ratio uncorrelated, correlated, normalization, and total uncertainty components, correspondingly. $u_1$, $u_2$, $u_3$, $u_4$, and $u_5$ are additional MERC (Medium Energy Range Correlations) uncertainty components as used in \GMA\textit{P}. All uncertainties are given in \%. \label{tab:U5U8-input}} \\
\hline \hline
\T      $E_n$, MeV & ~Ratio~$R_{8/5}$~~ & ~~$u_{\epsilon}$,~\%~ & ~$u_{\eta}$,~\%~ & $u_1$, \%
        & $u_2$, \% & $u_3$, \% & $u_4$, \% & $u_5$, \% & $u_{norm}$, \% & $u_{tot}$, \% \B \\
        \hline
\endfirsthead
\multicolumn{10}{c}{Continuation of Table~\ref{tab:U5U8-input}} \B \\
\hline \hline
\T      $E_n$, MeV & ~Ratio~$R_{8/5}$~~ & ~~$u_{\epsilon}$,~\%~ & ~$u_{\eta}$,~\%~ & $u_1$, \%
        & $u_2$, \% & $u_3$, \% & $u_4$, \% & $u_5$, \% & $u_{norm}$, \% & $u_{tot}$, \% \B \\
\hline
% Everything above this command will appear at the beginning of the table, in the first page.
\endhead
%Whatever you put before this command and below endfirsthead will be displayed at the top of the table in every page except the first one.
\hline \hline
\endfoot
%Similar to \endhead, what you put after \endhead and before this command will appear at the bottom of the table in every page except the last one.
\hline \hline
\endlastfoot
%Similar to endfisthead. The elements after \endfoot and before this command will be displayed at the bottom of the table but only in the last page where the table appears.
        \hline
        \multicolumn{10}{c}{~~~~~~~\protect{F.~Tovesson \etal\cite{Tovesson:2015}}, EXFOR=14402009, normalization uncertainty of 3\% was proposed by the authors} \B \\
        \hline
7.99 & 0.577 &  0.311 & 0.454 &  0.0 &  0.0 &   0.0 &   0.0 &   0.0 &  0.7 &    0.89  \\
8.27 & 0.576 &  0.315 & 0.461 &  0.0 &  0.0 &   0.0 &   0.0 &   0.0 &  0.7 &    0.90  \\
8.56 & 0.577 &  0.320 & 0.409 &  0.0 &  0.0 &   0.0 &   0.0 &   0.0 &  0.7 &    0.87  \\
8.86 & 0.577 &  0.326 & 0.375 &  0.0 &  0.0 &   0.0 &   0.0 &   0.0 &  0.7 &    0.86  \\
9.17 & 0.584 &  0.333 & 0.323 &  0.0 &  0.0 &   0.0 &   0.0 &   0.0 &  0.7 &    0.84  \\
9.50 & 0.581 &  0.340 & 0.601 &  0.0 &  0.0 &   0.0 &   0.0 &   0.0 &  0.7 &    0.98  \\
9.83 & 0.586 &  0.348 & 0.516 &  0.0 &  0.0 &   0.0 &   0.0 &   0.0 &  0.7 &    0.94  \\
10.2 & 0.586 &  0.357 & 0.386 &  0.0 &  0.0 &   0.0 &   0.0 &   0.0 &  0.7 &    0.88  \\
10.5 & 0.585 &  0.366 & 0.538 &  0.0 &  0.0 &   0.0 &   0.0 &   0.0 &  0.7 &    0.96  \\
10.9 & 0.591 &  0.374 & 0.447 &  0.0 &  0.0 &   0.0 &   0.0 &   0.0 &  0.7 &    0.91  \\
11.3 & 0.592 &  0.384 & 0.454 &  0.0 &  0.0 &   0.0 &   0.0 &   0.0 &  0.7 &    0.92  \\
11.7 & 0.599 &  0.392 & 0.668 &  0.0 &  0.0 &   0.0 &   0.0 &   0.0 &  0.7 &    1.04  \\
12.1 & 0.592 &  0.400 & 0.631 &  0.0 &  0.0 &   0.0 &   0.0 &   0.0 &  0.7 &    1.02  \\
12.5 & 0.570 &  0.406 & 0.484 &  0.0 &  0.0 &   0.0 &   0.0 &   0.0 &  0.7 &    0.94  \\
13.0 & 0.555 &  0.410 & 0.557 &  0.0 &  0.0 &   0.0 &   0.0 &   0.0 &  0.7 &    0.98  \\
13.4 & 0.551 &  0.409 & 0.526 &  0.0 &  0.0 &   0.0 &   0.0 &   0.0 &  0.7 &    0.97  \\
13.9 & 0.563 &  0.405 & 0.512 &  0.0 &  0.0 &   0.0 &   0.0 &   0.0 &  0.7 &    0.96  \\
14.4 & 0.575 &  0.402 & 0.461 &  0.0 &  0.0 &   0.0 &   0.0 &   0.0 &  0.7 &    0.93  \\
14.9 & 0.592 &  0.400 & 0.499 &  0.0 &  0.0 &   0.0 &   0.0 &   0.0 &  0.7 &    0.95  \\
15.4 & 0.602 &  0.400 & 0.531 &  0.0 &  0.0 &   0.0 &   0.0 &   0.0 &  0.7 &    0.97  \\
        \hline
        \multicolumn{10}{c}{~~~~~~~PPAC-TILT2 C.~Paradela \etal\cite{Paradela:2015}, n\_TOF} \B \\
        \hline
        \centering
7.586 & 0.6052 & 1.58 & 1.0   & 1.0  &  0.0 &  0.0  &  0.0  & 0.0   &  1.0 &    2.34  \\
8.318 & 0.5809 & 1.62 & 1.0   & 1.0  &  0.0 &  0.0  &  0.0  & 0.0   &  1.0 &    2.37  \\
9.120 & 0.5843 & 1.70 & 1.0   & 1.0  &  0.0 &  0.0  &  0.0  & 0.0   &  1.0 &    2.43  \\
10.00 & 0.6005 & 1.67 & 1.0   & 1.0  &  0.0 &  0.0  &  0.0  & 0.0   &  1.0 &    2.41  \\
10.96 & 0.5877 & 1.75 & 1.0   & 1.0  &  0.0 &  0.0  &  0.0  & 0.0   &  1.0 &    2.46  \\
12.02 & 0.5947 & 1.76 & 1.0   & 1.0  &  0.0 &  0.0  &  0.0  & 0.0   &  1.0 &    2.47  \\
13.18 & 0.5928 & 1.63 & 1.0   & 1.0  &  0.0 &  0.0  &  0.0  & 0.0   &  1.0 &    2.38  \\
14.45 & 0.6003 & 1.60 & 1.0   & 1.0  &  0.0 &  0.0  &  0.0  & 0.0   &  1.0 &    2.36  \\
15.85 & 0.6408 & 1.56 & 1.0   & 1.0  &  0.0 &  0.0  &  0.0  & 0.0   &  1.0 &    2.33  \\
        \hline
        \multicolumn{10}{c}{~~~~~~~PPAC-TILT1 C.~Paradela \etal\cite{Paradela:2015}, n\_TOF} \B \\
        \hline
7.586 & 0.5943 & 2.52 & 2.00  & 1.0  & 0.0 &  0.0 &  0.0 & 0.0 &  1.1 &    3.54  \\
8.318 & 0.5812 & 2.55 & 2.00  & 1.0  & 0.0 &  0.0 &  0.0 & 0.0 &  1.1 &    3.57  \\
9.120 & 0.5659 & 2.66 & 2.00  & 1.0  & 0.0 &  0.0 &  0.0 & 0.0 &  1.1 &    3.64  \\
10.00 & 0.5938 & 2.69 & 2.00  & 1.0  & 0.0 &  0.0 &  0.0 & 0.0 &  1.1 &    3.70  \\
10.96 & 0.6023 & 2.79 & 2.00  & 1.0  & 0.0 &  0.0 &  0.0 & 0.0 &  1.1 &    3.74  \\
12.02 & 0.5947 & 2.88 & 2.00  & 1.0  & 0.0 &  0.0 &  0.0 & 0.0 &  1.1 &    3.81  \\
13.18 & 0.5928 & 2.85 & 2.00  & 1.0  & 0.0 &  0.0 &  0.0 & 0.0 &  1.1 &    3.78  \\
14.45 & 0.6003 & 2.69 & 2.00  & 1.0  & 0.0 &  0.0 &  0.0 & 0.0 &  1.1 &    3.67  \\
15.85 & 0.6408 & 2.58 & 2.00  & 1.0  & 0.0 &  0.0 &  0.0 & 0.0 &  1.1 &    3.59  \\
        \hline
        \multicolumn{10}{c}{~~~~~~~PPAC-PERP C.~Paradela \etal\cite{Paradela:2015}, n\_TOF} \B \\
        \hline
7.586 & 0.5793 & 1.35 &  3.00 &  1.0  & 0.0 &  0.0 &  0.0 & 0.0 &  1.1 &   3.61  \\
8.318 & 0.5689 & 1.38 &  3.00 &  1.0  & 0.0 &  0.0 &  0.0 & 0.0 &  1.1 &   3.62  \\
9.120 & 0.5683 & 1.42 &  3.00 &  1.0  & 0.0 &  0.0 &  0.0 & 0.0 &  1.1 &   3.64  \\
10.00 & 0.5782 & 1.48 &  3.00 &  1.0  & 0.0 &  0.0 &  0.0 & 0.0 &  1.1 &   3.66  \\
10.96 & 0.5869 & 1.52 &  3.00 &  1.0  & 0.0 &  0.0 &  0.0 & 0.0 &  1.1 &   3.67  \\
12.02 & 0.6005 & 1.56 &  3.00 &  1.0  & 0.0 &  0.0 &  0.0 & 0.0 &  1.1 &   3.70  \\
13.18 & 0.5763 & 1.52 &  3.00 &  1.0  & 0.0 &  0.0 &  0.0 & 0.0 &  1.1 &   3.71  \\
14.45 & 0.5958 & 1.43 &  3.00 &  1.0  & 0.0 &  0.0 &  0.0 & 0.0 &  1.1 &   3.64  \\
15.85 & 0.6136 & 1.39 &  3.00 &  1.0  & 0.0 &  0.0 &  0.0 & 0.0 &  1.1 &   3.55  \\
        \hline
        \multicolumn{10}{c}{~~~~~~~FIC  C.~Paradela \etal\cite{Paradela:2015}, n\_TOF} \B \\
        \hline
7.499 & 0.6105 & 2.17 &  1.0  &   3.0 &  0.0 &  0.0 &  0.0 & 0.0 &    2.0 &  4.32 \\
8.414 & 0.5626 & 2.22 &  1.0  &   3.0 &  0.0 &  0.0 &  0.0 & 0.0 &    2.0 &  4.35 \\
9.441 & 0.5550 & 2.19 &  1.0  &   3.0 &  0.0 &  0.0 &  0.0 & 0.0 &    2.0 &  4.34 \\
10.59 & 0.5486 & 2.37 &  1.0  &   3.0 &  0.0 &  0.0 &  0.0 & 0.0 &    2.0 &  4.43 \\
11.88 & 0.5729 & 2.47 &  1.0  &   3.0 &  0.0 &  0.0 &  0.0 & 0.0 &    2.0 &  4.48 \\
13.33 & 0.5587 & 2.46 &  1.0  &   3.0 &  0.0 &  0.0 &  0.0 & 0.0 &    2.0 &  4.48 \\
14.96 & 0.5750 & 2.36 &  1.0  &   3.0 &  0.0 &  0.0 &  0.0 & 0.0 &    2.0 &  4.42 \\
16.79 & 0.6291 & 2.26 &  1.0  &   3.0 &  0.0 &  0.0 &  0.0 & 0.0 &    2.0 &  4.37 \\
        \hline
        \multicolumn{10}{c}{~~~~~~~J.W.~Behrens \etal\cite{Behrens:1977}} \B \\
        \hline
 7.648 & 0.5751 & 1.1 &  0.3 &  0.1 & 0.2 &  0.2 &  0.0 & 0.0 &   0.7 &    1.4  \\
 7.930 & 0.5607 & 1.1 &  0.3 &  0.1 & 0.2 &  0.2 &  0.0 & 0.0 &   0.7 &    1.4  \\
 8.229 & 0.5649 & 1.1 &  0.3 &  0.1 & 0.2 &  0.2 &  0.0 & 0.0 &   0.7 &    1.4  \\
 8.545 & 0.5587 & 1.2 &  0.3 &  0.1 & 0.2 &  0.2 &  0.0 & 0.0 &   0.7 &    1.5  \\
 8.879 & 0.5671 & 1.2 &  0.3 &  0.2 & 0.2 &  0.2 &  0.0 & 0.0 &   0.7 &    1.5  \\
 9.234 & 0.5633 & 1.2 &  0.2 &  0.2 & 0.2 &  0.2 &  0.0 & 0.0 &   0.7 &    1.4  \\
 9.610 & 0.5762 & 1.2 &  0.2 &  0.2 & 0.2 &  0.2 &  0.0 & 0.0 &   0.7 &    1.4  \\
 10.01 & 0.5626 & 1.3 &  0.2 &  0.2 & 0.2 &  0.2 &  0.0 & 0.0 &   0.7 &    1.5  \\
 10.44 & 0.5735 & 1.3 &  0.2 &  0.2 & 0.2 &  0.2 &  0.0 & 0.0 &   0.7 &    1.5  \\
 10.89 & 0.5711 & 1.3 &  0.2 &  0.2 & 0.2 &  0.2 &  0.0 & 0.0 &   0.7 &    1.5  \\
 11.37 & 0.5767 & 1.4 &  0.2 &  0.2 & 0.2 &  0.2 &  0.0 & 0.0 &   0.7 &    1.6  \\
 11.89 & 0.5627 & 1.4 &  0.2 &  0.2 & 0.2 &  0.2 &  0.0 & 0.0 &   0.7 &    1.6  \\
 12.44 & 0.5414 & 1.4 &  0.2 &  0.2 & 0.2 &  0.2 &  0.0 & 0.0 &   0.7 &    1.6  \\
 13.04 & 0.5367 & 1.5 &  0.2 &  0.2 & 0.2 &  0.2 &  0.0 & 0.0 &   0.7 &    1.7  \\
 13.67 & 0.5543 & 1.5 &  0.2 &  0.2 & 0.2 &  0.2 &  0.0 & 0.0 &   0.7 &    1.7  \\
 14.36 & 0.5588 & 1.6 &  0.2 &  0.2 & 0.2 &  0.2 &  0.0 & 0.0 &   0.7 &    1.8  \\
 15.10 & 0.5800 & 1.6 &  0.2 &  0.2 & 0.2 &  0.2 &  0.0 & 0.0 &   0.7 &    1.8  \\
 15.89 & 0.5973 & 1.7 &  0.2 &  0.2 & 0.2 &  0.2 &  0.0 & 0.0 &   0.7 &    1.9  \\
         \hline
         \multicolumn{10}{c}{~~~~~~~F.C.~Difilippo \etal\cite{Difilippo:1978}} \B \\
         \hline
 7.650 & 0.5700 & 0.6 &  1.0 &  2.0 & 0.0 &  0.0 &  0.0 & 0.0 & 1.0 &    2.5  \\
 7.930 & 0.5670 & 0.5 &  1.0 &  2.0 & 0.0 &  0.0 &  0.0 & 0.0 & 1.0 &    2.5  \\
 8.230 & 0.5700 & 0.6 &  1.0 &  2.0 & 0.0 &  0.0 &  0.0 & 0.0 & 1.0 &    2.5  \\
 8.545 & 0.5630 & 0.6 &  1.0 &  2.0 & 0.0 &  0.0 &  0.0 & 0.0 & 1.0 &    2.5  \\
 8.880 & 0.5700 & 0.6 &  1.0 &  1.9 & 0.0 &  0.0 &  0.0 & 0.0 & 1.0 &    2.4  \\
 9.235 & 0.5750 & 0.6 &  1.0 &  1.9 & 0.0 &  0.0 &  0.0 & 0.0 & 1.0 &    2.4  \\
 9.615 & 0.5730 & 0.7 &  1.0 &  1.9 & 0.0 &  0.0 &  0.0 & 0.0 & 1.0 &    2.5  \\
 10.00 & 0.5690 & 0.7 &  1.0 &  1.9 & 0.0 &  0.0 &  0.0 & 0.0 & 1.0 &    2.5  \\
 10.45 & 0.5720 & 0.8 &  1.0 &  1.9 & 0.0 &  0.0 &  0.0 & 0.0 & 1.0 &    2.5  \\
 10.90 & 0.5780 & 0.8 &  1.0 &  1.8 & 0.0 &  0.0 &  0.0 & 0.0 & 1.0 &    2.4  \\
 11.35 & 0.5850 & 0.9 &  1.0 &  1.8 & 0.0 &  0.0 &  0.0 & 0.0 & 1.0 &    2.5  \\
 11.90 & 0.5600 & 0.9 &  1.0 &  1.8 & 0.0 &  0.0 &  0.0 & 0.0 & 1.0 &    2.5  \\
 12.45 & 0.5410 & 0.9 &  1.0 &  1.7 & 0.0 &  0.0 &  0.0 & 0.0 & 1.0 &    2.4  \\
 13.00 & 0.5320 & 0.9 &  1.0 &  1.7 & 0.0 &  0.0 &  0.0 & 0.0 & 1.0 &    2.4  \\
 13.65 & 0.5340 & 1.0 &  1.0 &  1.7 & 0.0 &  0.0 &  0.0 & 0.0 & 1.0 &    2.4  \\
 14.35 & 0.5510 & 1.0 &  1.0 &  1.6 & 0.0 &  0.0 &  0.0 & 0.0 & 1.0 &    2.4  \\
 15.10 & 0.5740 & 1.0 &  1.0 &  1.6 & 0.0 &  0.0 &  0.0 & 0.0 & 1.0 &    2.4  \\
        \hline
        \multicolumn{10}{c}{~~~~~~~S.~Cierjacks \etal\cite{Cierjacks:1976} (SHAPE)} \B \\
        \hline
 7.443 & 0.5584 & 3.2 &  0.4 & 0.5 & 0.4 &  0.5 &  0.3 & 0.5 & 0.0 &    3.4  \\
 7.748 & 0.5543 & 3.0 &  0.4 & 0.5 & 0.4 &  0.5 &  0.3 & 0.5 & 0.0 &    3.2  \\
 8.050 & 0.5697 & 3.1 &  0.4 & 0.5 & 0.4 &  0.5 &  0.3 & 0.5 & 0.0 &    3.3  \\
 8.346 & 0.5503 & 2.9 &  0.4 & 0.5 & 0.4 &  0.5 &  0.3 & 0.5 & 0.0 &    3.1  \\
 8.635 & 0.5604 & 2.9 &  0.4 & 0.5 & 0.4 &  0.5 &  0.3 & 0.5 & 0.0 &    3.1  \\
 8.939 & 0.5564 & 2.8 &  0.4 & 0.5 & 0.4 &  0.5 &  0.3 & 0.5 & 0.0 &    3.0  \\
 9.259 & 0.5645 & 2.8 &  0.4 & 0.5 & 0.4 &  0.5 &  0.3 & 0.5 & 0.0 &    3.0  \\
 9.539 & 0.5432 & 2.8 &  0.4 & 0.5 & 0.4 &  0.5 &  0.3 & 0.5 & 0.0 &    3.0  \\
 9.833 & 0.5727 & 2.7 &  0.4 & 0.5 & 0.5 &  0.5 &  0.3 & 0.5 & 0.0 &    2.9  \\
 10.14 & 0.5685 & 2.6 &  0.4 & 0.5 & 0.5 &  0.5 &  0.3 & 0.5 & 0.0 &    2.8  \\
 10.46 & 0.5849 & 2.7 &  0.4 & 0.5 & 0.5 &  0.5 &  0.3 & 0.5 & 0.0 &    2.9  \\
 10.73 & 0.5697 & 2.8 &  0.4 & 0.5 & 0.5 &  0.5 &  0.3 & 0.5 & 0.0 &    3.0  \\
 11.05 & 0.5727 & 2.6 &  0.4 & 0.5 & 0.4 &  0.5 &  0.3 & 0.5 & 0.0 &    2.8  \\
 11.34 & 0.5920 & 2.6 &  0.4 & 0.5 & 0.4 &  0.5 &  0.3 & 0.5 & 0.0 &    2.8  \\
 11.64 & 0.5747 & 2.6 &  0.4 & 0.5 & 0.4 &  0.5 &  0.3 & 0.5 & 0.0 &    2.8  \\
 11.96 & 0.5859 & 2.5 &  0.4 & 0.5 & 0.4 &  0.5 &  0.3 & 0.5 & 0.0 &    2.7  \\
 12.16 & 0.5828 & 3.0 &  0.4 & 0.5 & 0.4 &  0.5 &  0.3 & 0.5 & 0.0 &    3.2  \\
 12.37 & 0.5564 & 3.0 &  0.4 & 0.5 & 0.4 &  0.5 &  0.3 & 0.5 & 0.0 &    3.2  \\
 12.59 & 0.5401 & 2.9 &  0.4 & 0.5 & 0.5 &  0.5 &  0.3 & 0.5 & 0.0 &    3.1  \\
 12.81 & 0.5238 & 2.8 &  0.4 & 0.5 & 0.5 &  0.5 &  0.3 & 0.5 & 0.0 &    3.0  \\
 13.03 & 0.5422 & 2.7 &  0.4 & 0.5 & 0.5 &  0.5 &  0.3 & 0.5 & 0.0 &    2.9  \\
 13.27 & 0.5432 & 2.6 &  0.4 & 0.5 & 0.5 &  0.5 &  0.3 & 0.5 & 0.0 &    2.8  \\
 13.50 & 0.5594 & 2.6 &  0.4 & 0.5 & 0.4 &  0.5 &  0.3 & 0.5 & 0.0 &    2.8  \\
 13.75 & 0.5371 & 2.5 &  0.4 & 0.5 & 0.4 &  0.5 &  0.3 & 0.5 & 0.0 &    2.8  \\
 14.00 & 0.5594 & 2.4 &  0.4 & 0.5 & 0.4 &  0.5 &  0.3 & 0.5 & 0.0 &    2.7  \\
 14.26 & 0.5614 & 2.4 &  0.4 & 0.5 & 0.4 &  0.5 &  0.3 & 0.5 & 0.0 &    2.7  \\
 14.53 & 0.5523 & 2.4 &  0.4 & 0.5 & 0.4 &  0.5 &  0.3 & 0.5 & 0.0 &    2.7  \\
 14.80 & 0.5768 & 2.3 &  0.4 & 0.5 & 0.4 &  0.5 &  0.3 & 0.5 & 0.0 &    2.6  \\
 15.09 & 0.5940 & 2.2 &  0.4 & 0.5 & 0.5 &  0.5 &  0.3 & 0.5 & 0.0 &    2.5  \\
        \hline
        \multicolumn{10}{c}{~~~~~~~M.S.~Coates \etal\cite{Coates:1975} (SHAPE)} \B \\
        \hline
  7.617 & 0.6113 & 1.9 &  1.0 & 2.5 & 0.5 & 1.0 &  0.0 & 0.0 & 0.0 &    3.5  \\
  7.800 & 0.5866 & 1.9 &  1.0 & 2.5 & 0.5 & 1.0 &  0.0 & 0.0 & 0.0 &    3.5  \\
  7.989 & 0.5908 & 1.9 &  1.0 & 2.5 & 0.5 & 1.0 &  0.0 & 0.0 & 0.0 &    3.5  \\
  8.186 & 0.5785 & 1.9 &  1.0 & 2.5 & 0.5 & 1.0 &  0.0 & 0.0 & 0.0 &    3.5  \\
  8.389 & 0.5693 & 1.9 &  1.0 & 2.5 & 0.5 & 1.0 &  0.0 & 0.0 & 0.0 &    3.5  \\
  8.601 & 0.5826 & 1.8 &  1.0 & 2.5 & 0.5 & 1.0 &  0.0 & 0.0 & 0.0 &    3.4  \\
  8.821 & 0.5651 & 1.8 &  1.0 & 2.5 & 0.5 & 1.0 &  0.0 & 0.0 & 0.0 &    3.4  \\
  9.049 & 0.5774 & 1.8 &  1.0 & 2.5 & 0.5 & 1.0 &  0.0 & 0.0 & 0.0 &    3.4  \\
  9.286 & 0.5856 & 1.8 &  1.0 & 2.5 & 0.5 & 1.0 &  0.0 & 0.0 & 0.0 &    3.4  \\
  9.533 & 0.5723 & 1.8 &  1.0 & 2.5 & 0.5 & 1.0 &  0.0 & 0.0 & 0.0 &    3.4  \\
  9.790 & 0.5591 & 1.8 &  1.0 & 2.5 & 0.5 & 1.0 &  0.0 & 0.0 & 0.0 &    3.4  \\
  10.06 & 0.5856 & 1.8 &  1.0 & 2.5 & 0.5 & 1.0 &  0.0 & 0.0 & 0.0 &    3.4  \\
  10.33 & 0.5723 & 1.8 &  1.0 & 2.5 & 0.5 & 1.0 &  0.0 & 0.0 & 0.0 &    3.4  \\
  10.62 & 0.5845 & 1.8 &  1.0 & 2.5 & 0.5 & 1.0 &  0.0 & 0.0 & 0.0 &    3.4  \\
  10.93 & 0.5845 & 1.8 &  1.0 & 2.5 & 0.5 & 1.0 &  0.0 & 0.0 & 0.0 &    3.4  \\
  11.24 & 0.5918 & 1.8 &  1.0 & 2.5 & 0.5 & 1.0 &  0.0 & 0.0 & 0.0 &    3.4  \\
  11.57 & 0.5856 & 1.8 &  1.0 & 2.5 & 0.5 & 1.0 &  0.0 & 0.0 & 0.0 &    3.4  \\
  11.92 & 0.5743 & 1.8 &  1.0 & 2.5 & 0.5 & 1.0 &  0.0 & 0.0 & 0.0 &    3.4  \\
  12.28 & 0.5580 & 1.8 &  1.0 & 2.5 & 0.5 & 1.0 &  0.0 & 0.0 & 0.0 &    3.4  \\
  12.66 & 0.5621 & 1.8 &  1.0 & 2.5 & 0.5 & 1.0 &  0.0 & 0.0 & 0.0 &    3.4  \\
  13.05 & 0.5478 & 1.8 &  1.0 & 2.5 & 0.5 & 1.0 &  0.0 & 0.0 & 0.0 &    3.4  \\
  13.47 & 0.5457 & 1.7 &  1.0 & 2.5 & 0.5 & 1.0 &  0.0 & 0.0 & 0.0 &    3.4  \\
  13.90 & 0.5375 & 1.7 &  1.0 & 2.5 & 0.5 & 1.0 &  0.0 & 0.0 & 0.0 &    3.4  \\
  14.36 & 0.5293 & 1.7 &  1.0 & 2.5 & 0.5 & 1.0 &  0.0 & 0.0 & 0.0 &    3.4  \\
  14.83 & 0.5610 & 1.7 &  1.0 & 2.5 & 0.5 & 1.0 &  0.0 & 0.0 & 0.0 &    3.4  \\
  15.34 & 0.5498 & 1.7 &  1.0 & 2.5 & 0.5 & 1.0 &  0.0 & 0.0 & 0.0 &    3.4  \\
          \hline
          \multicolumn{10}{c}{~~~~~~~O.~Shcherbakov \etal\cite{Shcherbakov:2001}} \B \\
          \hline
  7.679 & 0.5699 & 0.6 &  1.5 & 1.0 & 0.8 & 1.5 &  0.5 & 0.2 & 0.8  &  2.7  \\
  7.932 & 0.5598 & 0.6 &  1.5 & 1.0 & 0.8 & 1.5 &  0.5 & 0.2 & 0.8  &  2.7  \\
  8.197 & 0.5626 & 0.6 &  1.5 & 1.0 & 0.8 & 1.5 &  0.5 & 0.2 & 0.8  &  2.7  \\
  8.477 & 0.5576 & 0.6 &  1.5 & 1.0 & 0.8 & 1.5 &  0.5 & 0.2 & 0.8  &  2.7  \\
  8.771 & 0.5600 & 0.6 &  1.5 & 1.0 & 0.8 & 1.5 &  0.5 & 0.2 & 0.8  &  2.7  \\
  9.080 & 0.5607 & 0.6 &  1.5 & 1.0 & 0.8 & 1.5 &  0.5 & 0.2 & 0.8  &  2.7  \\
  9.406 & 0.5647 & 0.6 &  1.5 & 1.0 & 0.8 & 1.5 &  0.5 & 0.2 & 0.8  &  2.7  \\
  9.751 & 0.5655 & 0.6 &  1.5 & 1.0 & 0.8 & 1.5 &  0.5 & 0.2 & 0.8  &  2.7  \\
  10.11 & 0.5671 & 0.7 &  0.5 & 1.0 & 0.8 & 1.0 &  0.5 & 0.2 & 0.8  &  2.1  \\
  10.50 & 0.5715 & 0.7 &  0.5 & 1.0 & 0.8 & 1.0 &  0.5 & 0.2 & 0.8  &  2.1  \\
  10.91 & 0.5768 & 0.7 &  0.5 & 1.0 & 0.8 & 1.0 &  0.5 & 0.2 & 0.8  &  2.1  \\
  11.34 & 0.5706 & 0.7 &  0.5 & 1.0 & 0.8 & 1.0 &  0.5 & 0.2 & 0.8  &  2.1  \\
  11.79 & 0.5673 & 0.7 &  0.5 & 1.0 & 0.8 & 1.0 &  0.5 & 0.2 & 0.8  &  2.1  \\
  12.28 & 0.5582 & 0.7 &  0.5 & 1.0 & 0.8 & 1.0 &  0.5 & 0.2 & 0.8  &  2.1  \\
  12.80 & 0.5449 & 0.7 &  0.5 & 1.0 & 0.8 & 1.0 &  0.5 & 0.2 & 0.8  &  2.1  \\
  13.35 & 0.5407 & 0.7 &  0.5 & 1.0 & 0.8 & 1.0 &  0.5 & 0.2 & 0.8  &  2.1  \\
  13.93 & 0.5470 & 0.7 &  0.5 & 1.0 & 0.8 & 1.0 &  0.5 & 0.2 & 0.8  &  2.1  \\
  14.56 & 0.5610 & 0.6 &  0.5 & 1.0 & 0.8 & 1.0 &  0.5 & 0.2 & 0.8  &  2.0  \\
  15.23 & 0.5886 & 0.6 &  0.5 & 1.0 & 0.8 & 1.0 &  0.5 & 0.2 & 0.8  &  2.0  \\
          \hline
          \multicolumn{10}{c}{~~~~~~~P.W.~Lisowski \etal\cite{Lisowski:1991}, priv. comm. (01-29-1997)} \B \\
          \hline
  7.441 & 0.5985 & 1.15 &  0.8 & 0.0 & 0.0 & 0.0 &  0.0 & 0.0 & 0.5 &   1.49  \\
  7.668 & 0.5706 & 1.17 &  0.8 & 0.0 & 0.0 & 0.0 &  0.0 & 0.0 & 0.5 &   1.50  \\
  7.902 & 0.5663 & 1.19 &  0.8 & 0.0 & 0.0 & 0.0 &  0.0 & 0.0 & 0.5 &   1.52  \\
  8.142 & 0.5677 & 1.17 &  0.8 & 0.0 & 0.0 & 0.0 &  0.0 & 0.0 & 0.5 &   1.50  \\
  8.390 & 0.5668 & 1.16 &  0.8 & 0.0 & 0.0 & 0.0 &  0.0 & 0.0 & 0.5 &   1.50  \\
  8.646 & 0.5767 & 1.16 &  0.8 & 0.0 & 0.0 & 0.0 &  0.0 & 0.0 & 0.5 &   1.50  \\
  8.909 & 0.5744 & 1.18 &  0.8 & 0.0 & 0.0 & 0.0 &  0.0 & 0.0 & 0.5 &   1.51  \\
  9.180 & 0.5786 & 1.20 &  0.8 & 0.0 & 0.0 & 0.0 &  0.0 & 0.0 & 0.5 &   1.53  \\
  9.460 & 0.5935 & 1.20 &  0.8 & 0.0 & 0.0 & 0.0 &  0.0 & 0.0 & 0.5 &   1.53  \\
  9.748 & 0.5782 & 1.23 &  0.8 & 0.0 & 0.0 & 0.0 &  0.0 & 0.0 & 0.5 &   1.55  \\
  10.04 & 0.5940 & 1.24 &  0.8 & 0.0 & 0.0 & 0.0 &  0.0 & 0.0 & 0.5 &   1.56  \\
  10.35 & 0.5746 & 1.28 &  0.8 & 0.0 & 0.0 & 0.0 &  0.0 & 0.0 & 0.5 &   1.59  \\
  10.67 & 0.5866 & 1.30 &  0.8 & 0.0 & 0.0 & 0.0 &  0.0 & 0.0 & 0.5 &   1.61  \\
  10.99 & 0.5818 & 1.34 &  0.8 & 0.0 & 0.0 & 0.0 &  0.0 & 0.0 & 0.5 &   1.64  \\
  11.32 & 0.5911 & 1.35 &  0.8 & 0.0 & 0.0 & 0.0 &  0.0 & 0.0 & 0.5 &   1.65  \\
  11.67 & 0.6019 & 1.38 &  0.8 & 0.0 & 0.0 & 0.0 &  0.0 & 0.0 & 0.5 &   1.67  \\
  12.03 & 0.5858 & 1.39 &  0.8 & 0.0 & 0.0 & 0.0 &  0.0 & 0.0 & 0.5 &   1.68  \\
  12.39 & 0.5823 & 1.40 &  0.8 & 0.0 & 0.0 & 0.0 &  0.0 & 0.0 & 0.5 &   1.69  \\
  12.77 & 0.5516 & 1.43 &  0.8 & 0.0 & 0.0 & 0.0 &  0.0 & 0.0 & 0.5 &   1.71  \\
  13.16 & 0.5357 & 1.44 &  0.8 & 0.0 & 0.0 & 0.0 &  0.0 & 0.0 & 0.5 &   1.72  \\
  13.56 & 0.5526 & 1.41 &  0.8 & 0.0 & 0.0 & 0.0 &  0.0 & 0.0 & 0.5 &   1.70  \\
  13.97 & 0.5597 & 1.40 &  0.8 & 0.0 & 0.0 & 0.0 &  0.0 & 0.0 & 0.5 &   1.69  \\
  14.40 & 0.5860 & 1.36 &  0.8 & 0.0 & 0.0 & 0.0 &  0.0 & 0.0 & 0.5 &   1.66  \\
  14.84 & 0.5740 & 1.38 &  0.8 & 0.0 & 0.0 & 0.0 &  0.0 & 0.0 & 0.5 &   1.67  \\
  15.29 & 0.6072 & 1.36 &  0.8 & 0.0 & 0.0 & 0.0 &  0.0 & 0.0 & 0.5 &   1.66  \\
\end{longtable*}

\subsection{Maximum Likelihood Estimation}\label{app:MLE}
Maximum likelihood estimation is an established method in statistics that is used to infer unknown parameters of a probability distribution from data.
Assume a probability distribution $\mathcal{L}(\mathcal{D}\,|\, \vec{\theta})$ that gives the likelihood for the realization of any potential measurement $\mathcal{D}$ under a specific choice of values for the distribution parameters $\vec{\theta}$.
In maximum likelihood estimation, the values of $\vec{\theta}$ are chosen to maximize the likelihood for the actually observed data $\mathcal{D}_\textrm{obs}$.
This estimation technique is general and not bound to a specific choice of probability distribution.

In nuclear data evaluation, the most commonly employed distribution is the multivariate normal distribution whose functional form is given by
\eqnarrow
\begin{multline}
\label{eq:mvn}
\mathcal{N}(\vec{y}\,|\,\vec{\mu(\theta)},\vec{C}) =
\frac{1}{\sqrt{(2\pi)^N \det\vec{C} }}
\times \\
\exp\left(
-\frac{1}{2}
(\vec{y}-\vec{\mu(\theta)})^T
\vec{C}^{-1}
(\vec{y}-\vec{\mu(\theta)})
\right) \,,
\end{multline}
where $N$ denotes the number of elements in $\vec{y}$.
This distribution is characterized by the the covariance matrix $\vec{C}$, and the center vector $\vec{\mu(\theta)}\equiv \vec{f(\vec{x} ; \vec{\theta})}$, which is derived from the model function $f(\vec{x} ; \vec{\theta})$.

To illustrate the maximum likelihood approach in combination with the multivariate normal distribution, let us consider the case of a known diagonal covariance matrix $\vec{C}$ with all elements $C_{ii} = u_i^2$ and an unknown center vector~$\vec{\mu}$.
Additionally, we impose the requirement that all elements in $\vec{\mu}$ are the same, \ie, $\mu_i = \mu$.
Given the actually observed $\vec{y}_\textrm{obs}$, the task is to find a vector~$\hat{\vec{\mu}}$ so that $\mathcal{N}(\vec{y}_\textrm{obs}\,|\,\hat{\vec{\mu}},\vec{C})$ yields the maximal value among all possible choices for $\vec{\mu}$.
In this particular case, the solution can be obtained analytically.
First, we take the natural logarithm of Eq.~\eqref{eq:mvn} as this transformation does not affect the position of the maximum,
\begin{multline}
\log \mathcal{N}(\vec{y}_\textrm{obs}\,|\,\vec{\mu},\vec{C}) =
-\frac{N}{2} \log(2\pi) - \frac{1}{2}\log(\det \vec{C}) \\
-\frac{1}{2} (\vec{y}_\textrm{obs} - \vec{\mu})^T
\vec{C}^{-1}
(\vec{y}_\textrm{obs} - \vec{\mu}).
\end{multline}
Using the fact that $\vec{C}$ is diagonal, and the other structural assumptions, the right hand side can be rewritten as
\begin{equation}
-\frac{N}{2} \log(2\pi) - N \log \delta
-\frac{1}{2} \sum_{i=1}^N
\frac{(y_{\textrm{obs},i} - \mu)^2}{\delta^2}
\,.
\end{equation}
Finally, to maximize this expression we calculate the derivative with respect to $\mu$ and see for which value of $\mu$ it vanishes.
It turns out that the arithmetic mean $\hat{\mu} = 1/N \sum_{i=1}^N y_{\textrm{obs},i}$ maximizes $\mathcal{N}(\vec{y}_\textrm{obs}\,|\,\vec{\mu},\vec{C})$.
This example illustrated that in certain scenarios analytical formulas exist to perform maximum likelihood estimation.

However, when the unknown distribution parameters appear in the covariance matrix, usually no analytical formulas exist and a numerical optimization routine is needed.
We discuss this case here in the context of the determination of a covariance matrix associated with consideration of contributions from \USUn.
We assume the relationship
$
\vec{y} =
\vec{\mu} +
\vec{\varepsilon} +
\vec{\delta}
$
between the measurements and the true values,
and that all random vectors on the right-hand side are multivariate normal and independent, \ie,
$\vec{\mu} \sim \mathcal{N}(\vec{\mu}_\textrm{prior}, \vec{C}_\textrm{prior})$,
$\vec{\varepsilon} \sim \mathcal{N}(\vec{0}, \vec{C}_{\vec{\varepsilon}})$, and
$\vec{\delta} \sim \mathcal{N}(\vec{0}, \vec{C}_{\vec{\delta}})$.
The specification of a distribution for $\vec{\mu}$ is equivalent to imposing prior knowledge in the Bayesian framework.

The sum of independent random vectors governed by multivariate normal distributions is also multivariate normal~$\vec{y} \sim \mathcal{N}(\vec{\mu}, \vec{C}_\textrm{tot})$
and its center vector and covariance matrix are given as the sum of the individual center vectors and covariance matrices, respectively,
\begin{equation}\label{eq:MLE}
\vec{\mu} = \vec{\mu}_\textrm{prior}
\;\;\textrm{and}\;\;
\vec{C}_\textrm{tot} =
\vec{C}_\textrm{prior} +
\vec{C}_{\varepsilon} +
\vec{C}_{\vec{\delta}} \,.
\end{equation}
If all variables are known except some elements in $\vec{C}_{\vec{\delta}}$, we can again use the maximum likelihood approach to estimate them.
If the vector~$\vec{y}$ contains the measurements, the unknown elements in $\vec{C}_{\vec{\delta}}$ have to be chosen in order to maximize $\mathcal{N}(\vec{y}
,|\, \vec{\mu}_\textrm{prior}, \vec{C}_\textrm{tot})$.
Because the probability distribution converted to logarithmic form has the maximum at the same location as the original one and the removal of constants and common factors does not change the location either, this maximization problem is equivalent to minimizing
\begin{equation}
\label{eq:objfunMLO}
\log \det \vec{C}_\textrm{tot} +
(\vec{y} - \vec{\mu}_\textrm{prior})^T
\vec{C}_\textrm{tot}^{-1}
(\vec{y} - \vec{\mu}_\textrm{prior})\,,
\end{equation}
\eqnormal
which can be done using a numerical optimization routine, such as the BFGS algorithm~\cite{Schnabel:2017}.
It is noteworthy that the first term on the right hand side is within a constant factor proportional to the differential entropy of a multivariate normal distribution, and the second term is the generalized $\chi^2$ value.
As entropy can be regarded as a measure of model complexity and the generalized $\chi^2$ value as a measure for goodness of fit, the minimization of Eq.~\eqref{eq:objfunMLO} aims to find a model that strikes a balance between simplicity and goodness of fit to the data.

\subsection{Applying the PUBs Methodology to Estimate \CF~\nub~Uncertainty Bounds}\label{app:PUB}

The Physical Uncertainty Bounds (PUBs) Method by Vaughan and Preston~\cite{Vaughan_PUB:2014} was applied to investigate whether the previous (0.13\%) or current (0.42\%) Standard \CF~\nub~uncertainties are more realistic given all known uncertainties of the sub-processes applying to the 15 \CF~\nub~measurements used for the evaluation.

\begin{table}[!thb]
\vspace{-3mm}
\caption{The conservative PUBs and minimal realistic uncertainties, $\delta^c_l$ and $\delta^m_l$, respectively, are listed for each sub-process relevant to a particular measurement type or across different measurement types. These values were extracted from Refs.~\cite{BoldemanProc,AxtonProc,Croft:2019} and EXFOR entries of the respective \CF~\nub~measurements~\cite{Aleksandrov,Asplund,AxtonExp,Bozorgmanesh,DeVolpi,Colvin,Diven,Edwards,Hopkins,Smith:1984,Spencer,White,Huan-Qiao},  considering how uncertainties would be reduced if measured multiple times with the same technique. }
\label{tab:PUBS_Cf}
\centering
\begin{tabular}{l|c|c}
\hline
\hline
Uncertainty source & $\delta^c_l$ (\%) & $\delta^m_l$ (\%)\\
\hline
  \multicolumn{3}{c}{Mn bath experiments only} \\
  \hline
  H/Mn ratio  & 0.29 & 0.15  \\
  Impurities  & 0.05 & 0.02 \\
  Fast neutron capture in O/S  & 0.1 & 0.06 \\
  neutron leakage  & 0.08 & 0.04  \\
  Source capture  & 0.03 & 0.02 \\
  Mn resonance  & 0.09 & 0.07  \\
  S/Mn ratio  & 0.12 & 0.04 \\
  Mn eff. (rand.)  & 0.09 & 0.05  \\
  Source activ.  & 0.1 & 0.05  \\
  neutron attenuation  & 0.1 & 0.07  \\
  S/H ratio  & 0.073 & 0.073 \\
  Solid angle  & 0.15 & 0.11  \\
  Fission det. eff.  & 0.1 & 0.05 \\
  Statistics  & 0.04 & 0.02  \\
  \CF self transfer  & 0.2  & 0.14  \\
  \hline
 \multicolumn{3}{c}{Scintillator experiments only} \\
 \hline
 Delayed $\gamma$ & 0.1 & 0.07  \\
 MC neutron capture sim. & 0.25 &  0.1 \\
 Det. eff. $\gamma$  & 0.2 &  0.1 \\
 neutron leakage (hole)   & 0.2 & 0.1 \\
 Statistical  & 0.1 & 0.07 \\
 Energy calibration  & 0.15 & 0.1 \\
 French effect   & 0.15 & 0.1 \\
 proton background   &  0.15 & 0.1 \\
   \hline
  \multicolumn{3}{c}{Boron pile experiments only} \\
  \hline
Statistics  & 0.09 & 0.09  \\
Anisotropy (pile eff.) & 0.285 & 0.2\\
neutron leakage & 0.3 & 0.3 \\
MC sim. det. eff. & 0.59 & 0.27  \\
Background & 0.03 & 0.02  \\
FC position & 0.05 & 0.05 \\
   \hline
  \multicolumn{3}{c}{Boron pile and scintillators only} \\
  \hline
  neutron after gate  & 0.1 & 0.05 \\
\hline
\multicolumn{3}{c}{Boron pile and Mn baths only} \\
\hline
   Mn bath calibr. & 0.2 & 0.2 \\
\hline
\multicolumn{3}{c}{Applying to all experiments} \\
\hline
  deadtime & 0.07 & 0.03  \\
  PFNS & 0.12 & 0.05 \\
  multipl. scatt./p escape & 0.11 & 0.08  \\

\hline
\hline
\end{tabular}
\end{table}

Implicitly this analysis can answer the question as to whether the previous Standard uncertainties were truly underestimated given the information available on the measurements.
It was already discussed above that statistical methods indicate that the \CF~\nub~measurements are statistically coherent and, hence, no \USU should apply in the traditional sense (\ie, unknown sources of uncertainties).
However, it was suspected that the uncertainties were underestimated and therefore \USU components were added to the current Neutron Standard uncertainties to account for that.
The PUBs analysis here will investigate whether this suspicion is well-founded and if the added \USU contribution (the only difference between previous and current standard uncertainties) accounted for that adequately.

To this end, we apply the PUBs methodology as outlined in Ref.~\cite{Neudecker_PUB:2018} for estimating bounds on the $^{239}$Pu(n,f) cross-sections as evaluated within the Neutron Standards project.
The only difference between the work here and Ref.~\cite{Neudecker_PUB:2018} is that \CF~\nub~is a scalar quantity and hence no functional form and correlation matrices need to be estimated.
In the first step, the physics sub-processes governing the \CF~\nub~measurements are identified.
The physics sub-processes depend distinctly on the measurement type (boron-pile versus scintillators versus Mn bath), as is highlighted in Table~\ref{tab:PUBS_Cf}.
For instance, a manganese versus sulfur ratio reaction process does not apply to boron pile or scintillator measurements.

\begin{figure}[!thb]
\vspace{-2mm}
\subfigure{\includegraphics[width=.99\columnwidth]{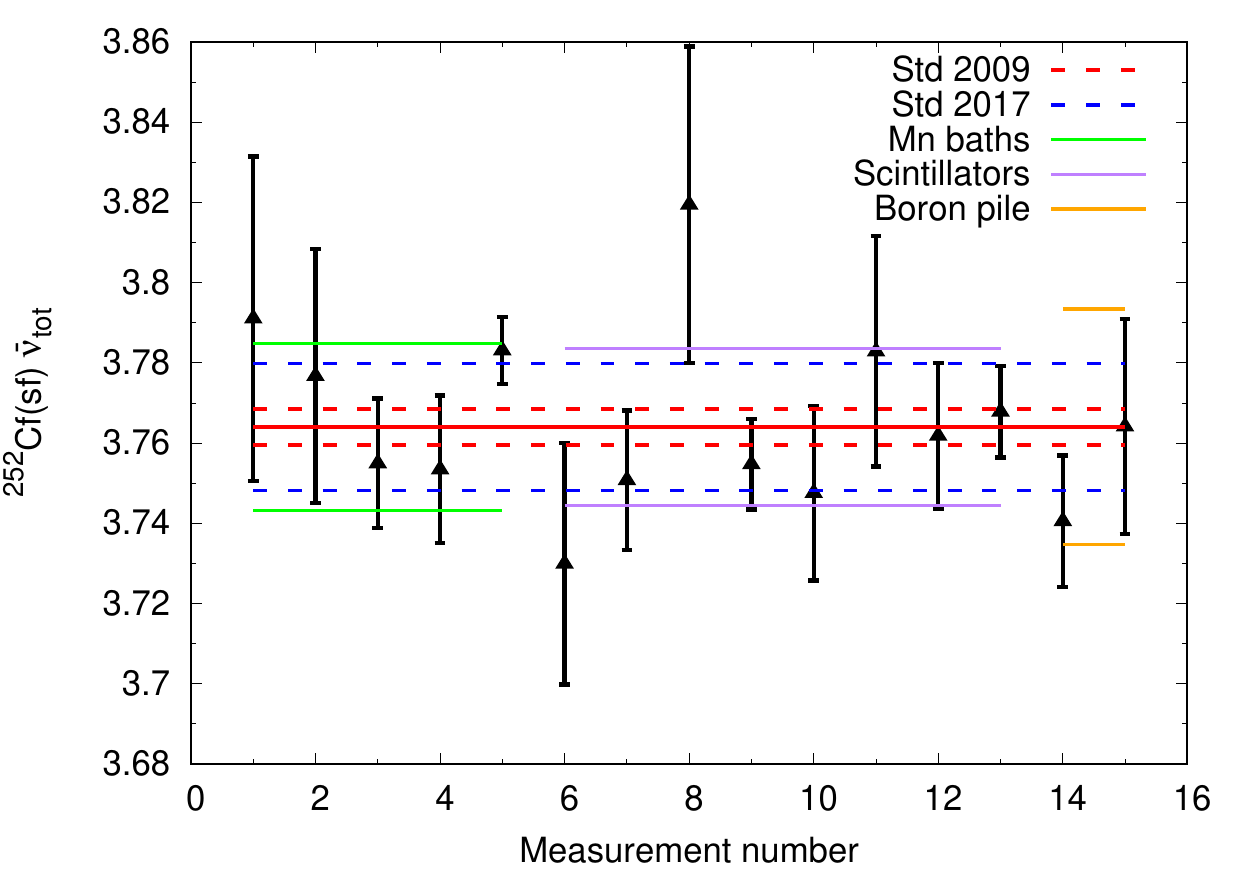}}
\subfigure{\includegraphics[width=.99\columnwidth]{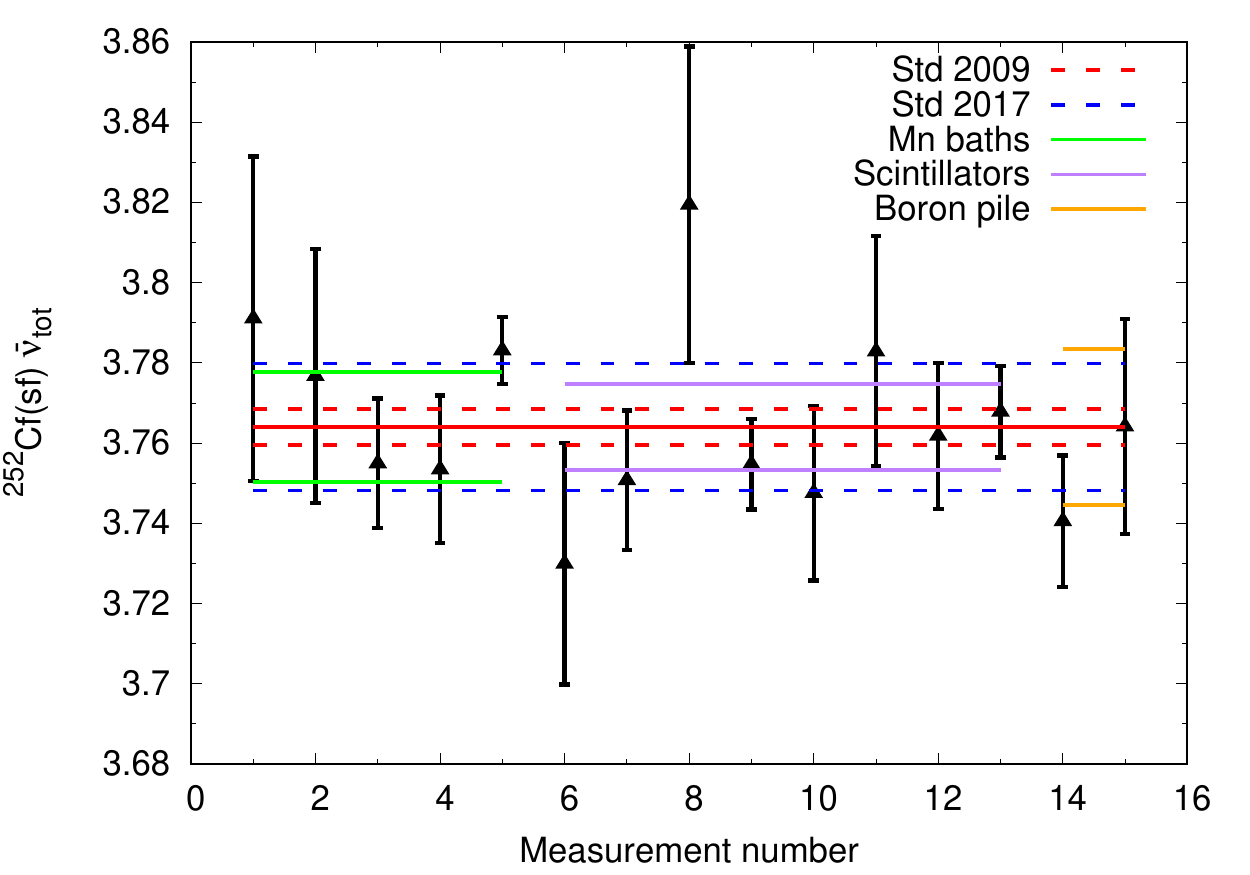}}
\vspace{-3mm}
\caption{(Color online) The conservative PUBs (upper panel) and minimally realistic (lower panel) uncertainties for the \CF~\nub~per measurement type are compared to the previous~\cite{Standards:2009} and current Standards~\cite{Standards:2018} uncertainties.}
\label{fig:PUBs_Cf_permeasurement}
\vspace{-3mm}
\end{figure}

\begin{figure}[!thbp]
%\vspace{+2mm}
\centering
\includegraphics[width=0.98\columnwidth]{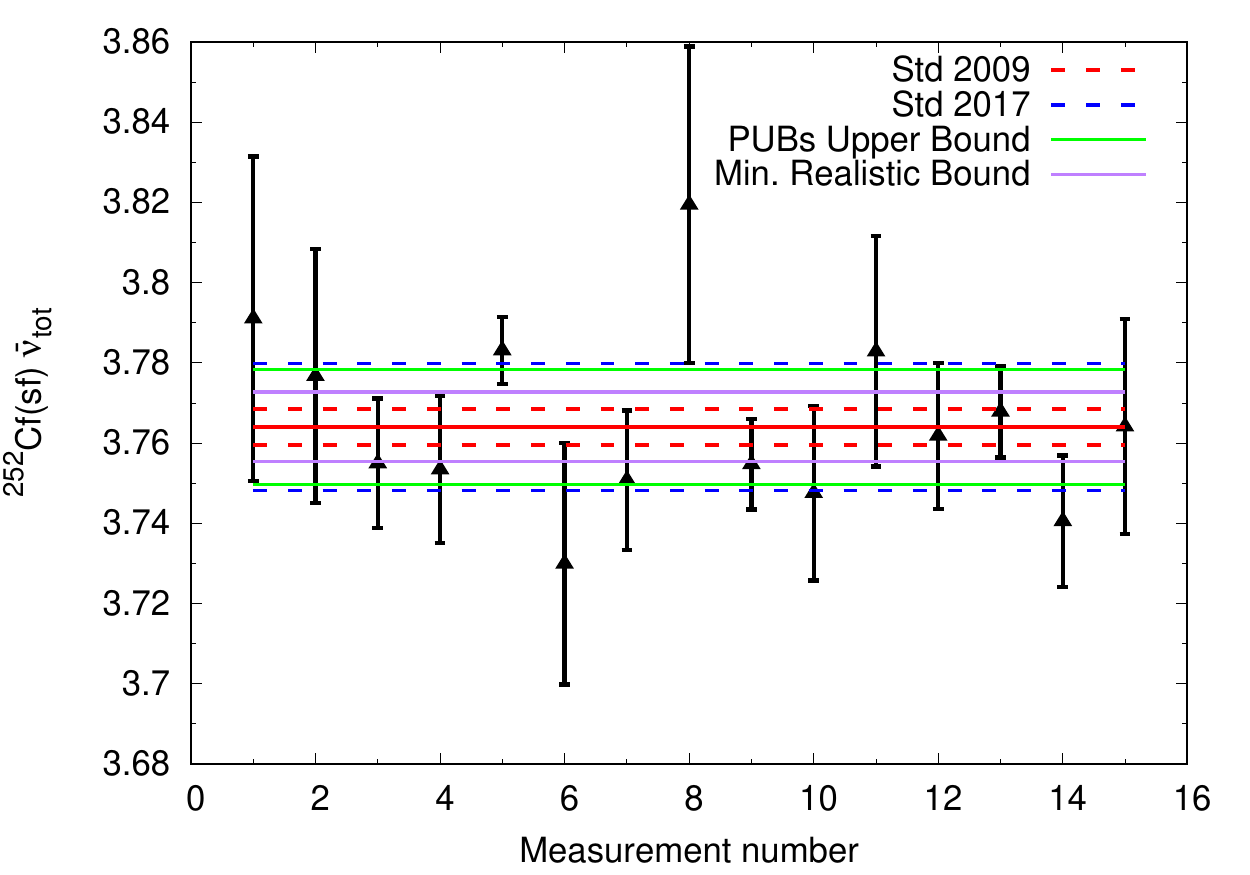}
\vspace{-2mm}
\caption{(Color online) The total minimal realistic and conservative PUBs uncertainties of \CF~\nub~are compared to those of the previous and current Standards. The previous Standards uncertainty~\cite{Standards:2009} is indicated to be underestimated while the current uncertainty~\cite{Standards:2018} is likely slightly overestimated.}
\label{fig:PUBs_CF_total}
\vspace{-2mm}
\end{figure}

In the second step, minimum realistic and conservative bounds are quantified for each sub-process by taking into account whether an uncertainty applies to all measurements (\eg, PFNS uncertainty) or is reduced because it is not fully correlated between measurements (\eg, statistical uncertainties).
Also, uncertainties applying to all types or only two out of three types of measurements are quantified.
A total minimum realistic and a conservative upper PUBs bound for each measurement type is quantified and shown in Fig.~\ref{fig:PUBs_Cf_permeasurement}.
It is noteworthy that the conservative upper PUBs bound encloses at least 66\% of the mid-points of the experimental data indicating that these bounds are realistic within their respective group, consistent with Clue 3.

The total minimum realistic (0.23\%) and conservative PUBs uncertainty (0.38\%), $\delta^{m}$ and $\delta^c$, respectively, are calculated by
\begin{equation}
\delta^{m/c} = \left(\sqrt{\sum\limits_{i,j=1}^{3} (\vec{C}^{m/c}})^{-1}_{ij}\right)^{-1}
\end{equation}
with a covariance matrix $\vec{C}^{m/c}$ populated by the conservative and minimal realistic variances $\mathrm{var}^k$ and covariances between uncertainties associated with the methods $k$ (manganese baths, scintillator and boron pile).

It is evident from Fig.~\ref{fig:PUBs_CF_total} that the previous Neutron Standards estimated uncertainty value of 0.13\% is clearly under-estimated, indicating that uncertainties of single experimental data sets and correlations between experiments are missing.
However, the current Neutron Standards uncertainty of 0.42\% lies above the conservative PUBs bound. Therefore, it is likely slightly over-estimated.
Given these results, it is obvious that a detailed uncertainty quantification of the \CF~\nub~and all pertinent correlations between uncertainties of different experiments is needed.
The work of Croft~\cite{Croft:2019} already took important steps into this direction.
However, questions were raised regarding whether missing uncertainties for single data sets were added and inter-experiment correlations were accounted for.


\begin{thebibliography}{999}

%% Aristotle
\bibitem{Aristotle} Aristotle (384 - 322 B.C.E), {\textit{Nicomachean Ethics}}.

%% Gauss least-squares method
\bibitem{Gauss} Carl Friedrich Gauss (1777-1855): At the relatively young age of 32, Gauss published his method of calculating orbits of celestial bodies. In doing so, he used the method of least squares, claiming to have knowledge of this fundamental mathematics approach as early as 1795, when he would have just been 18 years old. However, credit for inventing this method was in dispute at the time. Adrien-Marie Legendre first published a version of the method in 1805, but Gauss pushed his definition to the extreme in a way that Legendre did not do.

%% Smith 1991 book
\bibitem{Smith:1991} D.~Smith, \textit{Probability, Statistics, and Data Uncertainties in Nuclear Science and Technology}, American Nuclear Society, LaGrange Park, Illinois (1991).

%% Frank *******
\bibitem{cox:2006}
M.G.~Cox, C.~Ei{\o}, G.~Mana, and F.~Pennecchi, \ql The generalized weighted mean of correlated quantities,\qrs
{\sc Metrol. }\textbf{43}, S268--S275 (2006).

%% PPP reference
\bibitem{Peelle:1988} R.~Peelle, \ql Evaluating Nuclear Data Uncertainty: Progress, Pitfalls, and Prospects\qrs, see in~\cite{INDC192}, 68--71 (1988).

\bibitem{INDC192} \textit{Proc. of an IAEA Specialists' Meeting on Covariance Methods and Practices in the Field of Nuclear Data}, Rome, Italy, 17-19 November 1986\qrs, ed. V. Piksaikin, Report \textbf{INDC(NDS)-192/L}, IAEA, Vienna (1988).

\bibitem{D_Agostini:1994}
G.D.~Agostini, \ql On the use of the covariance matrix to fit correlated data,\qrs
{\sc Nucl. Instr. Meth. Phys. Res. }\textbf{A346}, 306--311 (1994).

\bibitem{Froehner:2003}
F.H.~Fr\"ohner, \ql Evaluation of data with systematic errors,\qrs
{\sc Nucl. Sci. \& Eng. }\textbf{145}, 342--353 (2003).

\bibitem{Neudecker:2012}
D.~Neudecker, R.~Fr\"uhwirth, and H.~Leeb, \ql Peelle's pertinent puzzle: A fake due to improper analysis,\qrs
{\sc Nucl. Sci. \& Eng. } \textbf{170}, 54--60 (2012).

\bibitem{Tanabashi:2018} M. Tanabashi \etal (Particle Data Group) \ql Review of Particle Physics,\qr
\textsc{Phys. Rev. }\textbf{D98}, 030001 (2018). Available on-line at \href{http://pdg.lbl.gov/2018/html/computer_read.html}{pdg.lbl.gov/2018/html/computer\_read.html}.

\bibitem{Badikov:2012}
S.A.~Badikov and V.P.~Chechev, \ql Procedure for statistical analysis of one-parameter discrepant
  experimental data,\qrs {\sc App. Rad. \& Isot. } \textbf{70}, 1850--1852 (2012).

\bibitem{Zhang:2006}
Nien~Fan Zhang, \ql The uncertainty associated with the weighted mean of measurement data,\qrs
{\sc Metrol. } \textbf{43}, 195--204 (2006).

\bibitem{Kossert:2004}
K.~Kossert and E.~G{\"u}nther, \ql {LSC} measurements of the half-life of $^{40}$K,\qrs
{\sc Appl. Rad. Isot. } \textbf{60}, 459--464 (2004).

\bibitem{Rukhin:2009}
A. L. Rukhin, \ql Weighted means statistics in interlaboratory studies,\qrs
{\sc Metrol. }\textbf{46}, 323--331 (2009).

\bibitem{Zimmerman:2012}
B.E.~Zimmerman, T.~Altzitzoglou, A.~Antohe, \etals, \ql Results of an international comparison for the activity measurement
  of $^{177}$Lu,\qrs {\sc App. Rad. \& Isot. }\textbf{70}, 1825--1830 (2012).

\bibitem{Ratel:2015}
G.~Ratel, C.~Michotte, and F.~Bochud, \ql Uncertainty of combined activity estimations,\qrs
{\sc Metrol. }\textbf{52}, S30--S41 (2015).

\bibitem{Lepy:2015}
M.C.~L{\'{e}}py, A.~Pearce, and O.~Sima, \ql Uncertainties in gamma-ray spectrometry,\qrs
{\sc Metrol. }\textbf{52}, S123--S145 (2015).

\bibitem{James:1992}
M.F.~James, R.W.~Mills, and D.R.~Weaver, , \ql The use of the normalized residual in averaging experimental data and
  in treating outliers,\qrs {\sc Nucl. Instr. Meth. Phys. Res. }\textbf{A313}, 277--282 (1992).

\bibitem{Willink:2002}
R.~Willink, \ql Statistical determination of a comparison reference value using
  hidden errors,\qrs {\sc Metrol. }\textbf{39}, 343--354 (2002).

\bibitem{Steele:2005}
A.G.~Steele, B.M.~Wood, and R.J.~Douglas, \ql Outlier rejection for the weighted-mean {KCRV},\qrs
{\sc Metrol. } \textbf{42}, 32--38 (2005).

\bibitem{Ellison:2018}
S.L.R.~Ellison, \ql An outlier-resistant indicator of anomalies among inter-laboratory comparison data with associated uncertainty,\qrs
{\sc Metrol. }\textbf{55}, 840--854 (2018).

%\bibitem{kendall:2009}
%A. Stuart and K. Ord, {\it Kendall's Advanced Theory of Statistics: Volume 1: Distribution Theory}, Wiley, 2009.
%
%\bibitem{Ratel:2005}
%G. Ratel, \ql Evaluation of the uncertainty of the degree of equivalence,\qrs
%{\sc Metrol. } \textbf{42}, 140--144 (2005).
%
%\bibitem{Rukhin:2019}
%A. L. Rukhin, \ql Homogeneous data clusters in interlaboratory studies,\qrs
%{\sc Metrol. } \textbf{56}, 035002 (2019).
%
%\bibitem{Willink:2007}
%R.~Willink, \ql A generalization of the welch{\textendash}satterthwaite formula for
%  use with correlated uncertainty components,\qrs {\sc Metrol. } \textbf{44}, 340--349 (2007).
%
%\bibitem{Nam:2008}
%Gyeonghee Nam, Chu-Shik Kang, Hun-Young So, and Jong Oh Choi, \ql An uncertainty evaluation for multiple measurements by {GUM}, {III}:
%  using a correlation coefficient,\qrs {\sc Accred. Qual. Assur. }\textbf{14}, 43--47 (2008).
%
%\bibitem{Cordero:2005}
%R. R. Cordero and G. Fuster, \ql Revisiting the problem of the evaluation of the uncertainty
%  associated with a single measurement,\qrs {\sc Metrologia}, 42(2):L15--L19 (2005).
%
%\bibitem{Silva:2018}
%J. Araujo~Da Silva and M.~Angelo Cirillo, \ql Selection criterion of work matrix as a function of limiting
%  estimates of the covariance matrix of correlated data in {GEE},\qrs {\sc Biomet. J. }\textbf{60} 979--990 (2018).
%
%\bibitem{Maples:2018}
%M.~P. Maples, D.~E. Reichart, N.~C. Konz, T.~A. Berger, A.~S. Trotter, J.~R.
%  Martin, D.~A. Dutton, M.~L. Paggen, R.~E. Joyner, and C.~P. Salemi, \ql Robust chauvenet outlier rejection,\qrs
%{\sc The Astroph. J. Suppl. Ser. }\textbf{238} 1:2 (2018).
%
%\bibitem{Tellinghuisen:1996}
%J. Tellinghuisen, \ql On the least-squares fitting of correlated data:a priorivsa {PosterioriWeighting},\qrs
%{\sc Journal of Molecular Spectroscopy}, 179(2):299--309 (1996).
%
%\bibitem{Froehner:1989}
%F.~H. Fr\"ohner, \ql Bayesian evaluation of discrepant experimental data.\qrs In {\it Maximum Entropy and Bayesian Methods}, pages 467--474.
%Springer Netherlands, 1989.

%% Evaluation methods review
\bibitem{Capote:2010} R.~Capote, D.L.~Smith and A.~Trkov, \ql Nuclear data evaluation methodology including estimates of covariances\qrs, \textsc{EPJ Web Conf. }\textbf{8}, 04001 (2010).

%% Smith and Otuka 2012 review paper
\bibitem{Smith-Otuka:2012} D.L.~Smith and N.~Otuka, \textit{Nucl. Data Sheets }\textbf{113}, 3006-–3053 (2012).

%
\bibitem{WPEC-SG24} E.~Bauge, R.~Capote, U.~Fisher \etals, \ql Covariance Data in the Fast Neutron Region\qrs. Final report of the \emph{NEA/WPEC-24}, Coordinator M. Herman, Report \textbf{NEA/NSC/WPEC/DOC(2010)427} (Nuclear Energy Agency, OECD, Paris, 2011).

%% GLSQ codes
\bibitem{Smith:1993} D.L.~Smith, \ql A Least-Squares Computational *Tool Kit*,\qrs Report \textbf{ANL/NDM-128}, Argonne National Laboratory, IL, USA (1993).

%
\bibitem{GLUCS} D.M.~Hetrick and C.Y.~Fu, \ql GLUCS: A generalized least-squares program for updating cross section evaluations with correlated data sets,\qrs~Oak Ridge National Laboratory, TN, USA (1980).

%
\bibitem{KALMAN} T.~Kawano and K.~Shibata, \ql KALMAN,\qrs~Report \textbf{JAERI-Data/Code 97-037}, JAERI, Tokai, Japan (1997).

%
\bibitem{Rising:2013} M.E.~Rising, P.~Talou, T. Kawano, and A.K. Prinja, ``Evaluation and Uncertainty Quantification of Prompt Fission Neutron Spectra of Uranium and Plutonium Isotopes,'' \textsc{ Nucl. Sci. Eng. }\textbf{175}, 81--93 (2013).

%
\bibitem{GANDR} D.W. Muir, \ql Global Assessment of Nuclear Data Requirements (GANDR project),\qrs~IAEA report (6 volumes), Vienna, Austria (2007). Available online at \url{https://www-nds.iaea.org/gandr/docs.html}.

\bibitem{Schnabel:2017} G. Schnabel, \ql Fitting and Analysis Technique for Inconsistent Nuclear Data,\qrs
Proc. of Conference MC2017, 2017.

%% GLSQ fit using Pade approximants
\bibitem{Pade:1892} H.E. Pad\'e, \ql Sur la Repr\'esentation Approch\'ee d' une Fonction par des Fractions Rationnelles,\qr Supplement to \textsc{Ann. Sci. l'\'Ecole Norm. Sup. }, Series 3, Vol. \textbf{9}, 3--93 (1892).

\bibitem{Graves-Morris:1973} P.R. Graves-Morris (Ed.), \ql Pad\'e Approximants and their Applications,\qr Academic Press, New York (1973).

\bibitem{Baker:1975} G.A. Baker Jr., \ql Essentials of Pad\'e Approximants,\qr Academic Press, New York (1975).

\bibitem{Vinogradov:1987} V.N. Vinogradov, E.V. Gai and N.S. Rabotnov, \ql Analytical Approximation of Data in Nuclear and Neutron Physics,\qr Energoatomizdat, Moscow (1987); in Russian.

\bibitem{Badikov:1992}  S.A. Badikov, E.V. Gai, M.A. Guseynov and N.S. Rabotnov, \ql Nuclear data processing, evaluation, transformation and storage with Pad\'e-approximants,\qrs Proc. Int. Conf. Nuclear Data for Science and Technology, J\"ulich, Germany, 1991, ed: S.M. Qaim, 182--187 (1992) Springer-Verlag, Berlin.

%% Charged-particle monitor reactions
\bibitem{Hermanne:2018} A. Hermanne, A.V. Ignatyuk, R. Capote \etal \ql Reference cross sections for charged-particle monitor reactions,\qr \textsc{Nucl. Data Sheets }\textbf{148}, 338--382 (2018).

% Stochastic evaluation methods
\bibitem{UMCG} D.L.~Smith, \ql A Unified Monte Carlo Approach to Fast Neutron Cross Section Data Evaluation,\qrs~in \textsc{Proc. 8th Int. Topical Meeting on Nucl. Appl. and Utilization of Accelerators}, Pocatello, ID, USA, 29 July 29--2 August, p. 736 (2007).

%
\bibitem{UMCG1} R.~Capote and D.L.~Smith, \ql An Investigation of the Performance of the Unified Monte Carlo Method of Neutron Cross Section Data Evaluation,\qrs \textsc{Nucl. Data Sheets }\textbf{109}, 2768 (2008).

%
\bibitem{BFMC} E.~Bauge and P.~Dossantos-Uzarralde, \ql Evaluation of the covariance matrix of Pu-239 neutronic cross sections in the continuum using a Backward-Forward Monte Carlo method,\qr \textsc{J. Kor. Phys. Soc. }\textbf{59}, 1218 (2011).

%
\bibitem{UMCB} R.~Capote, D.L.~Smith, A.~Trkov, M.~Meghzifene, \ql A New Formulation of the Unified Monte Carlo Approach (UMC-B) and Cross-Section Evaluation for the Dosimetry Reaction $^{55}$Mn($n$,$\gamma$)$^{56}$Mn,\qr \textsc{J. ASTM Int. }\textbf{9}, 104115 (2012).

%
\bibitem{BMC} A.J. Koning, \ql Bayesian Monte Carlo method for nuclear data evaluation,\qrs~  \textsc{EPJ }\textbf{A51}, 184 (2015).

\bibitem{HFB-mass} S. Goriely and R. Capote, \ql Uncertainties of mass extrapolations in Hartree-Fock-Bogoliubov mass models,\qrs \textsc{ Phys. Rev. }\textbf{C89}, 054318 (2014).

%% TENDL-2017
\bibitem{TENDL-2017} A. J. Koning, D. Rochman, J.-Ch. Sublet, N. Dzysiuk, M. Fleming, and S. van der Marck, \ql TENDL: Complete Nuclear Data Library for Innovative Nuclear Science and Technology,\qr \textsc{Nucl. Data Sheets }\textbf{155}, 1-–55 (2019).

%
\bibitem{Moore:1965} G.E. Moore,\ql Cramming more components onto integrated circuits\qrs. \textsc{Electronics} \textbf{1965-04-19}.

%% W. Poenitz method for evaluating neutron cross-section standards
\bibitem{Poenitz:1981} W. Poenitz, \ql Evaluation Methods for Neutron Cross Section Standards,\qr \textit{Proc. Conf. on Nuclear Data Evaluation Methods and Procedures}, Report \textbf{BNL-NCS-51363}, Brookhaven National Laboratory, Upton, NY, USA, p.249--290 (1981).
%
\bibitem{PoenitzAumeier:1997} W.P. Poenitz and S.E. Aumeier, ``The Simultaneous Evaluation of the Standards and Other Cross Sections of Importance for Technology,'' Report \textbf{ANL/NDM-139}, Argonne National Laboratory, IL, USA (1997).,

%% ENDF/B-VI standards reference
\bibitem{ENDFB-VI-standards}  C.L. Dunford, Evaluated Nuclear Data File, ENDF/B-VI, \textit{Proc. Int. Conf. on Nuclear Data for Science and Tech.}, 13-17 May 1991, Forschungszentrum Juelich, Fed. Rep. of Germany, ed. S.M. Qaim, pp. 788--792, Springer Verlag, Heidelberg (1992).

\bibitem{Carlson:1993} A.D. Carlson, W.P. Poenitz, G.M. Hale, \etals, \ql The ENDF/B-VI Neutron Cross Section Measurement Standards,\qr Technical report \textbf{NISTIR-5177}, National Institute of Standards and Technology (1993); also ENDF-351, Brookhaven National Laboratory.

\bibitem{Pronyaev:2003} V.G. Pronyaev, IAEA, Private Communication (2003). See also~\cite{Standards:2009}.

\bibitem{Badikov:2007} S.A. Badikov, Z. Chen, A.D. Carlson \etals, \ql International Evaluation of Neutron Cross-section Standards,\qrs
Technical Report \textbf{STI/PUB/1291}, IAEA, Vienna (2007).

%% Standards 2009
\bibitem{Standards:2009} A. D. Carlson, V. G. Pronyaev, D. L. Smith \etals, \ql International Evaluation of Neutron Cross Section Standards,\qrs  \textsc{ Nucl. Data Sheets }\textbf{110}, 3215--3324~(2009).

%% Standards 2018
\bibitem{Standards:2018} A. D. Carlson, V.G. Pronyaev, R. Capote \etals, \ql Evaluation of Neutron Data Standards,\qrs  \textsc{ Nucl. Data Sheets} {\bf 148}, 142--187 (2018).

%% ENDF/B-VIII.0 evaluation
\bibitem{ENDFB-VIII:2018} D. Brown, \textit{et al.}, \ql ENDF/B-VIII.0: The $8^{th}$ Major Release of the Nuclear Reaction Data Library
with CIELO-project Cross Sections, New Standards and Thermal Scattering Data\qrs~\textit{Nucl. Data Sheets} {\bf{148}}, 1--142 (2018).

%
%% Chiba and Smith fix
\bibitem{Chiba-Smith:1991} S. Chiba and D. Smith, \ql A Suggested Procedure for Resolving an Anomaly in Least-Squares Data Analysis Known as \ql Peelle’s Pertinent Puzzle\qr and the General Implications for Nuclear Data Evaluation Report,\qr \textbf{ANL/NDM-121}, Argonne National Laboratory (1991).

%
\bibitem{EXFOR} N. Otuka, E. Dupont, V. Semkova, B. Pritychenko \etals, ``Towards a More Complete and Accurate Experimental Nuclear Reaction Data Library (EXFOR): International Collaboration Between Nuclear Reaction Data Centres (NRDC),'' \textsc{ Nucl. Data Sheets }\textbf{120}, 272--276~(2014). Data available online (\eg, at \href{http://www-nds.iaea.org/exfor/}{\textit{www-nds.iaea.org/exfor/}}).

\bibitem{ENSDF} Evaluated Nuclear Structure Data File (ENSDF). Available online at \href{www.nndc.bnl.gov/ensdf/}{\textit{www.nndc.bnl.gov/ensdf/}}. Developed and maintained by the International Network Of Nuclear Structure and Decay Data Evaluators (NSDD) (see \href{https://www-nds.iaea.org/nsdd/}{\textit{www-nds.iaea.org/nsdd/}}).

\bibitem{Livechart} {\bf LiveChart of Nuclides}, IAEA decay data retrieval code available online at  \href{https://www-nds.iaea.org/medical/monitor\_reactions.html}{\textit{www-nds.iaea.org/medical/monitor\_reactions.html}}.

\bibitem{NUDAT} {\bf NuDat}, Brookhaven National Laboratory, USA. Decay data retrieval code available online at  \href{https://www.nndc.bnl.gov/nudat2/}{\textit{www.nndc.bnl.gov/nudat2/}}.

%
\bibitem{Neudecker:2018} D. Neudecker, B. Hejnal, F. Tovesson \etals, \ql Template for estimating uncertainties of measured neutron-induced fission cross-sections,\qrs~\textsc{EPJ Nucl. Sci. Tech. }\textbf{4}, 21 (2018).

%
\bibitem{Schillebeecks:2012} P. Schillebeeckx \textit{et al.}, ``Determination of Resonance Parameters and their Covariances from Neutron Induced Reaction Cross Section Data'', \textsc{ Nucl. Data Sheets }\textbf{113}, 3054--3100 (2012).

%
\bibitem{Helgesson:2015} P. Helgesson, mid-term Ph.D. Thesis, University of Uppsala, Sweden (2015).

\bibitem{Neudecker_Template:2019}
D.~Neudecker, D.L.~Smith, F.~Tovesson, \etal, \ql Applying a Template of Uncertainties to Updating Uncertainties of $^{239}$Pu(n,f) Cross-section Data in the Neutron Data Standards Database\qrs, {\sc Nucl. Data Sheets}, this issue.

\bibitem{Helgesson:2017} P.~Helgesson, H.~Sj\"ostrand and D.~Rochman, \textquotedblleft Uncertainty-driven nuclear data evaluation including thermal (n,$\alpha$) applied to $^{59}$Ni,\textquotedblright \textsc{Nucl. Data Sheets }\textbf{145}, 1--24 (2017).

\bibitem{Staples:1995}  P. Staples \etals, \textsc{Nucl. Phys. }\textbf{A591}, 41 (1995); EXFOR 13982001.

\bibitem{Neudecker:2014} D. Neudecker, P. Talou, T. Kawano \etals, \ql Evaluation of the $^{239}$Pu Prompt Fission Neutron Spectrum Induced by Neutrons of 500~keV and Associated Covariances,\qr
Report \textbf{LA-UR-14-22817}, Los Alamos National Laboratory, NM, USA (2014).

%% Gai paper (Russian & English)
\bibitem{Gai:2007} E.~Gai, \ql Some algorithms for evaluating nuclear data and generating uncertainty covariance matrices\qrs, \textsc{ Vopr. Atom. Nauki i Tekh., Ser. Yad. Konst.} issue 1-2, 56 (2007); in Russian. Translated to English as IAEA report \textbf{INDC(NDS)-0750}, January 2018, IAEA, Vienna, Austria. Available online at \href{https://www-nds.iaea.org/publications/indc/indc-nds-0750.pdf}{\textit{www-nds.iaea.org/publications/indc/indc-nds-0750.pdf}}.

\bibitem{Gai:2007a} E.V. Gai, ``On the Problem of Ambiguity of the Evaluated Nuclear Data Uncertainties,''
\textsc{ Vopr. Atom. Nauki i Tekh., Ser. Yad. Konst.} issue 1--2, 45--55 (2007); in Russian.

%% Badikov and Gai paper
\bibitem{Badikov:2003} S.A. Badikov and E.V. Gai, ``Some Sources of the Underestimation of Evaluated Cross-section Uncertainties,'' Report \textbf{INDC(NDS)-438}, IAEA, Vienna 2003, pp. 117--129. Available online at \url{https://www-nds.iaea.org/publications/indc/indc-nds-0438.pdf}.

\bibitem{Pronyaev:2003a} V. G. Pronyaev, \ql Does model fit decrease the uncertainty of the data in comparison with a general non-model least squares fit?,\qr Report \textbf{INDC(NDS)-438}, IAEA, Vienna 2003, pp. 172--183. Available online at \url{https://www-nds.iaea.org/publications/indc/indc-nds-0438.pdf}.

%% Covariance Workshop 2008
\bibitem{CW2008} Workshop on Neutron Cross Section Covariances, Port Jeffereson, NY, USA, 24--28 June, 2008, ed. P. Oblo{\v z}insk\'{y}, \textit{Nucl. Data Sheets }{\bf{109}}, 2725-2922 (2008). Available online at \url{https://www.sciencedirect.com/journal/nuclear-data-sheets/vol/109/issue/12}.

%% Covariance Workshop 2011
\bibitem{CW2011} Second Workshop on Neutron Cross Section Covariances, Vienna, Austria, 14--16 September, 2011, eds. D. Neudecker and H. Leeb, \textit{EPJ Web of Conf. }{\bf{27}} (2012). Available online at \url{https://www.epj-conferences.org/articles/epjconf/abs/2012/09/contents/contents.html}.

%% Covariance Workshop 2014
\bibitem{CW2014} International Workshop on Nuclear Data Covariances, Santa Fe, NM, USA, 28 April--1 May, 2014, eds. P. Oblo{\v z}insk\'{y} and B. Pritychenko, \textit{Nucl. Data Sheets }{\bf{123}}, 1--238 (2015). Available online at \url{https://www.sciencedirect.com/journal/nuclear-data-sheets/vol/123/suppl/C}.

%% Covariance Workshop 2017

\bibitem{CW2017} Fourth International Workshop on Nuclear Data Covariances, Aix en Provence, France, 2--6 October, 2017, ed. Cyrille de Saint Jean, \textit{EPJ Nucl. Sc. Tech. }{\bf{4}}, (2018). Available online at \url{https://www.epj-n.org/articles/epjn/abs/2018/01/contents/contents.html}.

\bibitem{Neudecker_PUB:2018}
D.~Neudecker, M.C. White and D.E.~Vaughan, \ql Validating Nuclear Data Uncertainties Obtained from a Statistical Analysis of Experimental Data with the ``Physical Uncertainty Bounds'' Method,\qr Report \textbf{LA-UR-19-22384}, Los Alamos National Laboratory, NM, USA (2018).

\bibitem{Neudecker:2013}
D.~Neudecker, R.~Capote and H.~Leeb, \ql Impact of Model Defect and Experimental Uncertainties on Evaluated Output,\qr {\sc Nucl. Instrum. and Meth. in Phys. Res. }\textbf{A723}, 163--172 (2013).

\bibitem{Gai:2008}  E.V.~Gai and A.V.~Ignatyuk, ``Uncertainties and Covariances of the Fission Cross Sections and the Fission Neutron Multiplicities for Actinides,'' \textsc{Nucl. Data Sheets }\textbf{ 109}, 2890--2893 (2008).

\bibitem{Blokhin:2016} A.I. Blokhin, E.V. Gai, A.V. Ignatyuk \etals, ``New Version of Neutron
 Evaluated Data Library BROND-3,\qrs \textsc{Vopr. Atom. Nauki i Tech., Ser. Nucl. Const. }, issue 2, pp. 62--93 (2016); in Russian.

%% Capote-Neudecker Cf nubar paper
\bibitem{Capote-Neudecker:2018} R. Capote and D. Neudecker, \ql How accurately we know the standard $^{252}$Cf(sf) neutron multiplicity?,\qr American Nuclear Society meeting ANTPC November 2018, Orlando, FL, USA. Available online at arXiv 1908.00272 (2019), at \url{http://arxiv.org/abs/1908.00272}.

%% Report JCGM 100:2008
\bibitem{JCGM100} JCGM 100:2008, \ql Evaluation of measurement data: Guide to the expression of uncertainty in measurement,\qrs~Bureau International des Poids et Mesures (BIPM), Paris (2008). BIPM publication available at \url{https://www.bipm.org/utils/common/documents/jcgm/JCGM_100_2008_E.pdf} (2008).

\bibitem{Pronyaev:2017} V.G. Pronyaev, R. Capote, A. Trkov, G. Noguere, and A. Wallner, \ql New fit of thermal neutron constants (TNC) for $^{233,235}$U, $^{239,241}$Pu and \CF: Microscopic vs. Maxwellian data,\qrs~\textsc{EPJ Web of Conf. }\textbf{146}, 02045 (2017).

\bibitem{Divadeenam:1984} M. Divadeenam and J.R. Stehn, \ql A least-squares evaluation of thermal data for fissile nuclei,\qrs  \textsc{ Ann. Nucl. Energy} {\bf 11}, 375 (1984).
%
\bibitem{Axton:1986} E.J. Axton, ``Evaluation of the thermal constants of $^{233}$U, $^{235}$U, $^{239}$Pu and $^{241}$Pu, and the fission neutron yield of $^{252}$Cf,'' \textbf{Report GE/PH/01/86}, Central Bureau for Nuclear Measurements, Geel (1986).

\bibitem{Vaughan_PUB:2014}
D.E.~Vaughan and D.L.~Preston, \ql Physical Uncertainty Bounds (PUB),\qr Report \textbf{LA-UR-14-20441},  Los Alamos National Laboratory, NM, USA (2014).

%% Occam Razor
\bibitem{OccamRazor} Occam Razor, Wikipedia Free Encyclopedia (2011). See: \url{http://en.wikipedia.org/wiki/Occam’s Razor}.

% ratio data

\bibitem{Tovesson:2015} F. Tovesson, A. Laptev, T.S. Hill, \ql Fast Neutron-Induced Fission Cross Sections of $^{233,234,236,238}$U up to 200 MeV,\qrs
\textsc{Nucl. Sc. Eng. }\textbf{178}, 57 (2014); EXFOR 14402.

\bibitem{Paradela:2015} C. Paradela \etal (n\_TOF collaboration), \ql High-accuracy determination of the \UF/\UT fission cross section ratio up to $\approx 1$ GeV at n\_TOF at CERN,\qrs \textsc{Phys. Rev. }\textbf{C91}, 024602 (2015); EXFOR 23269002,  23269003,  23269004,  23269005.

\bibitem{Behrens:1977} J.W. Behrens and G.W. Carlson, \ql Measurements of the neutron-induced fission cross
            sections of $^{234}$U, $^{236}$U, and $^{238}$U relative to $^{235}$U
            from 0.1 to 30 MeV,\qrs \textsc{Nucl. Sc. Eng. }\textbf{63}, 250 (1977); EXFOR 10653.

\bibitem{Difilippo:1978} F.C. Difilippo, R.B. Perez, G. de~Saussure, D. Olsen, R. Ingle, \ql Measurement of the Uranium-238 to Uranium-235 Fission
             Cross Section Ratio Between 0.1 and 25~MeV,\qrs \textsc{Nucl. Sc. Eng. }\textbf{68}, 43 (1978); EXFOR 10635.

\bibitem{Cierjacks:1976} S. Cierjacks, B. Leugers, K. Kari \etals,
\ql Measurements of neutron induced fission cross section ratios at the Karlsruhe isochronous cyclotron,\qrs
\textit{Meet. Fast Neutr. Cross Sect. of U and Pu}, Argonne, IL, USA (1976).

\bibitem{Coates:1975} M.S. Coates, D.B. Gayther, N.J. Pattenden,
\ql A measurement of the U-238/U-235 fission cross- section ratio,\qrs
\textit{Conf. on Nucl. Cross-Sect. and Techn. }, Washington, D.C., Vol.2, p.568 (1975), USA.

\bibitem{Shcherbakov:2001} O. Shcherbakov, A. Donets, A. Evdokimov \etals,
\ql Neutron-induced fission of $^{233}$U, $^{238}$U, $^{232}$Th, $^{239}$Pu, $^{237}$Np, $^{nat}$Pb and $^{209}$Bi relative to $^{235}$U in the energy range 1--200~MeV,\qrs
\textsc{J. Nucl. Sc. Tech. (Japan) Supp.~2}, 230--233 (2002); Technical report \textbf{JINR-E3-2001-192}, 257 (2001), JINR, Dubna, Russia; EXFOR 41455.

\bibitem{Lisowski:1991} P.W. Lisowski, A. Gavron, W.E. Parker \etals, \ql Fission cross sections ratios for $^{233}$U, $^{234}$U, and $^{236}$U~relative to \UT~from 0.5 to 400~MeV,\qrs
in \textit{Proc. Int. Conf. on Nuclear Data for Science and Technology}, J\"ulich, Fed. Rep. of Germany, 13--17 May 1991, S. M. Qaim, (Ed.), pp. 732--733, Springer (1991); EXFOR~14011.

%% By Frank Gunsing
\bibitem{Otuka:2017} N.~Otuka, R. Capote., V. Semkova, T. Kawai, and G. Noguere, \ql Experiments in the EXFOR library for evaluation of thermal
neutron constants,\qrs~\textsc{EPJ Web of Conf. }\textbf{146}, 07005 (2017).

%% From Peter
\bibitem{BENN54} C.A. Bennett and N.L. Franklin, \textit{Statistical Analysis in Chemistry and the Chemical Industry}, Wiley, New York; Chapman $\&$ Hall, 1954.

\bibitem{BORE07} A.~Borella, G.~Aerts, F.~Gunsing, M.~Moxon, P.~Schillebeeckx, R.~Wynants,
\ql The use of $C_6D_6$ detectors for neutron induced capture cross-section measurements in the resonance region,\qrs \textsc{Nucl. Instr. Meth. }\textbf{A577}, 626--640 (2007).

\bibitem{SCHI12} P. Schillebeeckx, B. Becker, Y. Danon \etals, \ql Determination of Resonance Parameters and their Covariances from Neutron Induced Reaction Cross Section Data,\qrs \textsc{Nucl. Data Sheets }\textbf{113}, 3054--3100 (2012).

\bibitem{MASS14} C. Massimi, B. Becker, E. Dupont \etals, \ql Neutron capture cross section measurements for $^{197}$Au from 3.5 to 84 keV at GELINA,\qrs \textsc{EPJ }\textbf{50}, 124  (2014).

\bibitem{SCHI19} P. Schillebeeckx, S. Kopecky, and C. Paradela, \ql Evaluation of measurement uncertainties and covariances,\qrs JRC Technical Report, Geel, to be published.

\bibitem{ICSBEP:2016} ICSBEP 2016: International Handbook of Evaluated Criticality Safety Benchmark Experiments,  Nuclear Energy Agency, OECD, Paris (2016). See \url{http://www.oecd-nea.org/science/wpncs/icsbep/handbook.html}.

\bibitem{NJOY} R.E.~MacFarlane, D.W.~Muir, R.M.~Boicourt, A.C.~Kahler, \ql The NJOY Nuclear Data Processing System, Version 2012,\qrs  LANL Report {\bf LA-UR-12-27079}, Los Alamos, NM, USA (2012).

\bibitem{MCNP} \ql  MCNP---A General Monte Carlo Code for Neutron and Photon Transport, Version 5,\qrs  LANL Report \textbf{LA-UR-05-8617}, Los Alamos, NM, USA (2005).

%\bibitem{ENDFB70} M.B. Chadwick, P. Oblo{\v z}insk{\' y}, M. Herman {\it et al.}, ``ENDF/B-VII.0: Next Generation Nuclear Data Library for Nuclear Science and Technology,'' {\sc Nucl. Data Sheets} {\bf 107}, 2931--3060 (2006).

\bibitem{ENDFB71} M.B. Chadwick, M.W. Herman, P. Oblo{\v z}insk\'{y} \etals, \ql ENDF/B-VII.1 nuclear data for science and technology: cross sections, covariances, fission product yields and decay data,\qrs  \textsc{ Nucl. Data Sheets }\textbf{112}, 2887--2996 (2011).

%\bibitem{JEFF31} A. Koning \textit{et al.}, "The JEFF-3.1 Nuclear Data
%Library", \textsc{JEFF Report} \textbf{21}, OECD 2006, NEA No. 6190.

\bibitem{IAEA-CIELO} M.B. Chadwick, \textit{et al.}, \ql CIELO Collaboration Summary Results: International Evaluations of
Neutron Reactions on Uranium, Plutonium, Iron, Oxygen and Hydrogen\qrs~\textsc{Nucl. Data Sheets} {\bf 148}, 189--213 (2018).

\bibitem{Capote:2018-U} R. Capote, A. TRkov, M. Sin, \textit{et al.}, \ql IAEA CIELO Evaluation of Neutron-induced Reactions on $^{235}$U and $^{238}$U Targets,\qr
\textsc{Nucl. Data Sheets }{\bf 148}, 254--292 (2018).

\bibitem{JEFF32} JEFF Scientific Working Group, \ql \textit{Joint Evaluated Fission and Fusion File (JEFF) release 3.2}," OECD, NEA, March~5~(2014).

%\bibitem{JENDL4} K. Shibata \etals, \ql JENDL-4.0: A New Library for Nuclear Science and Engineering,\qrs  \textsc{ J. Nucl. Sci. Techn. }\textbf{48}, 1--30~(2011).

\bibitem{Mihalczo:2002} J.T. Mihalczo, J.J. Lynn, J.R. Taylor \etals, \ql Delayed Critical ORNL Unreflected Uranium (93.2) Metal Sphere and the Pure Unreflected Uranium (93.80) Sphere,\qrs textsc{Nucl. En. }\textbf{29}, 552--560 (2002).

%% From Allan
\bibitem{Poenitz:1970} W. P. Poenitz, \ql Interpretation and Intercomparison of Standard Cross Sections,\qrs \textit{Proc. EANDC Symp. on Neutron Standards and Flux Normalization }, \textbf{CONF-701002}, IL, Chicago, USA, p.338 (1970).

\bibitem{Giorginis:2006} G. Giorginis and V. Khryachkov, \ql The cross section of the $^{10}$B(n,$\alpha$)$^7$Li reaction measured in the MeV energy range,\qrs \textsc{Nucl. Instr. Meth. } \textbf{A562}, 737 (2006).

\bibitem{Zhang:2002} Guohui Zhang, Guoyou Tang, Jinxiang Chen \etals, \ql Differential Cross Section Measurement for the $^{10}$B(n,$\alpha$)$^7$Li Reaction,\qrs \textsc{Nucl. Sci. Eng. }\textbf{142}, 203 (2002).

\bibitem{Zhang:2011} ZHANG Guo-Hui, LIU Xiang, LIU Jia-Ming, XUE Zhi-Hua, WU Hao, CHEN Jin-Xiang, “Measurement of Cross Sections for the $^{10}$B(n,$\alpha$)$^7$Li Reaction at 4.0 and 5.0MeV Using an Asymmetrical Twin Gridded Ionization Chamber,\qrs \textsc{Chin. Phys. Lett. }\textbf{28}, 082801 (2011).

\bibitem{Stavisskii:1961}  Ju. Ja. Stavisskii and V.A. Tolstikov, \ql Radiative neutron cross-section for several isotopes in the energy range 0.03--2.5~MeV,\qrs \textsc{Atomn. En. }\textbf{10}, 508 (1961) (in Russian).

\bibitem{Tolstikov:1963}   V.A. Tolstikov, L.E. Sherman, and Ju.Ja. Stavisskii, \ql A measurement of the capture cross sections of \UF~and \TH~for 5--200 keV neutrons,\qrs \textsc{Atomn. En. }\textbf{15}, 1170 (1963) (in Russian).

\bibitem{Wisshak:2001}  K. Wisshak, F. Voss, and F. Kaeppeler, \ql Neutron Capture Cross Section of Th-232,\qrs \textsc{Nucl. Sci. Eng. }\textbf{137}, 183 (2001).

\bibitem{Borella:2006}   A. Borella, K. Volev, A. Brusegan \etals, \ql \TH($n$,$\gamma$) cross section from 4~keV to 140~keV,\qrs \textsc{Nucl. Sci. Eng. }\textbf{152}, 1 (2006).

\bibitem{Aerts:2006}  G. Aerts, U. Abbondanno, H. Alvarez \etal(n\_TOF Collaboration), \ql Neutron capture cross section of Th-232 measured at the n\_TOF facility at CERN in the unresolved resonance region up to 1~MeV,\qrs \textsc{Phys. Rev. }\textbf{C73}, 054610 (2006).

\bibitem{Sirakov:2008} I. Sirakov, R. Capote, F. Gunsing, P. Schillebeeckx, and A. Trkov, \ql An ENDF-6 compatible evaluation for neutron induced reactions of \TH~in the unresolved resonance region,\qr \textsc{Ann. Nucl. En. }\textbf{35}, 1223--1231 (2008).

\bibitem{IAEA-th-eval:2010} R. Capote, L. Leal, Liu Ping \etals,
\ql Evaluated nuclear data for nuclides within the thorium-uranium fuel cycle,\qr Technical report \textbf{STI/PUB/1435} (Vienna, International Atomic Energy Agency, 2010), ISBN 978–92–0–101010–0.

%% From Toni
\bibitem{PWD19} S. Pavetich {\it et~al.}, \ql AMS measurements of the reaction $^{35}$Cl($n$,$\gamma$)$^{36}$Cl\qr, \textsc{Phys. Rev. } {\bf C99}, 015801 (2019).

\bibitem{DHK09} I. Dillmann {\it et~al.}, \ql Determination of the stellar ($n$,$\gamma$) cross section of $^{40}$Ca with accelerator mass spectrometry\qrs, \textsc{Phys. Rev. }\textbf{C79}, 065805 (2009).
\bibitem{WBB17} A. Wallner {\it et~al.}, \ql Precise measurement of the thermal and stellar $^{54}$Fe(n,$\gamma $)$^{55}$Fe cross sections via AMS,\qr \textsc{Phys. Rev. }\textbf{C96}, 025808 (2017).

\bibitem{LRD17} P. Ludwig {\it et~al.}, \ql Measurement of the stellar $^{58}$Ni(n,$\gamma $)$^{59}$Ni cross section with accelerator mass spectrometry,\qr \textsc{Phys. Rev. }\textbf{C95}, 035803 (2017).

%% [1] Gy. Gyürky, Zs. Fülöp, F. Käppeler, G.G. Kiss, and A. Wallner, “The activation method for cross section measurements in nuclear astrophysics”, Europ. Phys. Journal A 55 (2019) 41.
% \bibitem{GFK19} Gy. Gy{\"u}rky {\it et~al.}, \textsc{Eur. Phys. J. }{\bf A55}, 41 (2019).

%% TOF:
\bibitem{35Cl-TOF} K.H. Guber, R.O. Sayer, T.E. Valentine \etals, \ql New Maxwellian averaged neutron capture cross-sections for Cl-35,37,\qr \textsc{Phys. Rev. }\textbf{C65}, 058801 (2002).

\bibitem{40Ca-TOF} A. de L. Musgrove \etals, \ql Resonant neutron capture in $^{40}$Ca,\qr \textsc{Nucl. Phys. } \textbf{A259}, 365 (1976);
A. de L. Musgrove \etals, \ql Odd-even effects in radiative neutron capture by $^{42}$Ca, $^{43}$Ca and $^{44}$Ca,\qr \textsc{Nucl. Phys. } \textbf{A279}, 317 (1977).

\bibitem{54Fe-TOF} G. Giubrone (the n\_TOF Collaboration), \textsc{Ph.D. thesis, Universidad de Valencia}, Spain (2014).

\bibitem{58Ni-TOF1} K.H. Guber, H. Derrien, L. Leal \etals, \ql Astrophysical reaction rates for Ni-58, Ni-60 (n, gamma) from new neutron capture cross section measurements,\qr \textsc{Phys. Rev. }\textbf{C82},  057601 (2010).

\bibitem{58Ni-TOF2} P. Zugec \etal (The n\_TOF collaboration), \ql Experimental neutron capture data of $^{58}$Ni from the CERN n\_TOF facility,\qr \textsc{Phys. Rev. }\textbf{C89}, 014605 (2014).

\bibitem{Croft:2019}
S.~Croft, A.~Favalli and R.D.~McElro, Jr., \ql A Review of the Prompt Neutron Nu-bar value for $^{252}$Cf Spontaneous Fission,\qr  {\sc Nucl. Instr. Meth. Phys. Res. }\textbf{A723}, accepted (2019).

\bibitem{BoldemanProc}
J.W.~Boldeman, \ql Review of $\overline{\nu}$ for $^{252}$Cf and Thermal Neutron Fission,\qr \textit{Proc. Neutron Standards and Applications Sympos. }, Gaithersburg, MD, USA, 182--192 (1977).

\bibitem{AxtonProc}
E.J.~Axton, \ql Accuracies and Correction in the Neutron Bath Techniques,\qr \textit{Proc. Neutron Standards and Applications Sympos. }, Gaithersburg, MD, USA, p. 237--243 (1977).

\bibitem{Aleksandrov}
B.M.~Aleksandrov, E.V.~Korolev, Ya.M.~Kramaro \etal, \ql Absolute measurements of NU(Cf-252) by means of manganese bath method,\qr \textit{All Union Conf.on Neutron Phys. }, Kiev, 15--19 Sep.  1980, Vol.~4, p.119 (1980).

\bibitem{Asplund}
I.~Asplund-Nilsson, H.~Conde and N.~Starfelt, \ql An absolute measurement of nu-bar of Cf-252,\qr {\sc Nucl. Sc. Eng. }{\bf 16}, 124 (1963).

\bibitem{AxtonExp}
E.J.~Axton and A.G.~Bardell, \ql Neutron yield from the spontaneous fission of Cf-252(nu),\qr {\sc Metrol. }{\bf 21}, 59 (1985).

\bibitem{Bozorgmanesh}
H.~Bozorgmanesh and G.F~Knoll, \ql Absolute Measurement of the Number of Neutrons per Spontaneous Fission of \CF,\qr {\sc Trans. Amer. Nucl. Soc. }{\bf 27}, 864 (1977).

\bibitem{DeVolpi}
A.~DeVolpi and K.G.~Porges, \ql Neutron yield of \CF~based on absolute measurement of the neutron rate and fission rate,\qr {\sc Phys. Rev. } {\bf C1}, 683 (1970).

\bibitem{Colvin}
D.W.~Colvin and M.G.~Sowerby, \ql Boron pile nu-bar measurements,\qr \textit{IAEA Phys. Chem. Fission Conf. }, Salzburg 1965, Vol.~2, p. 25 (1965).

\bibitem{Diven}
B.C.~Diven and J.C.~Hopkins, \ql Numbers of Prompt Neutrons per Fission for U233, U235, Pu239 and Cf252,\qr \textsc{Reactor Physics Sem. }, Vienna 1961, Vol. 1, p. 149 (1961).

\bibitem{Edwards}
G.~Edwards, D.J.S.~Findlay and E.W.~Lees, \ql Measurements Of Prompt nu-Bar and Variance for the Spontaneous Fission of Cf-252 and Pu-242,\qr {\sc Ann. Nucl. En. }{\bf 9}, 127 (1982).

\bibitem{Hopkins}
J.C.~Hopkins and B.C.~Diven, \ql Prompt neutrons from fission,\qr {\sc Nucl. Phys. }{\bf 48}, 433 (1963).

\bibitem{Smith:1984}
J.R.~Smith, S.D.~ Reeder and R.J.~Gehrke, \ql Absolute measurement of nu-bar for Cf-252,\qr \textsc{Electric Power Res. Inst., Nucl. Phys. Ser. }\textbf{3436}, Vol.(1) (1984).

\bibitem{Spencer}
R.R.~Spencer, R.~Gwin and R.~Ingle, \ql A Measurement of the Average Number of Prompt Neutrons from Spontaneous Fission of Californium-252,\qr {\sc Nucl. Sc. Eng. }{\bf 80}, 603 (1982).

\bibitem{White}
P.H.~White and E.J.~Axton, \ql Measurement of the number of neutrons per fission for Cf-252,\qr {\sc J. Nucl. En. }{\bf 22}, 73 (1968).

\bibitem{Huan-Qiao}
Z.~Huan-Qiao and L.~Zu-Huz, \ql The measurement of the average number of prompt neutrons and the distribution of prompt neutron numbers for Cf-252 spontaneous fission,\qr {\sc Chin.  J. Nucl. Phys. }{\bf 1}, 9 (1979).

%%%% Not used yet below
%Kadonis 1.0:
%\bibitem{Kadonis10} I. Dillmann and R. Plag, \ql KADoNiS: The Karlsruhe Astrophysical Database of Nucleosynthesis in Stars,\qrs online at \url{https://exp-astro.de/kadonis1.0/}.

\end{thebibliography}
\end{document}